\documentclass[useAMS,usenatbib]{mnras}
\usepackage{natbib}
\usepackage{amsmath}
\usepackage{amssymb}
\usepackage{graphicx}
\usepackage{multirow}
\usepackage{url}
\usepackage{enumitem}
\usepackage[table]{xcolor}
\usepackage{bold-extra}
\numberwithin{equation}{section}

\title[Local stellar radiation and dust depletion]{The effects of local stellar radiation and dust depletion on non-equilibrium interstellar chemistry}
\author[A. J. Richings et al.]{Alexander J. Richings$^{1}$\thanks{Email: alexander.j.richings@durham.ac.uk}, Claude-Andr\'{e} Faucher-Gigu\`{e}re$^{2}$, Alexander B. Gurvich$^{2}$, Joop Schaye$^{3}$, \newauthor 
and Christopher C. Hayward$^{4}$ \\
$^{1}$Institute for Computational Cosmology, Department of Physics, Durham University, South Road, Durham, DH1 3LE, United Kingdom\\
$^{2}$Department of Physics \& Astronomy and CIERA, Northwestern University, 1800 Sherman Ave, Evanston, IL 60201, USA\\
$^{3}$Leiden Observatory, Leiden University, PO Box 9513, NL-2300 RA Leiden, The Netherlands\\
$^{4}$Center for Computational Astrophysics, Flatiron Institute, 162 Fifth Avenue, New York, NY 10010, USA}

\begin{document}

%\date{\today}

\date{Accepted 2022 August 16. Received 2022 August 15; in original form 2022 August 03}

\pagerange{\pageref{firstpage}--\pageref{lastpage}} \pubyear{2022}

\maketitle

\label{firstpage}

\begin{abstract} 
Interstellar chemistry is important for galaxy formation, as it determines the rate at which gas can cool, and enables us to make predictions for observable spectroscopic lines from ions and molecules. We explore two central aspects of modelling the chemistry of the interstellar medium (ISM): (1) the effects of local stellar radiation, which ionises and heats the gas, and (2) the depletion of metals onto dust grains, which reduces the abundance of metals in the gas phase. We run high-resolution (400~M$_{\odot}$ per baryonic particle) simulations of isolated disc galaxies, from dwarfs to Milky Way-mass, using the \textsc{fire} galaxy formation models together with the \textsc{chimes} non-equilibrium chemistry and cooling module. In our fiducial model, we couple the chemistry to the stellar fluxes calculated from star particles using an approximate radiative transfer scheme, and we implement an empirical density-dependent prescription for metal depletion. For comparison, we also run simulations with a spatially uniform radiation field, and without metal depletion. Our fiducial model broadly reproduces observed trends in H\textsc{i} and H$_{2}$ mass with stellar mass, and in line luminosity versus star formation rate for [C\textsc{ii}]$_{158 \rm{\mu m}}$, [O\textsc{i}]$_{63 \rm{\mu m}}$, [O\textsc{iii}]$_{88 \rm{\mu m}}$, [N\textsc{ii}]$_{122 \rm{\mu m}}$ and H$\alpha_{6563 \text{\AA}}$. Our simulations with a uniform radiation field predict fainter luminosities, by up to an order of magnitude for [O\textsc{iii}]$_{88 \rm{\mu m}}$ and H$\alpha_{6563 \text{\AA}}$, while ignoring metal depletion increases the luminosity of carbon and oxygen lines by a factor $\approx \! 2$. However, the overall evolution of the galaxy is not strongly affected by local stellar fluxes or metal depletion, except in dwarf galaxies where the inclusion of local fluxes leads to weaker outflows and hence higher gas fractions. 
\end{abstract}

\begin{keywords}
    astrochemistry -- ISM: atoms -- ISM: molecules -- galaxies: evolution -- galaxies: ISM 
\end{keywords}

\section{Introduction}\label{intro_sect} 

The chemistry of ions and molecules in interstellar gas plays a vital role in galaxy formation. The rate at which gas can cool depends on the relative abundances of chemical species, as different species radiate away the thermal energy at different rates due to transitions between their excited states. Radiative cooling enables gas to condense onto dark matter halos and trigger the formation of stars and galaxies \citep[e.g.][]{rees77, white78, white91}, while heating due to photoionisation from the UV background can suppress galaxy formation at low masses after reionisation \citep[e.g.][]{efstathiou92, fauchergiguere11, benitezllambay17, benitezllambay20}. 

Furthermore, there is a huge wealth of spectroscopic observations that identify ions and molecules through their emission and absorption lines. Such observations probe a wide range of phases of interstellar gas, including cold, dense molecular clouds \citep[e.g.][]{leroy09, saintonge17, rosolowsky21}, gas ionised by star-forming regions and/or a central Active Galactic Nucleus (AGN; \citealt{baldwin81, kauffmann03, kewley06}), and diffuse, highly ionised plasmas in the Circum-Galactic Medium (CGM; \citealt{tripp08, tumlinson11, turner14, burchett19}). By studying the chemistry, we can connect these observations to the conditions of the gas that they trace, which is crucial for understanding the physical mechanisms that drive the formation and evolution of galaxies. 

Large-scale cosmological simulations of the Universe often treat gas cooling using pre-computed tables of the cooling rate that depend on, for example, temperature, density, metallicity and redshift. When tabulating the cooling rate in this way, it is common to assume that the chemical reactions have reached equilibrium, either for a collisionally ionised plasma (Collisional Ionisation Equilibrium, CIE; \citealt{cox69, sutherland93}), or under the influence of photoionisation from a background UV radiation field (Photo-Ionisation Equilibrium, PIE; \mbox{\citealt{wiersma09}}; \citealt{gnedin12, ploeckinger20}). This approach of using pre-computed cooling tables has been applied in many state-of-the-art cosmological simulations \citep[e.g.][]{dubois14, schaye15, tremmel17, pillepich18, lee21}. 

To connect these hydrodynamic simulations to observations, we can create synthetic spectra of emission and absorption lines from the simulation outputs in post-processing if we again assume that the chemical abundances are in equilibrium. These abundances can be computed either using the temperatures and densities of each gas particle or cell directly \citep[e.g][]{hummels17, katz19, nelson21, oppenheimer20, wijers20}, or using subgrid models that capture unresolved features important for the observational tracers \citep[e.g][]{rahmati13a, vallini13, narayanan14, hirschmann17, olsen21, tan21}. Subgrid approaches can also be used to predict observable line emission in semi-analytic models of galaxy formation \citep[e.g][]{lagos12, popping19, baugh22}. 

The above methods for modelling gas cooling and synthetic observations in simulations of galaxy formation all rely on the assumption that the chemistry has had sufficient time to reach equilibrium. While this is reasonable in many cases, it is not applicable in scenarios where the gas is evolving rapidly, for example when the cooling time is short \citep{sutherland93, gnat07, oppenheimer13a}, in the presence of turbulence \citep{gray17}, or if the UV radiation field is fluctuating \citep{oppenheimer13b, segers17, oppenheimer18}. 

To capture such non-equilibrium effects in hydrodynamic simulations, we need to follow the time-dependent evolution of ions and molecules using a chemical reaction network, which integrates the rate equations, together with the resulting cooling and heating rates that determine the temperature evolution, for each gas particle or cell. Several astrochemistry codes have been developed for this purpose in recent years. The \textsc{krome} package implements chemical networks that include hydrogen, deuterium, helium, and low-ionisation metal species at temperatures $< \! 10^{4} \, \rm{K}$ \citep{grassi14, bovino16}, and has been widely applied to hydrodynamic simulations on galactic scales (e.g. \mbox{\citealt{lupi18, lupi20}}; \citealt{sillero21}). The \textsc{grackle} library follows the non-equilibrium chemistry of hydrogen, deuterium and helium species \citep{smith17}, and has been applied to cosmological simulations such as the \textsc{agora} project \citep{kim14} and \textsc{simba} \citep{dave19}. Several studies have also explored molecular networks that include the formation and destruction of CO \citep[e.g][]{nelson97, glover10, glover12, richings14a, richings14b}, which have been applied to simulations of the turbulent Interstellar Medium (ISM) and molecular clouds \citep[e.g][]{walch15, seifried17, smith20, hu21} and whole galaxies \citep[e.g][]{hu16, richings16}. 

The main disadvantage of non-equilibrium chemical models is the high computational cost, which limits the complexity of the reaction network that can be included and/or the size and resolution of the simulations to which they can be applied. Nevertheless, ongoing advances in this field are producing faster chemistry codes, for example through algorithms that reduce the complexity of the chemical network \citep[e.g][]{tupper02, grassi12, grassi21}, or using neural networks to emulate the full time-dependent calculation \citep{holdship21}. 

When coupling a hydrodynamic simulation of galaxy formation to a chemical reaction network, one important aspect to consider is the UV radiation field, which ionises the gas and dissociates molecules, as well as providing a crucial source of heating. In the Inter-Galactic Medium (IGM), the radiation field is typically dominated by an extragalactic background, consisting of contributions from quasars and star-forming galaxies throughout the Universe \citep[e.g][]{haardt12, fauchergiguere20}. However, in the ISM regime local sources of radiation such as young stars become important \citep[e.g][]{mathis83, schaye06, rahmati13b}. We also need to consider how dense gas becomes shielded from radiation \citep[e.g.][]{federman79, vandishoeck86, visser09, wolfire10, fumagalli11, wolcottgreen11, rahmati13a}. 

Modelling the spatial variations of the radiation field within the ISM of a galaxy requires a treatment of the 3D radiative transfer of ionising and dissociating radiation. Radiative transfer codes also incorporate non-equilibrium chemical networks to capture the interaction between the radiation and gas chemistry \citep[e.g][]{pawlik11, rosdahl13, kannan19, chan21a, katz22}, but solving the full radiative transfer equations in this way adds additional computational expense, on top of the cost of the chemical network itself. Approximate methods have therefore been developed to account for the radiation from local sources and/or gas self-shielding \citep[e.g][]{clark12, richings14b, safranekshrader17, hopkins18a, ploeckinger20}. 

Dust grains are another important aspect of the thermo-chemistry, as they shield the gas from UV radiation and catalyse the formation of molecules such as H$_{2}$ on grain surfaces, as well as providing heating and cooling channels such as photoelectric heating. Dust also depletes metals from the gas phase \citep[e.g][]{jenkins09, decia16}, as metals that are locked up in dust grains cannot participate in gas-phase chemical reactions and thermal processes. 

The simplest approach is to assume the dust abundance scales linearly with the overall metallicity, however observations suggest that dust to metal ratios may not be constant \citep{remyruyer14, delooze20}. Alternatively, models have recently been developed to follow the formation and destruction of dust grains in hydrodynamic simulations \citep[e.g.][]{bekki15, mckinnon18, choban22}. 

In this work we couple the \textsc{chimes} non-equilibrium chemistry module \citep{richings14a, richings14b} to hydrodynamic simulations of isolated galaxies using the \textsc{fire-2} subgrid galaxy formation models \citep{hopkins18a}, to study the effects of local UV sources and dust depletion on the interstellar chemistry and their impact on the overall galaxy evolution and observable tracers of the ISM. 

The \textsc{chimes} reaction network covers a wide range of gas phases from cold ($\sim$$10 \, \rm{K}$), dense molecular clouds to hot ($\sim$$10^{9} \, \rm{K}$), highly ionised plasmas, and captures non-equilibrium effects in most of the chemical species (including metal ions) commonly detected in spectroscopic observations. This will enable us to confront our simulations with a great variety of observational data sets. 

The \textsc{fire-2} galaxy formation models have been developed to implement un-resolved physical processes that typically are not explicitly captured in hydrodynamic simulations, such as the formation of stars and the subsequent feedback of energy and momentum via supernovae, stellar winds, photoionisation of the surrounding gas, and stellar radiation pressure. By following individual feedback channels in this way, the \textsc{fire-2} models produce a realistic multiphase ISM down to scales of Giant Molecular Clouds and star-forming regions, which will be crucial for this work. Applied to cosmological simulations, the \textsc{fire} models have been shown to reproduce many properties of observed galaxy populations at both low and high redshift, including the mass-metallicity \citep{ma16}, stellar mass-halo mass \citep{hopkins18a}, size-kinematics \citep{elbadry18}, and Kennicutt Schmidt \citep{orr18} relations. 

The remainder of this paper is organised as follows. In Section~\ref{methods_sect} we describe our methods, including a summary of the \textsc{fire}-2 subgrid models (Section~\ref{physics_sect}) and \textsc{chimes} (Section~\ref{chemistry_sect}). We introduce the isolated galaxy simulations in Section~\ref{sim_sect}, including a description of the initial conditions (Section~\ref{IC_sect}) and results for the morphology and evolution of the simulated galaxies (Section~\ref{morph_sect}), the stellar fluxes predicted by our models (Section~\ref{sim_flux_sect}), and the dust properties (Section~\ref{dust_properties_sect}). In Section~\ref{HI_H2_sect} we explore the transition from atomic to molecular gas in our simulations and compare to observations, while in Section~\ref{emission_line_sect} we study emission line tracers of the total star formation rate. We summarise our conclusions in Section~\ref{conclusions_sect}. In Appendix~\ref{esc_fraction_sect} we summarise how we calibrated the escape fraction parameters from individual H\textsc{ii} regions, and we present a study of the numerical convergence of our results in Appendix~\ref{resolution_sect}. 

\section{Methods}\label{methods_sect}

The simulations in this paper were run with the gravity and hydrodynamics code \textsc{gizmo} \citep{hopkins15}, using the Lagrangian Meshless Finite Mass (MFM) method to solve the hydrodynamics equations. We include subgrid models for the physical processes relevant to galaxy formation that are not explicitly resolved. These are mostly based on the \textsc{fire}-2 simulation models from the \textsc{fire} project\footnote{See the project website at \url{http://fire.northwestern.edu}} \citep{hopkins18a}, as detailed in Section~\ref{physics_sect} below, except for radiative cooling, which we describe in Section~\ref{chemistry_sect}.

\subsection{\textsc{fire}-2 subgrid physics models}\label{physics_sect}

Gas particles can be turned into stars if they are above a density threshold of $n_{\rm{H}} = 10^{3} \, \rm{cm}^{-3}$ and are locally self-gravitating and Jeans-unstable. If these criteria are met for a given particle, it is turned into a star particle stochastically at a rate given by the particle mass over the free-fall time. The details of the star formation algorithm are described in appendix~C of \citet{hopkins18a}. Note that, unlike the default \textsc{fire}-2 model, we do not scale the star formation rate by the fraction of gas that is molecular/shielded. In \citet{hopkins18a}, this fraction is calculated according to the analytic approximation from \citet{krumholz11}. Our simulations follow the time-dependent molecular chemistry, but if we used this molecular fraction in the star formation model it may introduce additional time-dependent effects in the star formation rate. For example, when a gas cloud cools there will be a lag before it becomes fully molecular as it takes time for molecules to form. However, molecules may not necessarily be required before star formation can proceed \citep{glover12}, so we would not expect a corresponding lag in the star formation rate. Nevertheless, \citet{hopkins18a} found that the molecular/self-shielded criterion has little effect in the \textsc{fire}-2 model, as gas that meets the other criteria will typically be fully molecular anyway, so we omit this criterion altogether in this work.

Star particles inject energy, momentum and mass via stellar feedback as follows. The rate of type Ia supernovae (SNe) is calculated according to \citet{mannucci06} for both prompt and delayed populations, while the rates of type II SNe and stellar mass loss from OB and AGB winds are obtained from simple fits to the stellar evolution models from \textsc{starburst99} \citep{leitherer99} with a \citet{kroupa01} initial mass function (IMF; see appendix~A of \citealt{hopkins18a}). The SNe and stellar winds are implemented using a mechanical feedback scheme described in appendix~D of \citet{hopkins18a} (see also \mbox{\citealt{hopkins18b}}). Radiation pressure is coupled to the gas using the \textsc{lebron} approximate radiative transport algorithm, in which extinction of the stellar radiation is assumed to occur locally around the emitting star particle and absorbing gas particle, and is then transported between the two under the optically thin approximation. This algorithm is described in appendix~E of \citet{hopkins18a}; see also \citet{hopkins19}. Note that, in this work, we treat photoheating and H\textsc{ii} regions differently from the fiducial \textsc{fire}-2 model as they are coupled to the \textsc{chimes} chemistry solver; see Section~\ref{flux_sect} for details.

The \textsc{fire-2} models also track the enrichment and evolution of the 11 elements that are used in the \textsc{chimes} chemistry network (see Section~\ref{chemistry_sect}). Star particles inject metals via SNe and stellar mass loss, with type Ia SNe yields from \citet{iwamoto99}, type II SNe yields from \citet{nomoto06}, and OB/AGB stellar wind yields from \citet{vandenhoek97}, \citet{marigo01} and \citet{izzard04}. These yields are summarised in appendix~A of \citet{hopkins18a}. The turbulent diffusion of metal elements between gas particles is modelled as described in appendix~F3 of \citet{hopkins18a}; see also \citet{hopkins17}. 

\subsection{Non-equilibrium chemistry and cooling}\label{chemistry_sect}

We follow the non-equilibrium evolution of 157 ions and molecules important for gas cooling using the \textsc{chimes} chemistry and cooling module\footnote{\url{https://richings.bitbucket.io/chimes/home.html}} \citep{richings14a, richings14b}. This includes all ionisation states of H, He, C, N, O, Ne, Mg, Si, S, Ca and Fe; the negative ions H$^{-}$, C$^{-}$ and O$^{-}$; and the molecules H$_{2}$, H$_{2}^{+}$, H$_{3}^{+}$, OH, OH$^{+}$, H${_2}$O, H$_{2}$O$^{+}$, H$_{3}$O$^{+}$, O$_{2}$, O$_{2}^{+}$, C$_{2}$, CH, CH$_{2}$, CH$_{3}^{+}$, CH$^{+}$, CH$_{2}^{+}$, CO, CO$^{+}$, HCO$^{+}$, HOC$^{+}$.

Appendix~B of \citet{richings14a} contains a complete list of the chemical reactions in \textsc{chimes}, which includes collisional ionisation, recombination (in the gas phase and on the surface of dust grains), charge transfer, photoionisation (including Auger ionisation) and photodissociation, cosmic ray ionisation and dissociation, molecular hydrogen formation on dust grains, and gas-phase molecular creation and destruction channels.

The photochemistry reactions require the total UV intensity, which we calculate using the redshift zero extragalactic UV background \citep{fauchergiguere20} plus the local radiation from star particles in the galaxy, along with a local treatment for self-shielding. The methods used to compute the UV fluxes are described in Section~\ref{flux_sect}.

We assume a constant primary ionisation rate of H\textsc{i} due to cosmic rays of $\zeta_{\rm{HI}} = 1.8 \times 10^{-16} \, \rm{s}^{-1}$, which corresponds to the value in the Milky Way as inferred from observations of H$_{3}^{+}$ \citep{indriolo12}. The ionisation and dissociation rates of other species due to cosmic rays are then scaled relative to this value as described in section~2.3 of \citet{richings14a}. Since cosmic rays are produced in sites of active star formation via acceleration in supernova remnants, we might expect the cosmic ray rate to scale with the local star formation rate \citep[e.g.][]{ploeckinger20}. However, depending on the cosmic ray transport mechanism, the cosmic rays may rapidly escape from dense regions and produce a smoother distribution than predicted by such a scaling. As the aim of this work is to explore the effects of local stellar radiation, and given the uncertainties in how the cosmic ray ionisation rates will depend on local properties, we therefore decided to use a constant, uniform rate, rather than coupling it to the stellar fluxes. With this approach, we can be confident that the effects of the local fluxes will be driven by the UV radiation and not by variations in the cosmic ray rate. In a future work, we will explore coupling the cosmic ray ionisation rates in \textsc{chimes} to methods that follow the production and transport of cosmic rays in galaxy formation simulations \citep[e.g.][]{chan19, chan21b, hopkins21}. 

The reactions on the surface of dust grains utilise a density- and temperature-dependent dust abundance using an empirical model based on observed metal depletion factors. This model is also used to deplete the elemental abundances of metals from the gas phase, as any metals that are located in dust grains will be unavailable for the gas phase chemical reactions. The dust model is described in Section~\ref{depl_sect}.

The resulting ion and molecule abundances are used to calculate the radiative cooling and heating rates. The thermal processes included in \textsc{chimes} are summarised in table~1 of \citet{richings14a}, although the rate of photoelectric heating from dust grains has been updated to use equations 19 and 20 from \citet{wolfire03}. 

We thus obtain a set of coupled ordinary differential equations (ODEs) for the rate equations and the thermal energy equation, which we integrate for each particle over each hydrodynamic time-step. To accelerate the integration of these ODEs, we first calculate the solution from the explicit forward Euler method. If the relative change in the thermal energy and chemical abundances is less than 0.05 (excluding species with an abundance below $10^{-10}$, which are negligible), then we take the explicit solution. Otherwise, we integrate the ODEs using the implicit backward difference formula method and Newton iteration, as implemented in the \textsc{cvode} library from the \textsc{sundials}\footnote{\url{https://computing.llnl.gov/projects/sundials}} suite of differential and algebraic equation solvers, with a relative tolerance of $10^{-4}$ and an absolute tolerance of $10^{-10}$. We find that using the explicit solution in this way does not affect our results, but allows us to avoid the more expensive implicit solver for particles that are either close to equilibrium or are evolving very slowly.

We also include the turbulent diffusion of ions and molecules between gas particles, using the same subgrid model described in appendix~F3 of \citet{hopkins18a} for the diffusion of elemental abundances but applied to each species in the \textsc{chimes} network. 

\subsubsection{Local stellar fluxes}\label{flux_sect}

We follow the propagation of radiation from star particles in the simulation using an approximate radiative transport method based on the \textsc{lebron} algorithm used to model stellar radiation pressure in \textsc{fire} \citep{hopkins18a, hopkins19}. This approximation assumes that the radiation is only absorbed locally around the star particle and the receiving gas particle. The subsequent transport of radiation from the star to the gas particle is then treated in the optically thin limit, which allows us to utilise the gravity solver to propagate the radiation between particles. In this work, we implement radiation pressure using the standard \textsc{lebron} method as in \textsc{fire}. However, when coupling the radiation to \textsc{chimes}, we modify the method as follows.

Firstly, the standard \textsc{lebron} method for radiation pressure tracks three stellar fluxes from all star particles, in the infrared, optical and ultraviolet bands. However, for the photochemistry we track the radiation in eight separate stellar age bins, which allows us to accurately capture the age-dependence of the UV spectra. The age bins are spaced logarithmically by 0.2 dex up to 10~Myr and by 1.0 dex above 10~Myr. Stars with an age $<$1~Myr are placed into a single bin, as are stars with an age $>$100~Myr. We further divide the radiation from each age bin into the non-ionising Far-UV band (FUV; 6$-$13.6~eV) and the ionising Extreme-UV band (EUV; $>$13.6~eV). This gives us 16 stellar fluxes in total for the photochemistry (the optical and infrared bands are not required for the photochemical reactions). 

\begin{figure}
\centering
\mbox{
	\includegraphics[width=84mm]{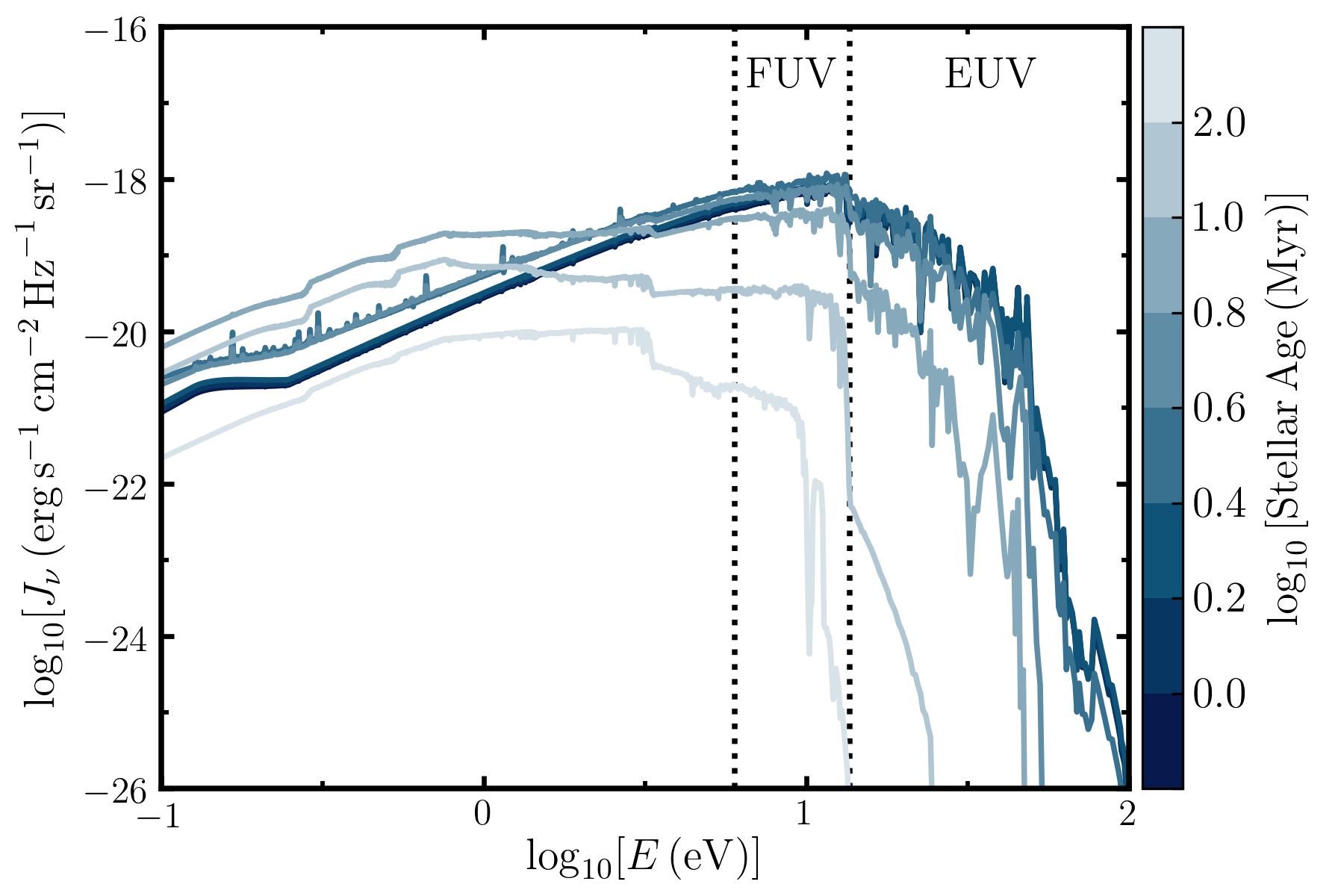}}
\vspace{-0.15 in}
\caption{Stellar spectra in the eight stellar age bins used to track the fluxes from star particles, calculated using \textsc{starburst99} stellar evolution models. These spectra are used to compute the average photoionisation cross sections in each age bin.
\vspace{-0.15 in}} 
\label{SB99_fig}
\end{figure}

We calculate the average cross sections of the photochemical reactions for each stellar age bin using spectra from \textsc{starburst99} models \citep{leitherer99} with a \citet{kroupa01} IMF, the Geneva stellar evolution tracks with a rotation velocity of 0.4 times the break-up velocity, and a metallicity $Z = 0.014$. These are the same models that were used for the rates of type II SNe and mass loss from OB and AGB winds in \textsc{fire}. By including only models at a fixed metallicity we do not account for the metallicity dependence of the stellar spectra, as this would require us to track even more fluxes. However, this will only affect the dwarf galaxies in our sample, as the galaxies at higher masses are close to solar\footnote{Throughout this paper we use the solar elemental abundances from Table~1 of \citet{wiersma09}, where the total solar metallicity is $\rm{Z}_{\odot} = 0.0129$, unless stated otherwise. However, the \textsc{starburst99} models used the \citep{asplund09} solar abundances.} metallicity (see Section~\ref{IC_sect}). Fig.~\ref{SB99_fig} shows the \textsc{starburst99} spectra for each age bin.

The luminosity of each star particle in the FUV and EUV bands, $L_{i}(t)$, are calculated as a function of stellar age, $t$, and current mass, $M_{\ast}(t)$ (accounting for mass loss). We use an analytic fitting function, which we fit to the outputs from the \textsc{starburst99} models with a spacing of 0.01~Myr. This fitting function is given in equation \ref{fuv_euv_eqn} below, and the fit parameters are shown in Table~\ref{SB99_fit_pars}.

\begin{equation}\label{fuv_euv_eqn} 
  \frac{L_{i}(t)}{\rm{photon} \, \rm{s}^{-1}} =        
  \begin{cases}
    \left( \frac{M_{\ast}(t)}{\rm{M}_{\odot}} \right) \! \exp \! \left[p_{1} \! + \! p_{2} \left( \frac{t}{\rm{Myr}} \right)^{p_{3}} \right] & \! t \! < \! 3.7 \, \rm{Myr} \\ 
    p_{4} \left( \frac{M_{\ast}(t)}{\rm{M}_{\odot}} \right) \left( \frac{p_{5}}{t} \right)^{p_{6}} & \\
    \, \, \, \, \, \, \, \, \times \left[ 1 + \left( \frac{t}{p_{5}}\right)^{p_{7}} \right]^{p_{8}} & \rm{otherwise}. \\
  \end{cases}
\end{equation}

\begin{table}
\begin{minipage}{84mm}
\centering
\caption{Best fit parameters for the fitting functions to the FUV and EUV luminosities (see equation~\ref{fuv_euv_eqn}).}
\label{SB99_fit_pars}
\begin{tabular}{ccc}
  \hline
  Parameter & FUV & EUV \\
\hline
$p_{1}$ & 108.1 & 107.2 \\ 
$p_{2}$ & 0.17 & 0.11 \\ 
$p_{3}$ & 0.92 & 0.97 \\
$p_{4}$ & 6.4 $\times$ 10$^{37}$ & 3.3 $\times$ 10$^{21}$ \\ 
$p_{5}$ & 1.77 $\times$ 10$^{6}$ Myr & 6.89 $\times$ 10$^{5}$ Myr \\ 
$p_{6}$ & 1.67 & 4.79 \\ 
$p_{7}$ & 28.2 & 1.12 \\ 
$p_{8}$ & 1.65 & $-$1.7 $\times$ 10$^{4}$ \\ 
\hline
\end{tabular}
\end{minipage}
\end{table}

The stellar luminosities are then attenuated by local absorption around the star particle. In the fiducial \textsc{lebron} method used in \textsc{fire}, the infrared, optical and UV fluxes used for the radiation pressure are shielded due to dust based on a local estimate of the gas column density around the star particle. However, as the ionising radiation is also absorbed by H\textsc{i}, the \textsc{fire} model treats the effects of ionising radiation separately based on a Str\"{o}mgren argument, where neighbouring gas particles are flagged as H\textsc{ii} regions and ionised, starting with the nearest neighbour, until the available ionising photon budget from the star particle is used up (see Appendix~E of \citealt{hopkins18a} for details). H\textsc{ii} region particles are then prevented from cooling below 10$^{4} \, \rm{K}$. 

This approach assumes that all ionising radiation is confined within H\textsc{ii} regions. However for this work we are also interested in the diffuse ionising radiation that escapes these regions, and the effects it has on the interstellar chemistry. For the stellar fluxes used in the photochemistry calculations, we therefore modify the attenuation around star particles, in both the FUV and EUV bands, by introducing two free parameters, $f_{\rm{FUV}}^{\rm{esc}}$ and $f_{\rm{EUV}}^{\rm{esc}}$. These parameters represent the escape fractions of radiation in the FUV and EUV bands, respectively, from H\textsc{ii} regions. The stellar luminosities are then reduced by these fractions, before being propagated in the optically thin limit through the gravity tree to the receiving gas particles as described in Appendix~E of \citet{hopkins18a}.

We also identify neighbouring gas particles in the H\textsc{ii} region of each star particle, using the same Str\"{o}mgren method as in the fiducial \textsc{lebron} model. However, rather than impose a temperature floor of 10$^{4} \, \rm{K}$, we instead disable shielding in H\textsc{ii} region particles, so that they receive the full flux from the local star particle. This allows us to follow the non-equilibrium chemistry and temperature evolution of the H\textsc{ii} regions.

To determine the values of the escape fraction parameters, we note that, in our simulations, the median flux incident on each gas particle scales linearly with the star formation rate surface density averaged over the galaxy disc, albeit with a scatter of $\pm$0.5~dex (see Section~\ref{sim_flux_sect}), while the normalisation of this scaling depends on the assumed escape fraction. We therefore calibrated these parameters so that the scaling relation between the median stellar fluxes and disc-averaged star formation rate surface density reproduces the observed Milky Way FUV and EUV fluxes from \citet{black87} at the star formation rate surface density of the Milky Way. We thus find escape fractions of $f_{\rm{FUV}}^{\rm{esc}} = 0.1$ and $f_{\rm{EUV}}^{\rm{esc}} = 0.05$. The details of this calibration can be found in Appendix~\ref{esc_fraction_sect}.

\citet{diemer18} used a similar approach of calibrating the escape fraction based on the relation between UV flux and star formation rate surface density normalised to values in the Milky Way, which they used to model the atomic to molecular transition in the \textsc{illustris-tng} cosmological simulations in post-processing. Based on this calibration, they found an escape fraction at a wavelength of $1000 \, \rm{\AA}$ (in the Lyman-Werner band) of 0.1, which agrees with our value in the FUV band. 

There are many examples of numerical studies in the literature that have modelled the escape fraction of ionising and non-ionising radiation from H\textsc{ii} regions \citep[e.g.][]{dale12, howard17, rahner17, kim19}. They find that the escape fractions vary widely, from zero to nearly unity, depending on the age of the H\textsc{ii} region and local conditions such as density and the initial mass of the stellar birth cloud. While our model using constant escape fractions reproduces the observed strength of the diffuse FUV and EUV radiation field, based on constraints from the Milky Way, we do not capture variations in the escape fractions. In the future, this could be improved by developing a subgrid model for the escape fractions as a function of resolved properties of the H\textsc{ii} regions in the simulations.

Finally, for gas particles that do not lie within an H\textsc{ii} region the incident radiation is further attenuated due to shielding by the local gas cloud. We calculate the shielding length, $L_{\rm{sh}}$, based on a Sobolev-like approximation using the density gradient as follows:

\begin{equation}\label{sobolev_eqn}
  L_{\rm{sh}} = \frac{1}{2} \left( \frac{\rho}{\nabla \rho} + h_{\rm{inter}} \right),
\end{equation}
where $\rho$ is the gas density and $h_{\rm{inter}}$ is the mean inter-particle spacing. The first term accounts for the size of the resolved gas cloud around the particle, while the second term accounts for the size of the particle itself. The local column density of a given species, $i$, is given by $N_{i} = n_{i} L_{\rm{sh}}$, where $n_{i}$ is the density of species $i$. We then suppress the photochemical rates as a function of the local column densities of H\textsc{i}, H$_{2}$, He\textsc{i}, He\textsc{ii}, CO and dust, using the methods described in \citet{richings14b}.

Equation~\ref{sobolev_eqn} treats the shielding using a single, average shielding length (and hence column density) for the local gas cloud. However, as the photochemical rates are typically dominated by the low column density sight lines through the cloud, we caution that this approximation will tend to overestimate the shielding, which could lead to higher molecular abundances.

To study the impact of using a local treatment for stellar fluxes, we also repeat our simulations with a uniform Interstellar Radiation Field (ISRF). In these runs, we scale the normalisation of the radiation field by the star formation rate over the preceding 10~Myr averaged over the disc of the galaxy, $\Sigma_{\rm{SFR,} \, \rm{disc}}$. Throughout this paper we define the galaxy disc as a cylinder with a radius of $6 R_{\rm{exp}}$, where $R_{\rm{exp}}$ is the initial exponential scale radius of the stellar disc component, and extending to $\pm$1.2$R_{\rm{exp}}$ (i.e. 20 per cent of the radius of the cylinder) above and below the mid-plane. 

The flux incident on each gas particle in the FUV and EUV bands is then scaled from the Milky Way values as follows:
\begin{align}
  \mathcal{F}_{\rm{FUV,} \, \rm{uniform}} &= \mathcal{F}_{\rm{FUV,} \, \rm{MW}} \frac{\Sigma_{\rm{SFR,} \, \rm{disc}}}{\Sigma_{\rm{SFR,} \, \rm{MW}}}, \label{uni_fuv_flux} \\
  \mathcal{F}_{\rm{EUV,} \, \rm{uniform}} &= \mathcal{F}_{\rm{EUV,} \, \rm{MW}} \frac{\Sigma_{\rm{SFR,} \, \rm{disc}}}{\Sigma_{\rm{SFR,} \, \rm{MW}}}, \label{uni_euv_flux}
\end{align}
where $\mathcal{F}_{\rm{FUV,} \, \rm{MW}} = 1.7 \times 10^{8} \, \rm{photon} \, \rm{cm}^{-2} \, \rm{s}^{-1}$ and $\mathcal{F}_{\rm{EUV,} \, \rm{MW}} = 1.1 \times 10^{7} \, \rm{photon} \, \rm{cm}^{-2} \, \rm{s}^{-1}$ are the FUV and EUV fluxes in the local solar neighbourhood of the Milky Way, respectively \citep{black87}, and $\Sigma_{\rm{SFR,} \, \rm{MW}} = 4 \times 10^{-3} \, \rm{M}_{\odot} \, \rm{kpc}^{-2}$ is the star formation rate surface density in the Milky Way \citep[e.g.][]{robertson08}. 

This time-dependent radiation field is applied uniformly to all gas particles. Local shielding around the receiving gas particles is implemented using the Sobolev-like shielding length in equation~\ref{sobolev_eqn} as before. The shape of the UV spectrum is obtained by averaging the \textsc{starburst99} spectra shown in Fig.~\ref{SB99_fig}, assuming a constant star formation rate. We also disable the H\textsc{ii} region prescription in these runs, as this is a local effect of the stellar fluxes. 

\subsubsection{Depletion of metals onto dust grains}\label{depl_sect}

The non-equilibrium chemistry solver requires as input the gas-phase abundances of each element in the reaction network. The simulations track the total elemental abundances, but for some elements a fraction of the total abundance will be in dust grains and therefore will not contribute to the gas-phase chemistry. We therefore need to determine the fraction of each element in dust grains, and reduce the gas-phase abundances accordingly.

\citet{jenkins09} determined the fraction of metals that are depleted onto dust grains in the solar neighbourhood on an element-by-element basis by measuring the column densities of 17 metals and neutral hydrogen along 243 sight lines in the Milky Way (although not all sight lines include measurements for all elements). By assuming that the total metal abundances in the solar neighbourhood are at their solar values (for which they used the solar abundances from \citealt{lodders03}), they inferred that any discrepancies between the measured and solar abundances were due to depletion onto dust grains. They parameterised the overall strength of dust depletion along a given sight line according to a parameter $F_{\ast}$, which was normalised such that, in this Milky Way sample of sight lines, $F_{\ast}$ varied between values of 0 (the least depleted sight line, not including sight lines with neutral hydrogen column densities $<$10$^{19.5} \, \rm{cm}^{-2}$ which they exclude due to potential contamination from ionised hydrogen) and 1 (the $\zeta$ Oph sight line). The fraction of each individual element that remains in the gas phase can then be expressed as a linear function of $F_{\ast}$ as follows, from equation~10 of \citet{jenkins09}:

\begin{equation}\label{depl_eqn}
  \log_{10} [ M_{X}^{\rm{gas}} / M_{X}^{\rm{tot}} ] = B_{X} + A_{X} (F_{\ast} - z_{X}),
\end{equation}
where $M_{X}^{\rm{gas}}$ and $M_{X}^{\rm{tot}}$ are the gas-phase and total masses of element $X$, respectively. The best-fit linear coefficients $A_{X}$, $B_{X}$ and $z_{X}$ for each element are given in Table~4 of \citet{jenkins09}.

\citet{jenkins09} also showed that $F_{\ast}$ is closely correlated with the average neutral hydrogen density along the line of sight between the observer and the background source, $\langle n_{\rm{H}} \rangle$ (see the left-hand panel of Fig.~16 in \citealt{jenkins09}). They find the best-fit relation is: 

\begin{equation}\label{Fast_eqn} 
  F_{\ast} = 0.772 + 0.461 \log_{10} \langle n_{\rm{H}} \rangle.
\end{equation}

\citet{decia16} expanded on the results of \citet{jenkins09} by adding a sample of 70 damped Lyman-$\alpha$ absorbers (DLAs) observed in quasar spectra, in addition to the Milky Way sight lines. This allowed them to extend the linear fits of the depletion factors to $F_{\ast} < 0$ (i.e. systems with weaker overall dust depletion than seen in the solar neighbourhood).

For our simulations, we implement an empirical model for the depletion of metals onto dust grains based on these observations as follows. We use equation~\ref{Fast_eqn} to calculate the overall strength of dust depletion, $F_{\ast}$, for each gas particle as a function of density. We assume that the particle's total hydrogen density, $n_{\rm{H, \, tot}}$, is approximately equal to $\langle n_{\rm{H}} \rangle$ in equation~\ref{Fast_eqn}, which is the average neutral hydrogen density along the line of sight to the background source in the observations. However, this will overestimate the strength of depletion because observationally the true density at which the depletion occurs will tend to be higher than the average density along the line of sight. 

At high densities ($n_{\rm{H, \, tot}} \! > \! 3.12 \, \rm{cm}^{-3}$), we limit $F_{\ast}$ to be no greater than unity, corresponding to the strongest overall dust depletion strength observed in the Milky Way sight lines. It is possible that $F_{\ast}$ may exceed unity in dense environments, however it is uncertain how to extrapolate the observed relations to this regime. We also impose a temperature cut such that, above $10^{6} \, \rm{K}$, all metals are in the gas phase, as we expect dust grains will be rapidly destroyed via sputtering above this temperature \citep[e.g.][]{tsai95}.

\begin{table}
\begin{minipage}{84mm}
\centering
\caption{Linear fit coefficients used in equation~\ref{depl_eqn} for the depletion of metals onto dust grains.}
\label{depl_coeff}
\begin{tabular}{ccccc}
  \hline
  Element & $A_{X}$ & $B_{X}$ & $z_{X}$ & Ref.\footnote{References: J09 \citep{jenkins09}; DC16 \citep{decia16}.} \\
  \hline
  C & $-0.101$ & $-0.193$ & $0.803$ & J09 \\
  N & $0.0$ & $-0.109$ & $0.55$ & J09 \\
  O & $-0.101$ & $-0.02$ & $-1.50$ & DC16 \\ 
  Mg & $-0.412$ & $-0.03$ & $-1.50$ & DC16 \\ 
  Si & $-0.426$ & $-0.03$ & $-1.50$ & DC16 \\
  S & $-0.189$ & $-0.04$ & $-1.50$ & DC16 \\ 
  Fe & $-0.851$ & $-0.01$ & $-1.50$ & DC16 \\
\hline
\end{tabular}
\vspace{-0.2 in}
\end{minipage}
\end{table}

We then obtain the depletion factors of individual elements using equation~\ref{depl_eqn}, with linear fit coefficients derived from the fits of \citet{decia16} where available. They fit the depletion factors as a function of [Zn/Fe], which is related to $F_{\ast}$ by $F_{\ast} = 1.48 [\rm{Zn} / \rm{Fe}] - 1.50$, so we convert the fit coefficients reported in Table~3 of \citet{decia16} to the coefficients $A_{X}$, $B_{X}$ and $z_{X}$ used in equation~\ref{depl_eqn}. For elements not included in \citet{decia16}, we instead use the linear fits from \citet{jenkins09}. We summarise the fit coefficients used in this work in Table~\ref{depl_coeff}. The fits for some of these elements are uncertain due to limited observational data. For example, the \citet{jenkins09} sample contains only a handful of carbon depletion measurements based on weak-line transitions of C\textsc{ii}. However, \citet{sofia11} find that the gas-phase column densities of carbon measured from strong-line transitions of C\textsc{ii} are a factor $\approx$2 lower than those measured from weak-line transitions. This would result in stronger depletion of carbon than expected from these fits by a factor $\approx$2. Some elements in the \textsc{chimes} network do not appear in Table~\ref{depl_coeff} as they are not depleted onto dust grains.

We use the resulting depletion factors to reduce the gas-phase abundance of each element. We also sum the mass of each element in dust grains, using all 17 elements in \citet{decia16} and/or \citet{jenkins09}, to determine the total dust abundance. We use this to scale the rate of reactions that occur on the surface of dust grains (e.g. the formation of H$_{2}$, and grain surface recombination reactions), and thermal processes involving dust grains, such as photoelectric heating. However, we only scale the rates by the total dust abundance and we do not consider varying the grain size distributions that were originally assumed in the calculation of the rates for these processes (which used either the \citealt{mathis77} or the \citealt{weingartner01} distributions; see \citealt{richings14a} and references therein for details of how these rates were calculated). 

\begin{figure}
\centering
\mbox{
	\includegraphics[width=84mm]{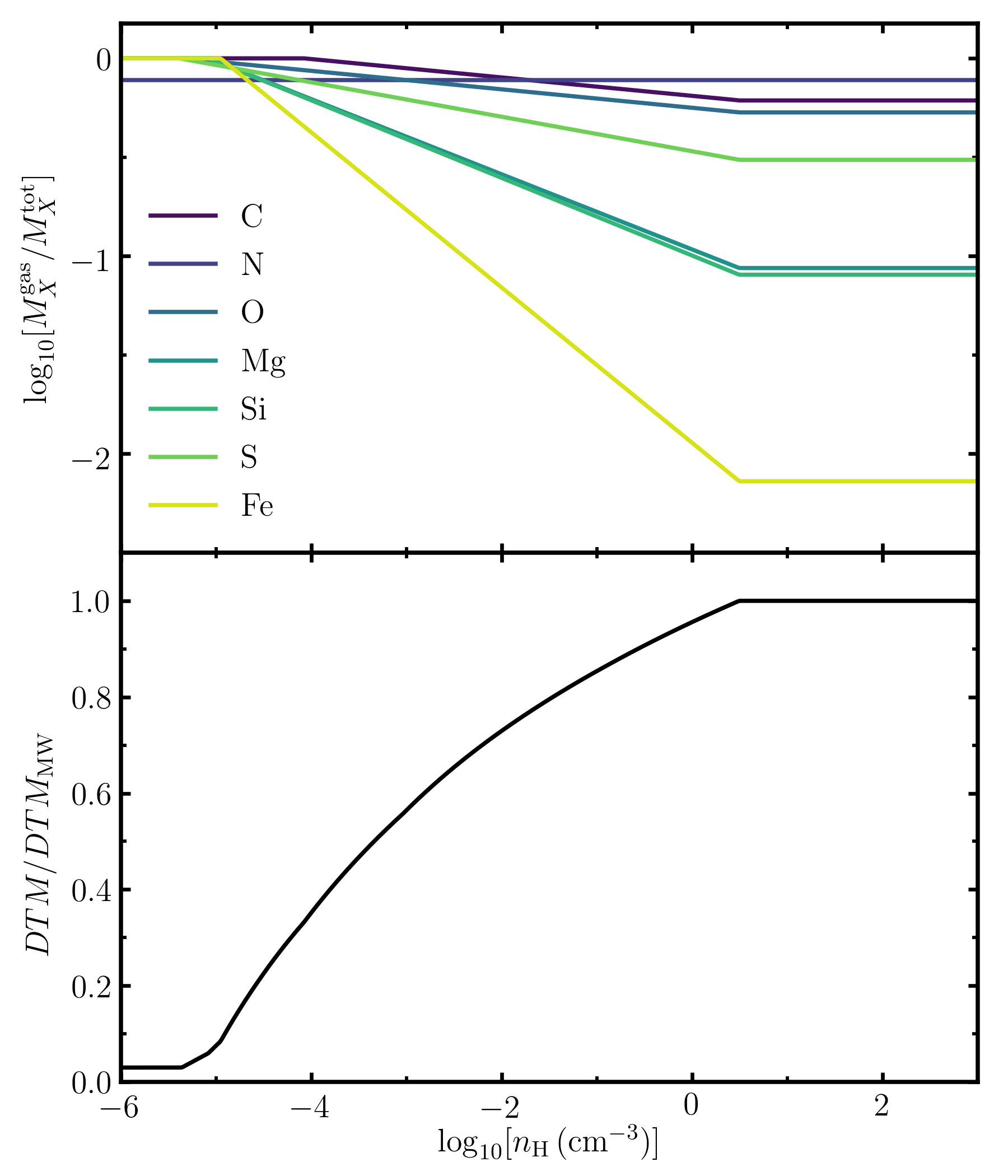}}
\vspace{-0.15 in}
\caption{\textit{Top Panel:} Mass fraction of each element in the gas phase ($M_{X}^{\rm{gas}} / M_{X}^{\rm{tot}}$) plotted as a function of hydrogen density ($n_{\rm{H}}$), using an empirical model based on the observed depletion factors of \citet{jenkins09} and \citet{decia16}. \textit{Bottom panel:} Dust to metals mass ratio ($DTM$) obtained by summing the depletion factors of all metals, relative to the maximum Milky Way dust to metals ratio ($DTM_{\rm{MW}} = 0.485$). 
\vspace{-0.15 in}} 
\label{depl_fig}
\end{figure}

The top panel of Fig.~\ref{depl_fig} shows the mass fraction of each element in the gas phase as a function of hydrogen density. We see that at the highest densities carbon, nitrogen and oxygen are reduced by up to approximately a factor of two, while iron exhibits the strongest depletion as it is reduced by more than two orders of magnitude. In the bottom panel of Fig.~\ref{depl_fig} we show the total dust to metals mass ratio ($DTM$), normalised to the dust to metals ratio along the sight lines with the strongest dust depletion in the Milky Way ($DTM_{\rm{MW}} = 0.485$), corresponding to $F_{\ast} = 1$.

This empirical model for the depletion of metals onto dust grains is based on observations in the Milky Way and DLAs. However, it does not explicitly follow the formation and destruction mechanisms that govern the abundance of dust grains. Other studies have developed numerical models that capture these processes \citep[e.g.][]{asano13, bekki15, hirashita15, mckinnon18, choban22}, which allows for a more complex evolution of the dust grain population. In this work we only consider the empirical model, as it is the simplest implementation that reproduces observed depletion factors. However, in the future it would be interesting to compare how the different approaches to modelling dust grains impact the non-equilibrium interstellar chemistry.

To study the effects of dust depletion on the non-equilibrium chemistry, we also repeat each simulation with a constant dust to metals ratio equal to the maximum Milky Way value ($DTM_{\rm{MW}}$), but without reducing the gas-phase element abundances by the corresponding depletion factor, so that the chemistry solver uses the total elemental abundances. This approach is inconsistent, as all metals are in the gas phase but dust grains are also present, so some metals are counted twice. However, such an approach has been used in previous studies \citep[e.g.][]{richings16}, so this will allow us to quantify the uncertainties that are introduced if the depletion of metals onto dust grains is not correctly accounted for.

\section{Simulations}\label{sim_sect}

\subsection{Initial conditions}\label{IC_sect}

We simulate a series of isolated disc galaxies, with initial conditions created using the \textsc{MakeDisk} code \citep{springel05} as follows. The model galaxies consist of a rotating disc of gas and stars along with a central stellar bulge, embedded within a live dark matter halo. The halo and stellar bulge are spherical, with a \citet{hernquist90} radial density profile. The stellar and gaseous components follow an exponential radial profile. The vertical structure of the stellar disc follows that of an isothermal sheet, with a constant scale height that we set to 0.1 times the radial exponential scale length. For the gas disc, the vertical profile is computed to be in hydrostatic equilibrium for the given gravitational potential, at a temperature of $10^{4} \, \rm{K}$.

The parameters of the galaxy models are chosen according to redshift zero scaling relations, to represent typical disc galaxies in the nearby Universe. We consider galaxies with halo masses ranging from dwarfs, with $M_{200, \, \rm{crit}} = 10^{10} \, \rm{M}_{\odot}$, to Milky Way-mass galaxies with $M_{200, \, \rm{crit}} = 10^{12} \, \rm{M}_{\odot}$. The concentration parameter of the dark matter halo is calculated as a function of $M_{200, \, \rm{crit}}$ using the redshift zero mass-concentration relation from \citet{duffy08}, using their full halo sample. The total stellar mass is calculated using the abundance matching model of \citet{moster13}, which we modify according to \citet{sawala15} to account for the inefficiency of galaxy formation at low halo masses.

\begin{figure}
\centering
\mbox{
	\includegraphics[width=84mm]{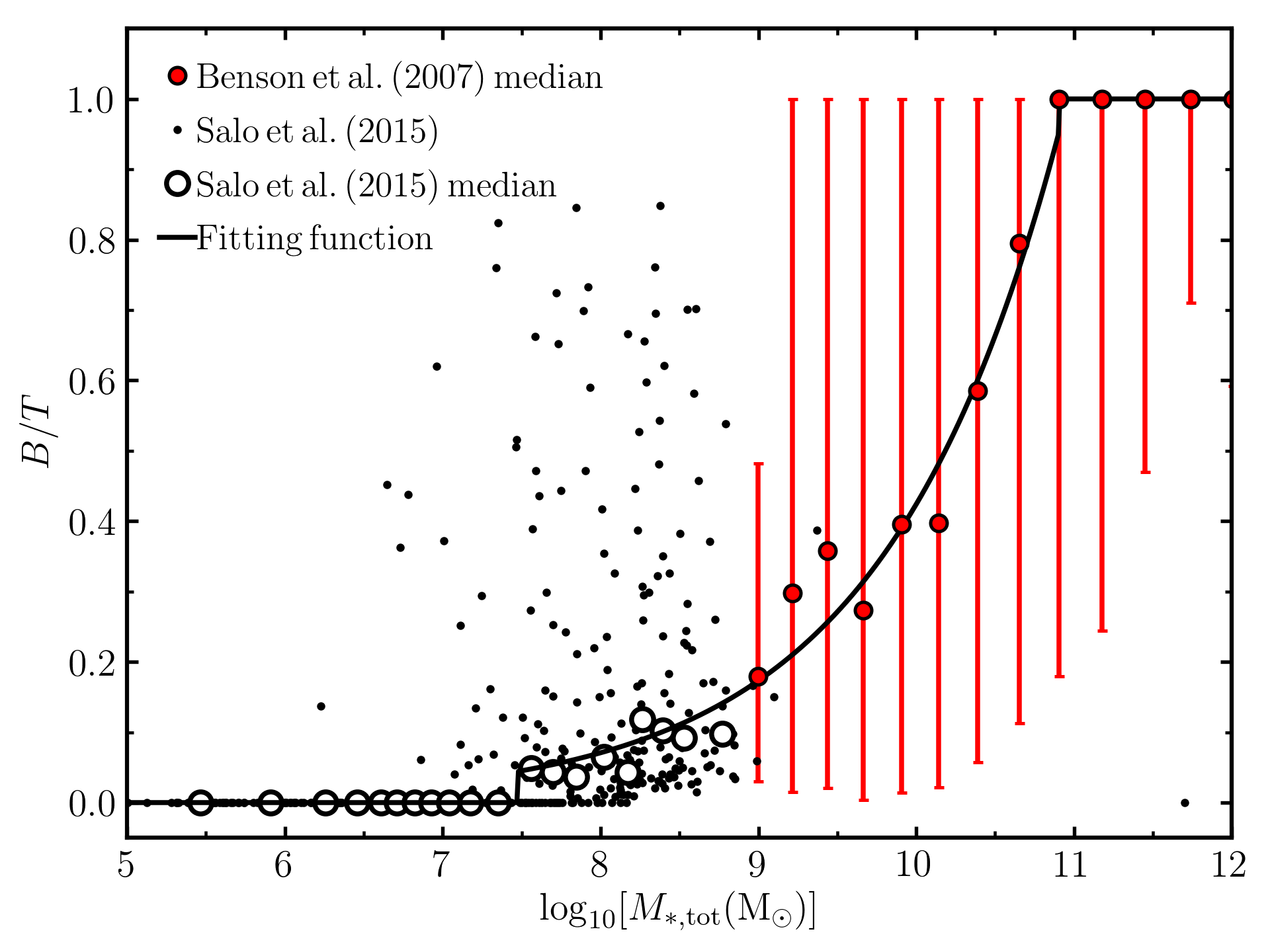}}
\vspace{-0.15 in}
\caption{The ratio of bulge to total stellar mass ($B/T$) versus total stellar mass ($M_{\ast, \rm{tot}}$). The red data points show the median $B/T$ ratio in bins of $M_{\ast}$, along with the tenth to ninetieth percentiles, in a sample of galaxies from the SDSS survey \citep{benson07}. The black points show individual galaxies in the S$^{4}$G survey \citep{salo15}, while the empty black circles show the median $B/T$ ratio in bins of total stellar mass for the S$^{4}$G sample. The blue curve shows the best-fit power-law relation, given in equation~\ref{bulge_eqn}. 
\vspace{-0.15 in}} 
\label{bulge_fig}
\end{figure}

To divide the stellar mass between the bulge and disc components, we need to determine the ratio of the bulge stellar mass to total stellar mass ($B/T$). In Fig.~\ref{bulge_fig}, the red data points show the median and tenth to ninetieth percentiles of the $B/T$ ratio in bins of total stellar mass from a sample of galaxies in the Sloan Digital Sky Survey (SDSS; \citealt{benson07}). The black data points show $B/T$ for individual galaxies in the \textit{Spitzer} Survey of Stellar Structure in Galaxies (S$^{4}$G; \citealt{salo15}), while the open black circles show the median $B/T$ ratio in bins of stellar mass for the S$^{4}$G sample. We fit a power-law function to the median $B/T$ ratios versus total stellar mass, using the S$^{4}$G and SDSS samples at stellar masses below and above $10^{9} \, \rm{M}_{\odot}$, respectively. We enforce $B/T = 0.0$ at $M_{\ast, \rm{tot}} \! < \! 3.0 \times 10^{7} \, \rm{M}_{\odot}$, and $B/T = 1.0$ at $M_{\ast, \rm{tot}} \! > \! 8.0 \times 10^{10} \, \rm{M}_{\odot}$. The resulting best-fit power-law relation (black curve in Fig.~\ref{bulge_fig}) is: 

\begin{equation}\label{bulge_eqn} 
  B/T =
  \begin{cases}
    \! 0.0 & \frac{M_{\ast, \rm{tot}}}{\rm{M}_{\odot}} \! < \! 3 \! \times \! 10^{7} \\
    \! 0.424 \! \left( \frac{M_{\ast, \rm{tot}}}{10^{10} \, \rm{M}_{\odot}} \right)^{0.3887} \! & 3 \! \times \! 10^{7} \! \leq \! \frac{\mathit{M}_{\ast, \rm{tot}}}{\rm{M}_{\odot}} \! \leq \! 8 \! \times \! 10^{10} \\
    \! 1.0 & \rm{otherwise}. \\
  \end{cases}
\end{equation}

\citet{lange16} study the relation between the stellar half-light radius and stellar mass of the bulge and disc components in galaxies from the Galaxy And Mass Assembly (GAMA) survey. We use their best-fit power-law relations for their final redshift zero disc and spheroid samples (see Table~1 of \citealt{lange16}) to calculate the stellar half-light radii of the disc and bulge, respectively, in our galaxy models. We assume that these are equal to the half-mass radii, $R_{1/2}$. From the half-mass radius, we calculate the exponential scale length of the disc as $R_{\rm{exp}} = R_{1/2} / 1.68$, and the scale radius of the \citet{hernquist90} profile as $a = R_{1/2} / (1 + \sqrt{2})$. Both the stellar and gaseous discs use the same scale length. 

\begin{figure}
\centering
\mbox{
	\includegraphics[width=84mm]{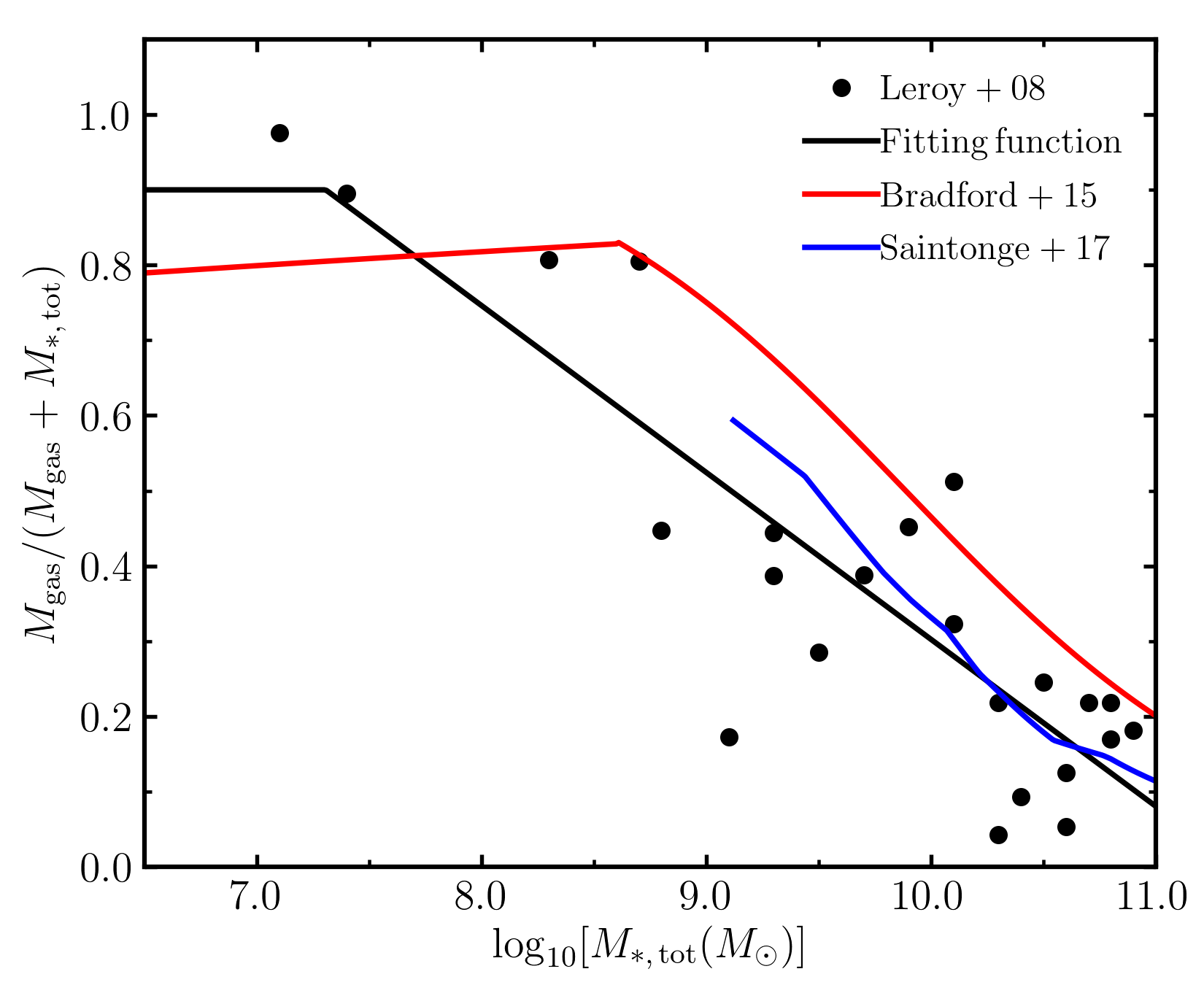}}
\vspace{-0.15 in}
\caption{The ratio of galaxy gas mass, $M_{\rm{gas}}$, to total galaxy baryonic mass, $M_{\rm{gas}} + M_{\rm{\ast, \, \rm{tot}}}$, plotted against total stellar mass for the galaxies in the THINGS survey (\citealt{leroy08}; black points), a sample of low-mass isolated galaxies in SDSS (\citealt{bradford15}; red curve), and the xCOLD GASS survey (\citealt{saintonge17}; blue curve). The black curve shows the best-fit scaling relation fit to the \citet{leroy08} data (see equation~\ref{fgas_eqn}).
\vspace{-0.15 in}} 
\label{fgas_fig}
\end{figure}

\begin{table*}
\begin{minipage}{168mm}
\centering
\caption{Model galaxy parameters: total halo mass ($M_{200, \, \rm{crit}}$), dark matter halo concentration parameter ($c_{200}$), total stellar mass ($M_{\ast, \, \rm{tot}}$), bulge to total ratio ($B/T$), \citet{hernquist90} scale radius of the bulge ($R_{\rm{bulge}}$), exponential scale length of the stellar and gaseous discs ($R_{\rm{disc}}$), total galaxy gas fraction ($f_{\rm{gas}}$), initial metallicity ($Z_{\rm{init}}$), and initial mass of gas and star particles ($m_{\rm{b}}$).}
\label{galaxy_pars}
\begin{tabular}{lccccccccc}
  \hline
  Name & $M_{200, \, \rm{crit}}$ & $c_{200}$ & $M_{\ast, \, \rm{tot}}$ & $B/T$ & $R_{\rm{bulge}}$ & $R_{\rm{disc}}$ & $f_{\rm{gas}}$ & $Z_{\rm{init}}$ & $m_{\rm{b}}$ \\
   & (M$_{\odot}$) & & (M$_{\odot}$) & & (kpc) & (kpc) & & (Z$_{\odot}$) & (M$_{\odot}$) \\
  \hline
  \multicolumn{10}{c}{\textbf{Fiducial, Uniform ISRF, and No Depletion}} \\ 
  \hline
  m1e10 & $10^{10}$ & 9.9 & $6.6 \times 10^{6}$ & 0.0 & N/A & 0.41 & 0.90 & 0.06 & 400 \\
  m3e10 & $3 \times 10^{10}$ & 8.9 & $8.9 \times 10^{7}$ & 0.07 & 0.12 & 0.82 & 0.77 & 0.3 & 400 \\
  m1e11 & $10^{11}$ & 7.9 & $1.4 \times 10^{9}$ & 0.20 & 0.33 & 1.68 & 0.49 & 0.8 & 400 \\
  m3e11 & $3 \times 10^{11}$ & 7.1 & $1.1 \times 10^{10}$ & 0.43 & 0.70 & 2.66 & 0.30 & 1.1 & 400 \\
  m3e11\_lowGas & $3 \times 10^{11}$ & 7.1 & $1.1 \times 10^{10}$ & 0.43 & 0.70 & 2.66 & 0.10 & 1.1 & 400 \\
  m3e11\_hiGas & $3 \times 10^{11}$ & 7.1 & $1.1 \times 10^{10}$ & 0.43 & 0.70 & 2.66 & 0.50 & 1.1 & 400 \\
  m1e12 & $10^{12}$ & 6.4 & $3.1 \times 10^{10}$ & 0.66 & 1.03 & 3.10 & 0.19 & 1.2 & 400 \\
  \hline
  \multicolumn{10}{c}{\textbf{Resolution tests, Fiducial only}} \\ 
  \hline
  m1e10\_lowRes08 & $10^{10}$ & 9.9 & $6.6 \times 10^{6}$ & 0.0 & N/A & 0.41 & 0.90 & 0.06 & 3200 \\
  m3e10\_lowRes08 & $3 \times 10^{10}$ & 8.9 & $8.9 \times 10^{7}$ & 0.07 & 0.12 & 0.82 & 0.77 & 0.3 & 3200 \\
  m1e11\_lowRes08 & $10^{11}$ & 7.9 & $1.4 \times 10^{9}$ & 0.20 & 0.33 & 1.68 & 0.49 & 0.8 & 3200 \\
  m3e11\_lowRes08 & $3 \times 10^{11}$ & 7.1 & $1.1 \times 10^{10}$ & 0.43 & 0.70 & 2.66 & 0.30 & 1.1 & 3200 \\
  m1e12\_lowRes08 & $10^{12}$ & 6.4 & $3.1 \times 10^{10}$ & 0.66 & 1.03 & 3.10 & 0.19 & 1.2 & 3200 \\
  m3e10\_hiRes08 & $3 \times 10^{10}$ & 8.9 & $8.9 \times 10^{7}$ & 0.07 & 0.12 & 0.82 & 0.77 & 0.3 & 50 \\
  \hline
  \vspace{-0.3 in}
\end{tabular}
\end{minipage}
\end{table*}

To calculate the gas fractions for our model galaxies, we use observed galaxy H\textsc{i} and H$_{2}$ masses from The H\textsc{i} Nearby Galaxy Survey (THINGS; \citealt{leroy08}). The black data points in Fig.~\ref{fgas_fig} show the gas fraction, $f_{\rm{gas}} = M_{\rm{gas}} / (M_{\rm{gas}} + M_{\ast, \, \rm{tot}})$, plotted against total stellar mass in the THINGS galaxies. The gas mass is the sum of the atomic and molecular masses, $M_{\rm{gas}} = M_{\rm{HI}} + M_{\rm{H}2}$, and includes a factor 1.36 correction for helium. Using the \citet{leroy08} data, we fit the following function to the gas fraction versus stellar mass (black curve in Fig.~\ref{fgas_fig}):

\begin{align}\label{fgas_eqn} 
  f_{\rm{gas}} &= \frac{M_{\rm{gas}}}{M_{\rm{gas}} + M_{\ast, \, \rm{tot}}} \\
   &= \!
    \begin{cases}
      \! 0.9 & \! \! \! \! \frac{M_{\ast, \, \rm{tot}}}{\rm{M}_{\odot}} \! < 2 \! \times \! 10^{7} \\
      \! 0.524 \! - \! 0.222 \! \log_{10} \! \left( \frac{M_{\ast, \, \rm{tot}}}{10^{9} \, \rm{M_{\odot}}} \right) & \! \! \! \! 2 \! \times \! 10^{7} \! < \! \frac{M_{\ast, \, \rm{tot}}}{\rm{M}_{\odot}} \! < \! 2 \! \times \! 10^{11} \\
      \! 0.0 & \! \! \! \! \rm{otherwise}.
    \end{cases}
\end{align}
We cap the gas fraction in the fitting function to be no greater than 0.9, at stellar masses $M_{\ast, \, \rm{tot}}$$<$$2 \times 10^{7} \, \rm{M_{\odot}}$, and by definition it can be no less than zero. We use this fitting function to determine the gas fraction for our galaxy models, given the total stellar mass calculated above. For the model with a halo mass $M_{200, \, \rm{crit}} = 3 \times 10^{11} \, \rm{M}_{\odot}$ (m3e11), we also run two additional models with a gas fraction reduced/increased by 20 per cent from the best-fit scaling relation (m3e11\_lowGas and m3e11\_hiGas, respectively). This will allow us to explore the effects of different gas fractions at fixed halo mass. 

For comparison, the red curve in Fig.~\ref{fgas_fig} shows the best-fit scaling relation for a sample of low-mass isolated galaxies in SDSS from \citet{bradford15}, taken from the first two rows of their Table~3 (see also their Fig.~5). The gas masses in this SDSS sample were determined from H\textsc{i} observations, including a correction for helium, but do not include the molecular component. We also plot the gas fractions in the xCOLD GASS sample (\citealt{saintonge17}; blue curve), using the H\textsc{i} and H$_{2}$ masses in bins of stellar mass from their Fig.~13.

Finally, we set the initial metallicity of the gas and stars in our model galaxies according to the mass-metallicity relation of SDSS galaxies from \citet{andrews13}. The relative abundances between different metal elements are assumed to be solar, with the initial Helium abundance scaled between primordial and solar according to the total metallicity. Our simulations include the injection of metals from winds and supernovae (see Section~\ref{physics_sect}), so the metallicity in our model galaxies will increase over time. The idealised nature of these model galaxies means that we do not include cosmological accretion of primordial or low-metallicity gas onto the galaxy. We therefore might not expect the evolution to maintain the observed mass-metallicity scaling relation. However, we find that, as we only evolve each galaxy for 800~Myr in total (see below), the change in metallicity is relatively small, and the galaxies do not evolve far from this relation by the end of the simulation. 

The parameters of our seven galaxy models are summarised in Table~\ref{galaxy_pars}. As discussed in Sections~\ref{flux_sect} and \ref{depl_sect}, we run each galaxy model three times: first with the fiducial model, including the prescriptions for local stellar fluxes and dust depletion; second with a uniform ISRF, in which a uniform radiation field is applied to all gas particles; and third with no depletion, in which we use a constant dust to metals ratio and we do not reduce the gas phase metal abundances to account for depletion onto dust grains.

In these runs, we use a mass resolution of $400 \, \rm{M}_{\odot}$ per particle for the gas and stars. The mass of dark matter particles is $1910 \, \rm{M}_{\odot}$, which corresponds to $(\Omega_{\rm{m}} - \Omega_{\rm{b}}) / \Omega_{\rm{b}}$ times the baryonic particle mass, where  $\Omega_{\rm{m}}$ and $\Omega_{\rm{b}}$ are the cosmological density parameters for the total matter and baryonic content of the Universe, respectively. We use a constant gravitational softening length of 2.8~pc and 1.6~pc for dark matter and star particles, respectively. The gas particles use an adaptive gravitational softening length equal to the mean inter-particle spacing, down to a minimum gas softening of 0.08~pc. At the star formation density threshold of $n_{\rm{H}} = 10^{3} \, \rm{cm}^{-3}$ the gas softening length is 2.2~pc. 

To test the importance of numerical resolution on our results, we also repeat some of the galaxy models with 8 times lower mass resolution, and we repeat the m3e10 dwarf galaxy with 8 times higher mass resolution. In each case, the gravitational softening lengths are scaled with $m_{\rm{b}}^{1/3}$, where $m_{\rm{b}}$ is the baryonic particle mass. We only run the resolution tests with the fiducial model.

At the beginning of each simulation, the gas disc rapidly cools from its initial temperature of $10^{4} \, \rm{K}$ and starts to form stars. However, there is a delay before the onset of stellar feedback, which is needed to regulate this process. This delay results in a strong initial burst of star formation, which disrupts the gas disc and in some cases can destroy it altogether. To alleviate this disruption and allow the disc to settle into a self-regulated steady state, we therefore modify the subgrid feedback models during the first 300~Myr of the simulation as follows. For the inital 150~Myr, we reduce the time-scales for supernova feedback by a factor of 100, and renormalise the rates by the same factor so that the total number of supernovae per unit mass of stars formed remains unchanged. This enables the stellar feedback to regulate the initial burst of star formation more rapidly. Then from 150 to 300~Myr we smoothly reduce the factor by which the supernova time-scales are suppressed, until they reach their fiducial value after 300~Myr. We use the resulting snapshot at 300~Myr as our initial conditions for the main runs, which are then run for a further 500~Myr. For all of the results presented in this paper, we denote the time $t = 0$ as starting from the snapshot at the end of the 300~Myr settling in period. We do not include any of the snapshots prior to this point in our analysis.

\subsection{Galaxy morphology and evolution}\label{morph_sect}

\begin{figure}
\centering
\mbox{
	\includegraphics[width=84mm]{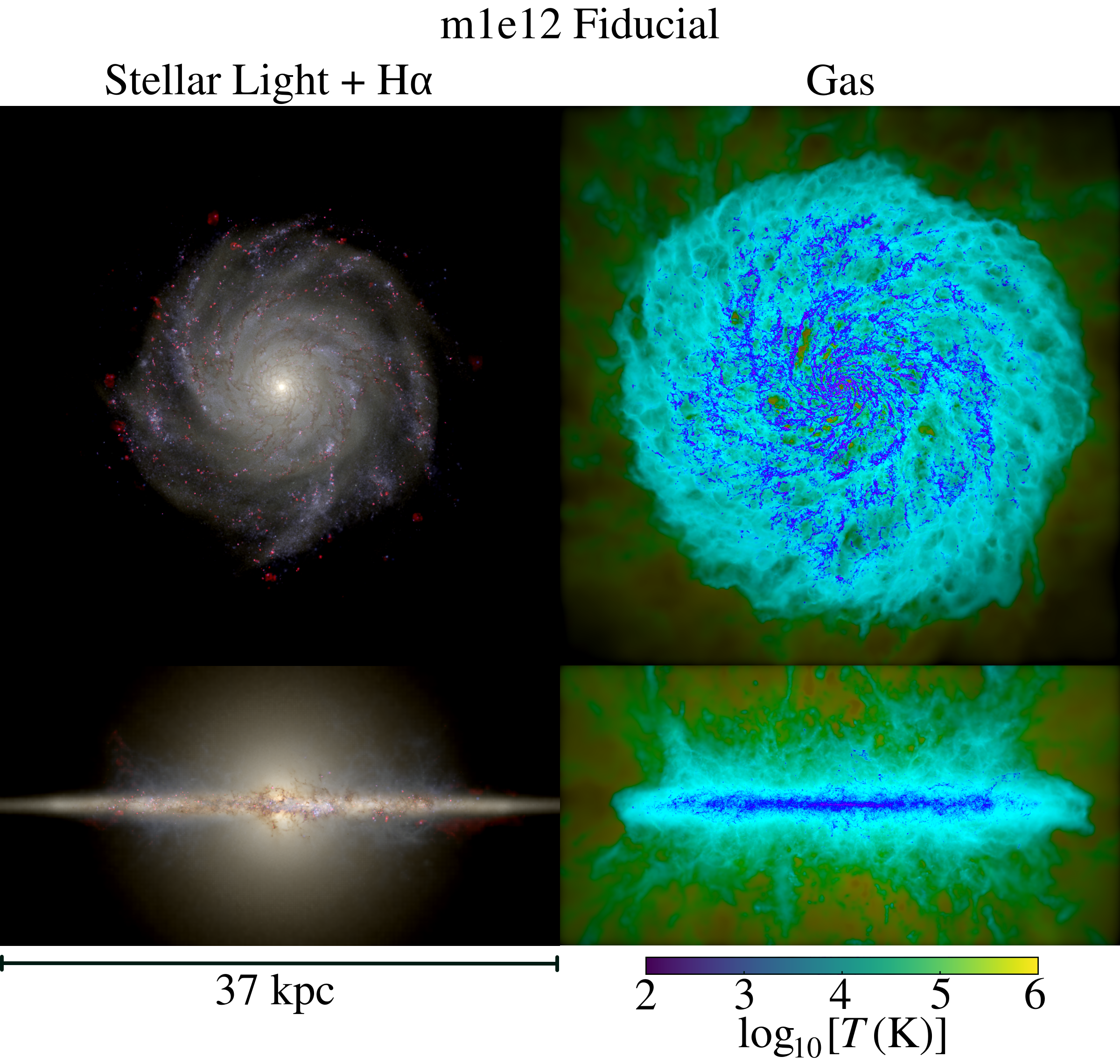}}
\vspace{-0.15 in}
\caption{\textit{Left panels:} Mock Hubble images of the stellar light, attenuated by dust, observed in the F336W, F555W and F814W filters, with H$\alpha$ emission superimposed in red, from the m1e12 fiducial simulation. \textit{Right panels:} Images of the gas distribution, with temperature indicated by the colour scale. The disc of the galaxy is viewed face-on (top panels) and edge-on (bottom panels). All panels show the galaxy at the end of the simulation after 500~Myr. The dust lanes seen in the mock Hubble images coincide with cold, dense gas structures, while H$\alpha$ emission traces the young stellar components. 
\vspace{-0.15 in}} 
\label{m1e12_morph_fig}
\end{figure}

\begin{figure}
\centering
\mbox{
	\includegraphics[width=84mm]{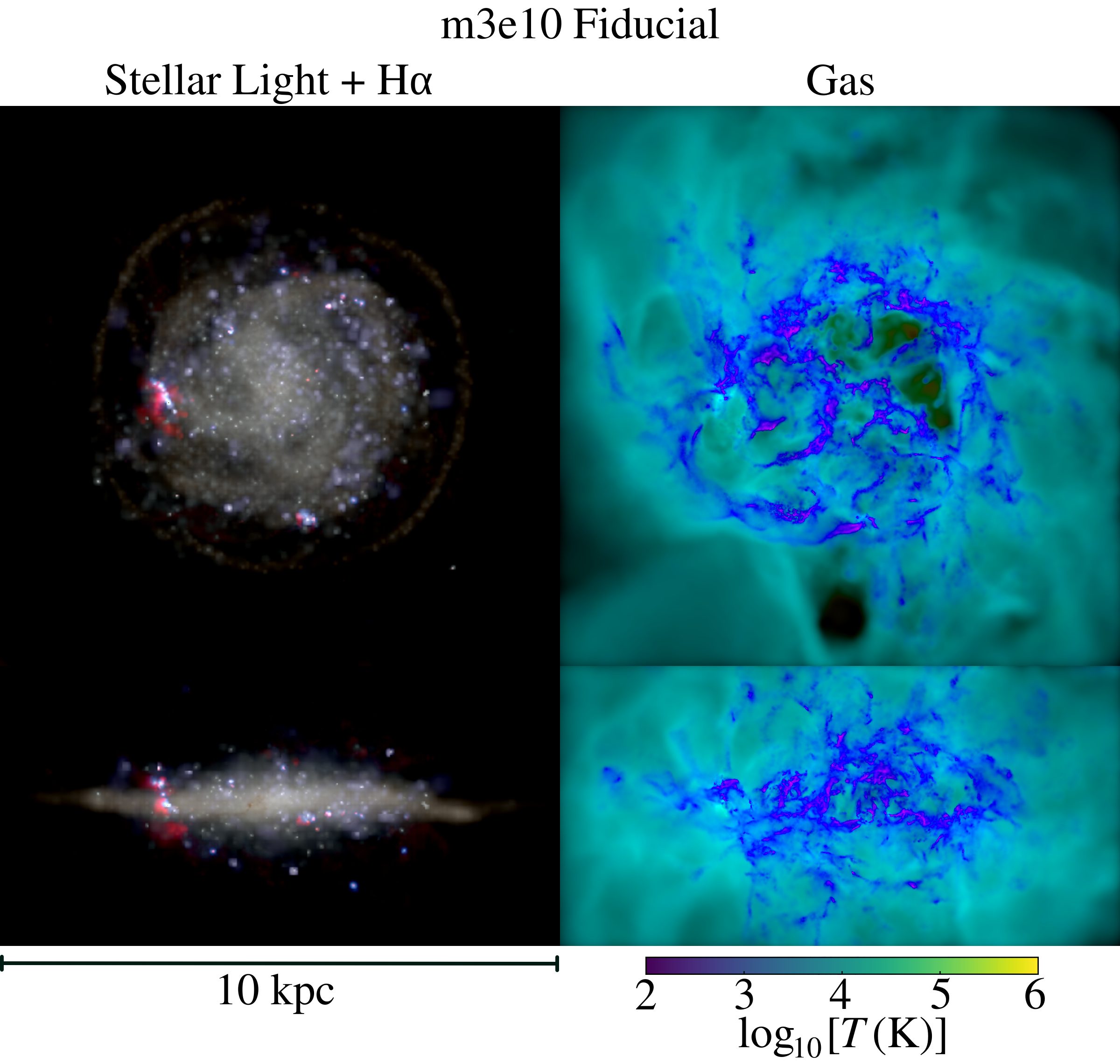}}
\vspace{-0.15 in}
\caption{As Fig.~\ref{m1e12_morph_fig} but for the m3e10 fiducial simulation. Compared to m1e12, the disc in the dwarf galaxy is less well defined, with a broader vertical extent relative to the radial extent. 
\vspace{-0.15 in}} 
\label{m3e10_morph_fig}
\end{figure}

The Milky Way-mass simulation (m1e12) using the fiducial model is shown at 500~Myr in Fig.~\ref{m1e12_morph_fig}. The left-hand panels show mock Hubble images of the stellar light, including dust attenuation calculated using dust abundances from our fiducial dust depletion model. These images use the Hubble filters F336W, F555W and F814W. We have superimposed images of the continuum-subtracted H$\alpha$ emission line in red. These mock observations were created in post-processing using the publicly available radiative transfer code \textsc{radmc-3d}\footnote{\url{http://www.ita.uni-heidelberg.de/~dullemond/software/radmc-3d/}} \citep{dullemond12}. We describe how we post-process the simulation outputs with \textsc{radmc-3d} in more detail in Section~\ref{line_emission_method_sect}. 

The H$\alpha$ emission generally coincides with regions containing young, blue stars. These H$\alpha$-emitting regions are somewhat more extended at large galactic radii, due to the lower gas densities compared to the galactic centre, which results in larger Str\"{o}mgren radii. We also see prominent dust lanes, particularly along the flocculent spiral arm structures. 

The right-hand panels of Fig.~\ref{m1e12_morph_fig} show the distribution of gas. The brightness of each pixel indicates the gas density, while the gas temperature is shown by the colour scale. The cold ($\sim$100~K), dense gas, shown in dark blue, forms a thin ($\approx$600~pc) disc, and is arranged in flocculent spiral structures that coincide with the dust lanes seen in the mock Hubble images. The warm ($10^{4} \, \rm{K}$), diffuse phase of the ISM forms a somewhat thicker disc, with a vertical extent of $\approx$2$-$3~kpc, while the outflows are heated to temperatures $\gtrsim$10$^{6} \, \rm{K}$. 

Fig.~\ref{m3e10_morph_fig} shows mock Hubble images (left-hand panels) and the gas distribution (right-hand panels) in the m3e10 dwarf galaxy after 500~Myr, with the fiducial model. Compared to m1e12, the disc is less well defined in the dwarf galaxy. In particular, the cold gas phase is not confined to a thin disc, but instead exhibits strong, turbulent motions that lead to a broader vertical structure. This is due to the weaker gravitational potential of the dwarf galaxy, which makes it more difficult to retain gas that is driven out by stellar feedback. 

The overall morphology of gas and stars in the simulations using the fiducial model (Figs.~\ref{m1e12_morph_fig} and \ref{m3e10_morph_fig}) are similar to the corresponding runs using the uniform ISRF and no depletion models (not shown). The only significant difference is in the H$\alpha$ emission, which is much weaker and does not trace star-forming regions in the uniform ISRF model. We study the effects of the model variations on H$\alpha$ emission, and other emission line tracers of the star formation rate, in more detail in Section~\ref{emission_line_sect}. 

\begin{figure}
\centering
\mbox{
	\includegraphics[width=84mm]{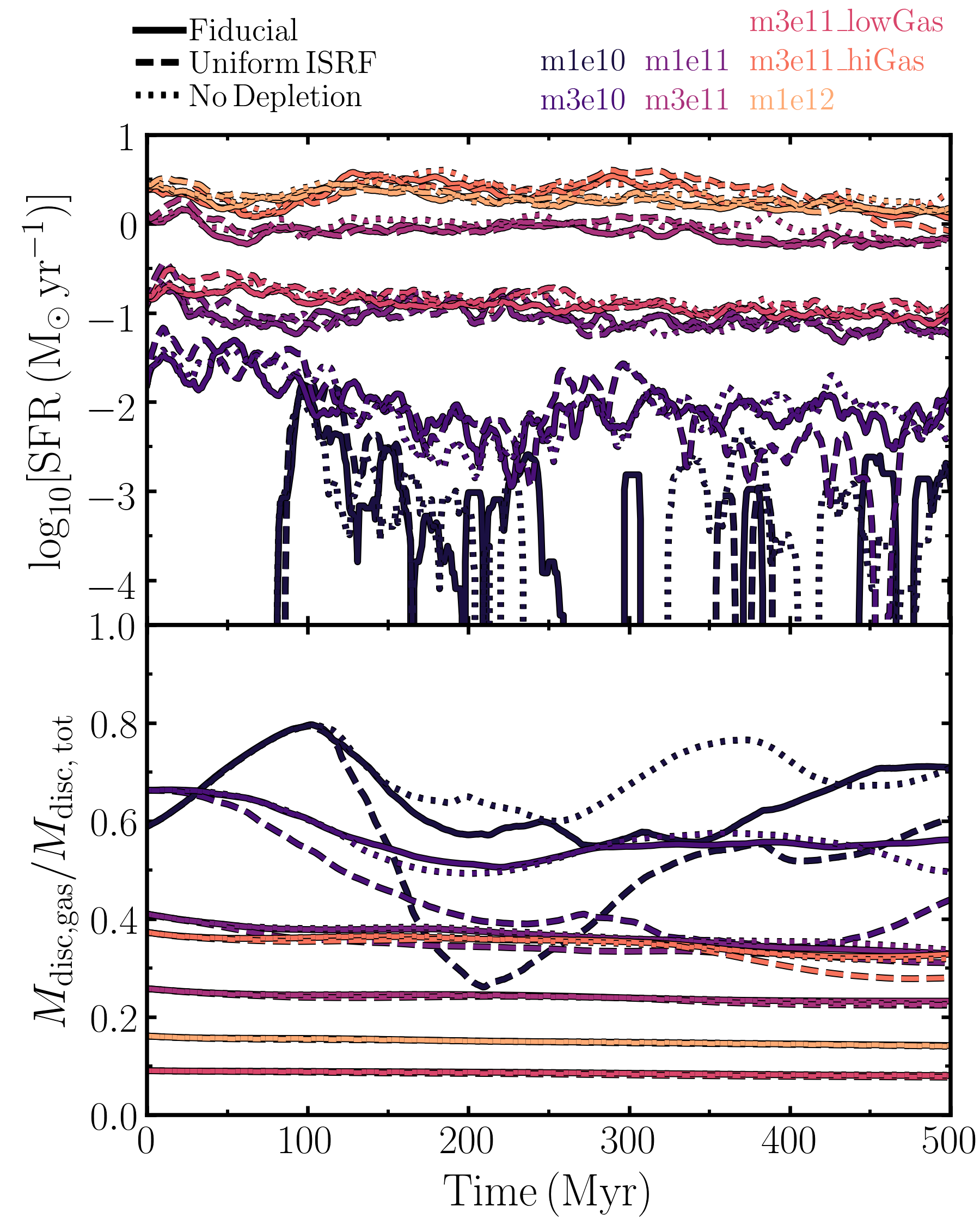}}
\vspace{-0.15 in}
\caption{\textit{Top panel:} Evolution of the total star formation rate in the galaxy disc, averaged over the preceding 10~Myr. \textit{Bottom panel:} Evolution of the gas mass fraction of the galaxy disc. The solid, dashed and dotted line styles indicate simulations run with the fiducial, uniform ISRF and no depletion models, while the line colours show the different galaxies. In the high-mass galaxies ($M_{200, \, \rm{crit}}$$\geq$10$^{11} \, \rm{M}_{\odot}$) there is very little difference between the three model variations. However, in the dwarf galaxies the uniform ISRF model drives stronger outflows, resulting in lower disc gas fractions.
\vspace{-0.15 in}} 
\label{SFH_fig}
\end{figure}

To compare the evolution of the simulated galaxies with the three model variations, Fig.~\ref{SFH_fig} shows the total star formation rate (top panel) and disc gas fraction (bottom panel) versus time. We calculate the star formation rate from the total mass of stars in the galaxy disc that formed in the preceding 10~Myr, using the initial mass of each star particle (before stellar mass loss). See the final paragraph of Section~\ref{flux_sect} for the definition of the galaxy disc. The fiducial, uniform ISRF and no depletion models are shown by the solid, dashed and dotted curves respectively, while the line colours indicate the different galaxies. 

In the highest-mass galaxies, with halo masses $\geq$10$^{11} \, \rm{M}_{\odot}$, the star formation rate and disc gas fraction are very similar in the three models. However, the dwarf galaxies exhibit deviations in disc gas fraction between different models. In particular, the dwarf galaxies run with the uniform ISRF model have lower gas fractions than the fiducial model by up to $\approx$30 per cent. The uniform ISRF model does not include H\textsc{ii} regions, so this may suggest that removing this early stellar feedback channel (which acts preferentially in star-forming regions) leads to stronger outflows in dwarf galaxies, thereby reducing the disc gas fraction. \citet{hopkins20} also found that the removal of radiative stellar feedback processes leads to more violent star formation histories, which would be consistent with the trends that we see in our dwarf galaxies. However it is difficult to draw strong conclusions as we only have two dwarf galaxies in our sample. Nevertheless, this reduction in disc gas fraction in the dwarf galaxies does not lead to a reduction in the star formation rate. The total mass of stars formed over 500~Myr is actually higher with the Uniform IRSF than the fiducial model, by 27 and 20 per cent in m1e10 and m3e10 respectively. 

We therefore conclude that, apart from the lower gas fractions and higher total mass of stars formed in dwarf galaxies with the uniform ISRF, the three model variations do not have a strong impact on the overall evolution of the galaxy. 

We compare the evolution of star formation rate and disc gas fraction in simulations at different numerical resolutions in Appendix~\ref{resolution_sect}. While the high-mass galaxies ($M_{200}$$\geq$$10^{11} \, \rm{M}_{\odot}$) show good numerical convergence, the evolution in dwarf galaxies can vary significantly between runs at different resolutions. As we discuss further in Appendix~\ref{resolution_sect}, this may be due to stochastic variations between runs due to the bursty nature of star formation in the dwarf galaxy regime \citep[e.g.][]{fauchergiguere18}. 

\subsection{Stellar fluxes}\label{sim_flux_sect} 

\begin{figure}
\centering
\mbox{
	\includegraphics[width=84mm]{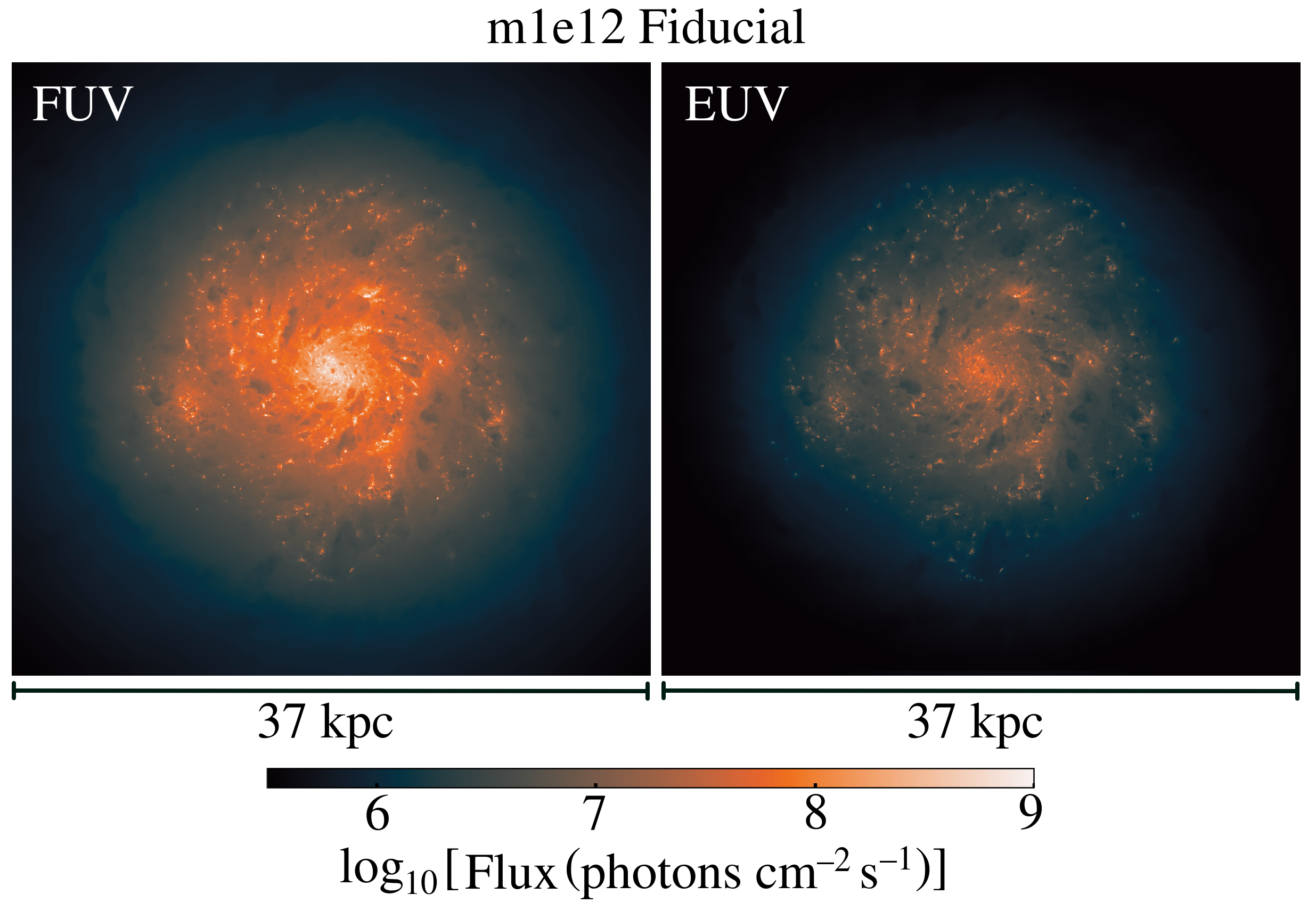}}
\vspace{-0.15 in}
\caption{Images of the incident stellar fluxes received by gas particles in the FUV (left panel) and EUV (right panel) bands summed over all stellar age bins, before self-shielding due to the local gas particle is applied. We show the final snapshot, after 500~Myr, from m1e12 using the fiducial model. The stellar fluxes vary by more than three orders of magnitude over the disc of the galaxy and, unsurprisingly, correlate with star forming regions. 
\vspace{-0.15 in}} 
\label{m1e12_flux_fig}
\end{figure}

The images in Fig.~\ref{m1e12_flux_fig} show mass-weighted projections of the stellar fluxes received by gas particles in the final snapshot of m1e12, after 500~Myr, calculated using our fiducial model. The left and right panels show the total fluxes in the FUV and EUV bands, respectively, summed over all stellar age bins. These fluxes include attenuation by the escape fraction from the H\textsc{ii} region around the emitting star particle, but do not include self-shielding by the receiving gas particle. This represents the flux that would be incident at the edge of a gas cloud, before the local self-shielding of the cloud itself has been applied.

We see that the stellar fluxes vary by more than three orders of magnitude over the galaxy disc. The strongest fluxes are found in small regions ($\sim$10$-$100~pc across) that predominantly lie in the spiral arms. Unsurprisingly, these regions coincide with the locations of young stars (see the top-left panel of Fig.~\ref{m1e12_morph_fig}).

The simulations with a uniform ISRF do not capture these strong spatial variations. For comparison, we can calculate the fluxes that would have been used in the uniform ISRF model, using equations~\ref{uni_fuv_flux} and \ref{uni_euv_flux}. The star formation rate surface density over the whole disc, viewed face-on, in m1e12 after 500~Myr is $\Sigma_{\rm{SFR,} \, \rm{disc}} = 1.2 \times 10^{-3} \, \rm{M}_{\odot} \, \rm{kpc}^{-2}$. Hence the FUV and EUV fluxes in the uniform ISRF case would be $\log_{10} [\mathcal{F} \, (\rm{photons} \, \rm{cm}^{-2} \, \rm{s}^{-1})] = 7.7$ and $6.5$, respectively. 

\begin{figure}
\centering
\mbox{
	\includegraphics[width=84mm]{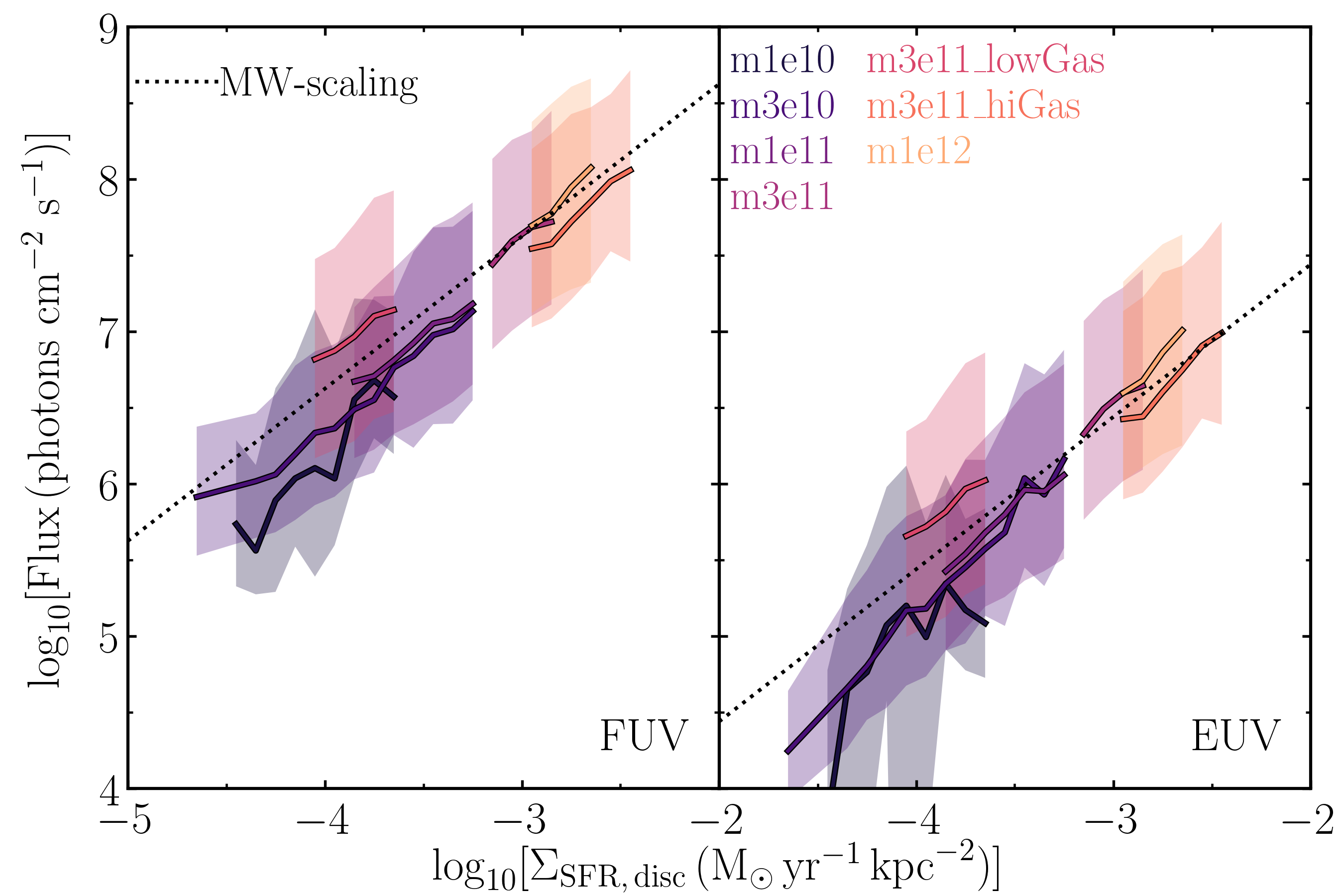}}
\vspace{-0.15 in}
\caption{Median stellar fluxes (solid curves) and tenth to ninetieth percentiles (shaded regions) received by gas particles in bins of star formation rate surface density averaged over the galaxy disc ($\Sigma_{\rm{SFR,} \, \rm{disc}}$). The left-hand and right-hand panels show the FUV and EUV bands, respectively. The different colours indicate different galaxy models. The dotted black line shows a linear scaling normalised to the Milky Way values (equations~\ref{uni_fuv_flux} and \ref{uni_euv_flux}). The median fluxes broadly follow a linear scaling, to within $\approx$0.5~dex.
\vspace{-0.15 in}} 
\label{flux_fullDisc_fig}
\end{figure}

The uniform ISRF model assumes that the stellar fluxes scale linearly with $\Sigma_{\rm{SFR,} \, \rm{disc}}$. Similar approaches have been used by other theoretical models, often using the local (rather than disc-averaged) star formation rate surface density \citep[e.g.][]{robertson08, lagos15}. It is therefore interesting to explore the extent to which the stellar fluxes in our fiducial model scale with the global and local star formation rate surface densities.

Fig.~\ref{flux_fullDisc_fig} shows the FUV (left-hand panels) and EUV (right-hand panels) fluxes incident on gas particles (before local gas self-shielding) that lie within the galaxy disc, combining snapshots at 10~Myr intervals. The solid curves show the median fluxes in bins of $\Sigma_{\rm{SFR,} \, \rm{disc}}$, while the shaded regions indicate the tenth to ninetieth percentile range. Different galaxy models are represented by different colours, from the dwarf galaxy m1e10 (dark purple) to the Milky Way-mass galaxy m1e12 (light orange), as shown in the legend. The dotted black line indicates a linear scaling between the fluxes and $\Sigma_{\rm{SFR,} \, \rm{disc}}$ normalised to the Milky Way values, as given by equations~\ref{uni_fuv_flux} and \ref{uni_euv_flux}.

We see that the median fluxes in all galaxy models broadly follow the linear relation, spanning two orders of magnitude in $\Sigma_{\rm{SFR,} \, \rm{disc}}$, although the dwarf galaxies are up to 0.5~dex below this relation. The tenth to ninetieth percentile spread in each galaxy extends to $\pm$0.5~dex about the median relation at fixed $\Sigma_{\rm{SFR,} \, \rm{disc}}$. 

In Appendix~\ref{esc_fraction_sect} we show that the normalisation of the linear scaling between the median stellar fluxes and $\Sigma_{\rm{SFR,} \, \rm{disc}}$ is determined by the escape fraction from H\textsc{ii} regions in each band. We use this to calibrate the escape fraction parameters in the fiducial model, such that the median fluxes in the simulations follow the same normalisation as the linear relation normalised to the Milky Way values (black dotted lines in Fig.~\ref{flux_fullDisc_fig}). For comparison, the FUV and EUV fluxes in the uniform ISRF model follow the linear Milky Way scaling relations, with no scatter. 

\begin{figure}
\centering
\mbox{
	\includegraphics[width=84mm]{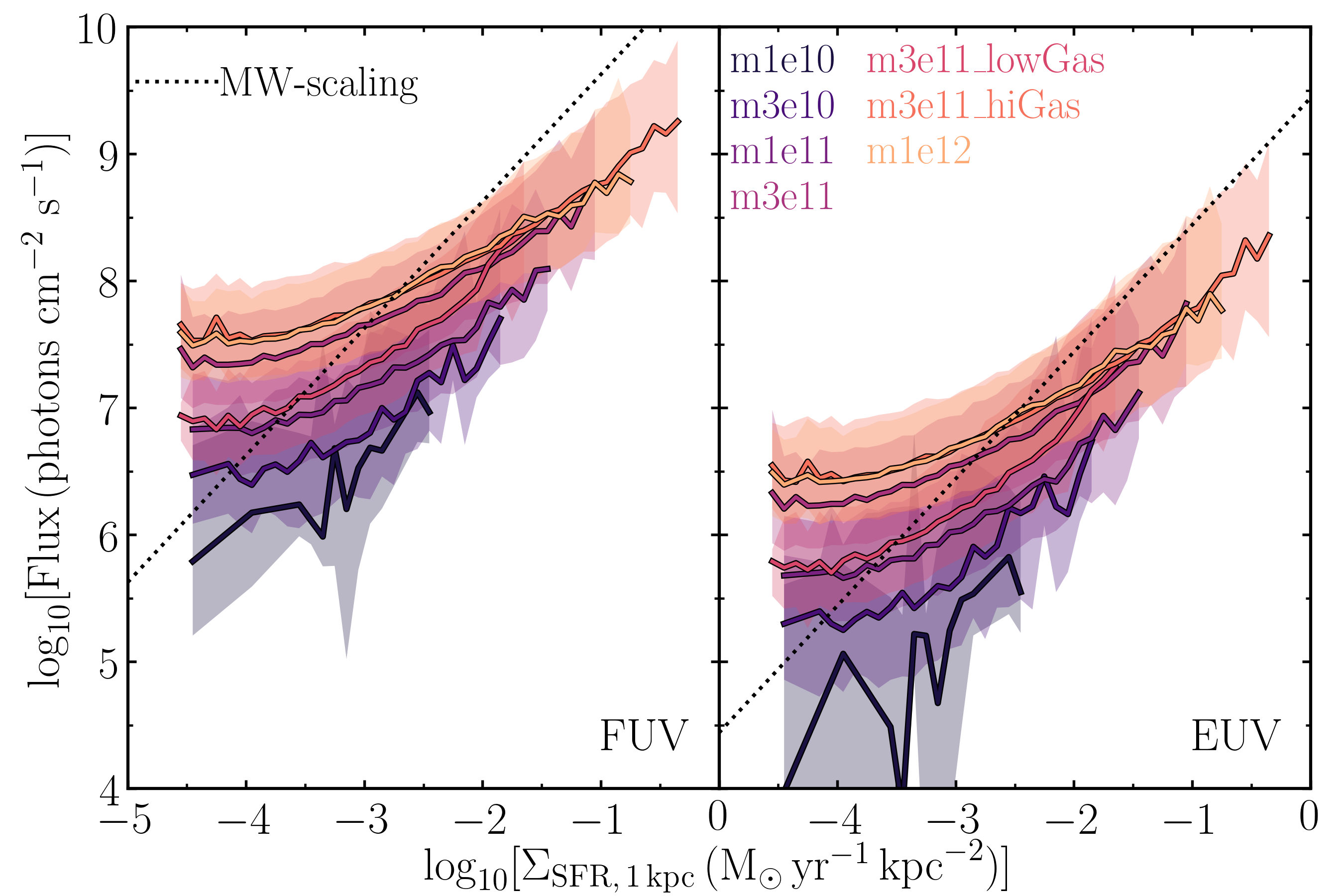}}
\vspace{-0.15 in}
\caption{As Fig.~\ref{flux_fullDisc_fig}, but for star formation rate surface densities averaged over 1~kpc regions ($\Sigma_{\rm{SFR,} \, 1 \, \rm{kpc}}$). The fluxes no longer follow a linear scaling, which suggests they are determined predominantly by star formation over the whole disc rather than just the local star forming properties. 
\vspace{-0.15 in}} 
\label{flux_1kpc_fig}
\end{figure}

The strong linear scaling between stellar flux and $\Sigma_{\rm{SFR,} \, \rm{disc}}$ seen in Fig.~\ref{flux_fullDisc_fig} suggests that the fluxes are driven by star formation over the whole disc of the galaxy. However, we might expect that the strongest contribution to the stellar fluxes comes from star formation in the local region. We therefore explored how the fluxes depend on the local star formation rate surface density, $\Sigma_{\rm{SFR,} \, 1 \, \rm{kpc}}$, calculated in two-dimensional cells 1~kpc across viewing the disc of the galaxy face-on. The median and tenth to ninetieth percentile fluxes in bins of $\Sigma_{\rm{SFR,} \, 1 \, \rm{kpc}}$ are shown in Fig.~\ref{flux_1kpc_fig}.

We see that the fluxes no longer follow a linear scaling with star formation surface density when averaged on 1~kpc scales. The slope of this relation flattens towards lower $\Sigma_{\rm{SFR,} \, 1 \, \rm{kpc}}$. This suggests that, in regions with relatively little star formation, additional contributions to the stellar fluxes from other regions of the galaxy with higher star formation rates dominate, which increases the total flux beyond what we would expect from local star formation alone. Thus the fluxes in our simulations are driven by star formation over the whole disc of the galaxy, and not just local star formation.

This conclusion may however by a consequence of the assumptions in the approximate \textsc{lebron} radiative transfer method, which assumes that radiation is only attenuated locally around the emitting star particle and the receiving gas particle. It therefore neglects absorption through the plane of the galaxy disc between widely separate regions, which might otherwise shield a gas cloud from young stars on the opposite side of the disc. We may therefore underestimate the spatial variations in the local stellar fluxes over the galaxy disc. To tackle this question more accurately would require a full 3D radiative transfer method coupled to the non-equilibrium chemistry network.

\subsection{Dust properties}\label{dust_properties_sect}

The empirical model for the depletion of metals onto dust grains described in Section~\ref{depl_sect} primarily aims to capture how the removal of metals from the gas phase affects the cooling and observable emission lines. However, it may also affect the total abundance of dust, as the dust to metals ratio in the fiducial model depends on gas density (see the bottom panel of Fig.~\ref{depl_fig}). In contrast, the simulations run with the no depletion model assume a constant dust to metals ratio. In this section, we compare the dust properties of our simulated galaxies with the fiducial and no depletion models to observations.

\begin{figure}
\centering
\mbox{
	\includegraphics[width=84mm]{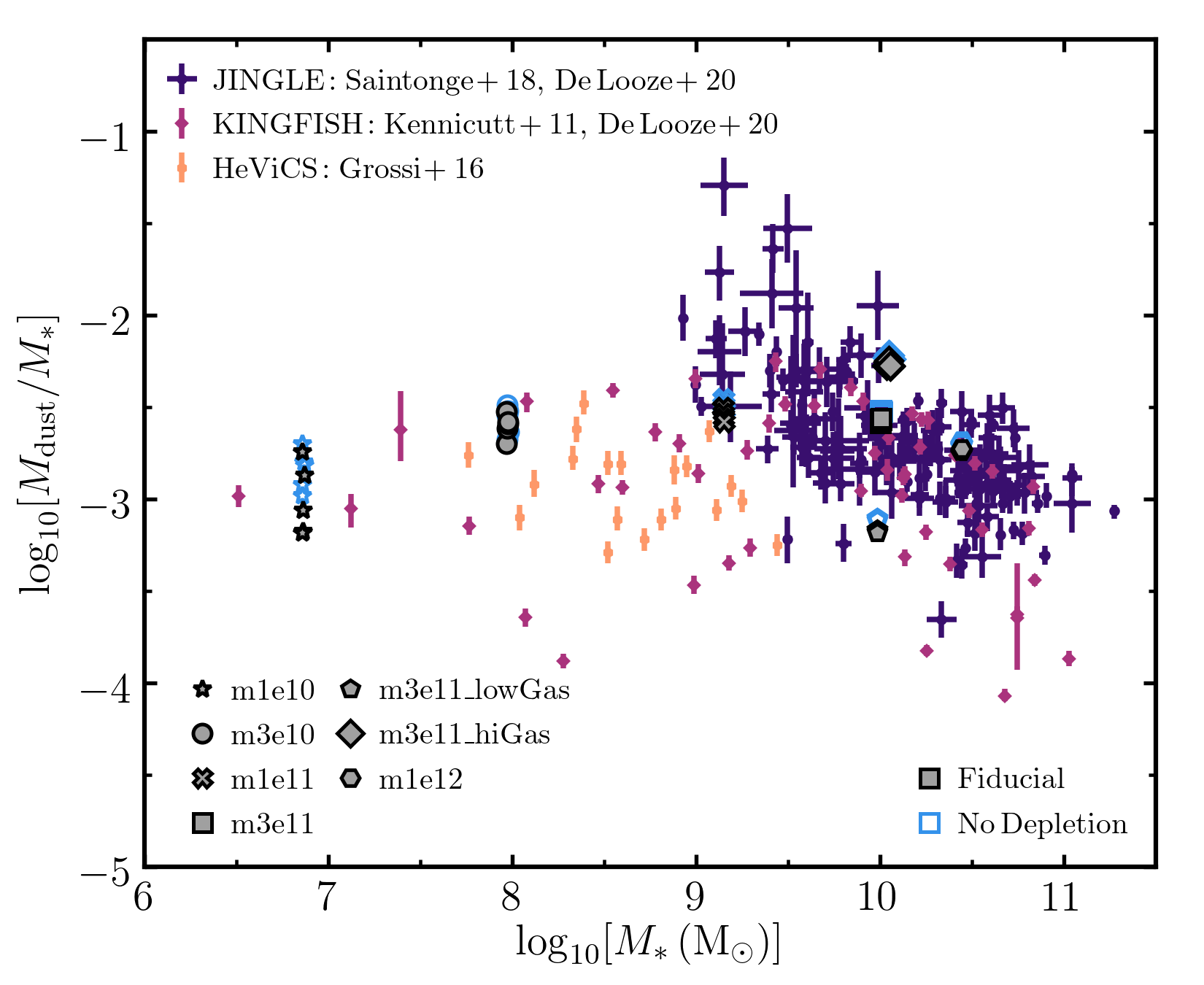}}
\vspace{-0.15 in}
\caption{The ratio of dust mass to stellar mass ($M_{\rm{dust}} / M_{\ast}$) versus stellar mass in our simulations with the fiducial model (grey symbols) and no depletion model (blue open symbols), compared to observations from the JINGLE (\citealt{saintonge18}; \citealt{delooze20}), KINGFISH (\citealt{kennicutt11}; \citealt{delooze20}), and HeViCS \citep{grossi16} surveys. The total dust content of the simulated galaxies are in good agreement with observed galaxies at the same stellar mass. There is little difference in dust mass between the fiducial and no depletion models. 
\vspace{-0.15 in}} 
\label{dust_mass_fig}
\end{figure}

Fig.~\ref{dust_mass_fig} compares the ratio of dust mass to stellar mass versus stellar mass in simulations with the fiducial model (grey symbols) and no depletion model (blue open symbols), calculated within the disc of the galaxy. For each simulation we show five snapshots at intervals of 100~Myr. 

We also plot observations from three galaxy surveys in the nearby Universe. The dark purple symbols show galaxies from the JINGLE survey \citep{saintonge18}, with dust masses measured by \citet{delooze20}. The stellar masses reported by \citet{saintonge18} assume a \citet{chabrier03} IMF. We convert these to a \citet{kroupa01} IMF, as used in the simulations, by multiplying by a factor of 1.06 (see equation~2 of \citealt{speagle14}). Galaxies from the KINGFISH survey \citep{kennicutt11} are shown by the light purple symbols, with stellar masses from \citet{hunt19} (also converted from a \citealt{chabrier03} to a \citealt{kroupa01} IMF as above), and dust masses from \citet{delooze20}. Finally, the light orange data points show a sample of 20 star-forming dwarf galaxies from \citet{grossi16}, which were selected from the \textit{Herschel} Virgo Cluster Survey (HeViCS; \citealt{davies12}). 

The dust to stellar mass ratios in the simulations overlap with observations at the same stellar mass, suggesting that we reproduce a realistic total abundance of dust. We also find little difference in the dust mass predicted by the fiducial and no depletion models. From the lower panel of Fig.~\ref{depl_fig}, we see that the dust to metals ratio ($DTM$) in the fiducial model only deviates significantly from the Milky Way value ($DTM_{\rm{MW}}$) at low densities. For example, $DTM / DTM_{\rm{MW}}$$>$0.5 at densities $\log_{10} [ n_{\rm{H}} \, (\rm{cm}^{-3})]$$>$$-3.3$ in our simulations (although as we note in Section~\ref{depl_sect}, our depletion model equates the particle density in the simulations to the average line of sight density in the observations, so we will underestimate the true local density for a given level of depletion). However, most of the gas mass in the galaxy disc is at much higher densities than this (see Section~\ref{Trho_sect}). Hence the empirical dust depletion model has little impact on the total dust abundance, although we will see in Section~\ref{emission_line_sect} that it does have a significant effect on observable emission lines arising from metals in the gas phase. 

Given that the dust depletion model does not have a strong effect on the dust to metals ratio, the total dust mass is determined by the total gas mass and metallicity. However, the scaling relations between stellar mass, gas fraction and metallicity were set in the initial conditions according to the redshift zero relations (see Section~\ref{IC_sect}). It may therefore seem unsurprising that we can reproduce the observed scaling between dust and stellar masses in the simulations.

\begin{figure}
\centering
\mbox{
	\includegraphics[width=84mm]{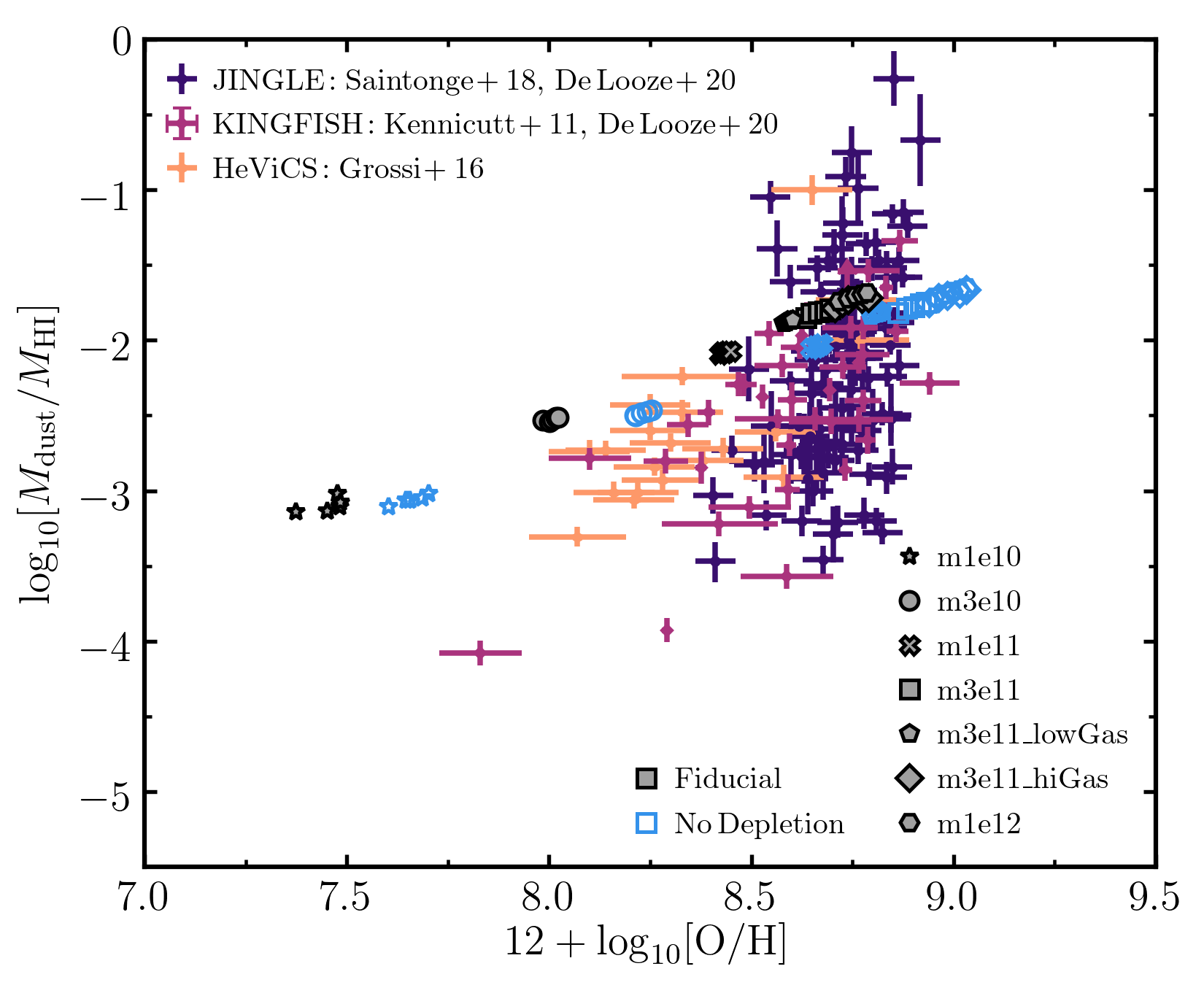}}
\vspace{-0.15 in}
\caption{The ratio of dust mass to H\textsc{i} mass ($M_{\rm{dust}} / M_{\rm{HI}}$) plotted against gas-phase oxygen abundance ($12 + \log_{10} [\rm{O/H}]$) in simulations with the fiducial model (grey symbols) and no depletion model (blue open symbols). The dark purple, light purple and light orange data points show observations from the JINGLE (\citealt{saintonge18}; \citealt{delooze20}), KINGFISH (\citealt{kennicutt11}; \citealt{delooze20}) and HeViCS \citep{grossi16} surveys, resepectively. The simulations closely follow a linear scaling between dust to H\textsc{i} ratio and metallicity, whereas the observations follow a super-linear relation with greater scatter. 
\vspace{-0.15 in}} 
\label{dust_fraction_fig}
\end{figure}

In Fig.~\ref{dust_fraction_fig} we plot the ratio of dust to H\textsc{i} mass ($M_{\rm{dust}} / M_{\rm{HI}}$) versus the average gas-phase oxygen abundance ($12 + \log_{10} [\rm{O/H}]$) over the galaxy disc in simulations with the fiducial (grey symbols) and no depletion (blue open symbols) models. We compare these to observations from JINGLE (dark purple symbols, with metallicities from \citealt{saintonge18} and H\textsc{i} masses from \citealt{durbala20}), KINGFISH (light purple symbols, with metallicities from \citealt{devis19} and H\textsc{i} masses from \citealt{remyruyer14}), and HeViCS (light orange symbols, with metallicities and H\textsc{i} masses from \citealt{grossi16}). The ratio $M_{\rm{dust}} / M_{\rm{HI}}$ can be used as a proxy for the dust to gas ratio \citep[e.g.][]{delooze20}, although it does not include the molecular gas phase. Nevertheless, it allows us to include galaxies with no molecular observations, and it avoids uncertainties in the conversion factors between observational tracers such as CO emission and total molecular mass \citep[e.g.][]{chiang21}, although there are still uncertainties in $M_{\rm{dust}}$ which can differ by up to a factor of 3 depending on the assumed dust emission model \citep[e.g.][]{chastenet21}. 

The simulations with the fiducial model follow a tight linear relation between $M_{\rm{dust}} / M_{\rm{HI}}$ and $12 + \log_{10} [\rm{O/H}]$, which is expected given that the dust to metals ratio in this model is almost constant, as discussed above. The no depletion model also follows a tight linear relation, but offset to higher metallicity by a factor $\approx$2. This offset arises because the $12 + \log_{10} [\rm{O/H}]$ abundance only includes oxygen in the gas phase, and so is reduced in the fiducial model due to depletion of oxygen onto dust grains. In the no depletion model this is not accounted for, and so oxygen atoms in dust grains are also counted in the gas phase.

At high metallicities, with $12 + \log_{10} [\rm{O/H}]$$>$8.5, the dust to gas ratios predicted by the simulations overlap with the observed ratios. However, the simulations do not reproduce the observed scatter in $M_{\rm{dust}} / M_{\rm{HI}}$ at fixed metallicity. This may be due to the idealised nature of our initial conditions, which we set according to redshift zero scaling relations between galaxy properties. Apart from the gas fraction, we do not consider the scatter in these scaling relations for the initial conditions, which may reflect the lack of scatter in dust to gas ratio.

Observational studies of the scaling between dust to gas ratio and metallicity have also noted a strong dependence on the galaxy's evolutionary stage \citep[e.g.][]{remyruyer14, delooze20}. Galaxies at an early stage of their evolution, which have not had time to process much of their gas reservoir into stars, have a lower fraction of their metals in dust grains. As our simulations do not capture a range of evolutionary histories for each set of galaxy parameters, this may also explain the lack of scatter in our results.

\citet{delooze20} find a best-fit relation between $M_{\rm{dust}} / M_{\rm{HI}}$ and $12 + \log_{10} [\rm{O/H}]$ with a logarithmic slope of 2.26$\pm$0.07, which is steeper than the linear relation that we find in our simulations with a slope of 1. \citet{remyruyer14} find that a linear relation provides a good fit to their observational data at high metallicities, $12 + \log_{10} [\rm{O/H}]$$\gtrsim$8, but they require a super-linear relation, with a slope of 2.02$\pm$0.28, at lower metallicities. \citet{devis19} also find a super-linear relation, with a slope of 2.15$\pm$0.11, while \citet{casasola20} observe a linear scaling. However, apart from \citet{delooze20}, these observational results include the molecular component in the dust to gas ratio. 

The shallower relation exhibited by our simulations compared to the JINGLE, KINGFISH and HeViCS samples results in dust to gas ratios that are up to an order of magnitude higher than observed at low metallicities (i.e. in the dwarf galaxies). This may suggest that the empirical dust model that we employed in our fiducial simulations, which relies on the correlation between depletion strength ($F_{\ast}$) and gas density ($n_{\rm{H}}$), may not extrapolate well to the dwarf galaxy regime. For example, in our model the dust to metals ratio saturates at the Milky Way value even in the dwarf galaxies. However, observations find lower dust to metals ratios in low-metallicity galaxies, which may be due to either less efficient growth of grains in the ISM or more efficient grain destruction in this regime \citep[e.g.][]{galliano21, priestley22}. This could explain the discrepancy in the M$_{\rm{dust}} / M_{\rm{HI}}$ ratios that we find in our dwarf galaxy simulations compared to the observational data.

Resolving this discrepancy may require live dust evolution models \citep[e.g][]{choban22}. However, \citet{mckinnon17} also find tensions between the predictions of their live dust evolution models for this relation and the observations (see their figure~8). In their fiducial model the slope of this relation is too flat, while their `M16' model reproduces the observed slope but with a normalisation that is too high. 

It might appear surprising that our simulated dwarf galaxies can reproduce the dust to stellar mass ratios seen in observations (Fig.~\ref{dust_mass_fig}) when the dust to gas ratio may be overestimated by an order of magnitude (Fig.~\ref{dust_fraction_fig}). If we instead plot $M_{\rm{dust}} / M_{\rm{HI}}$ versus stellar mass (not shown), we find that the simulated dwarf galaxies lie near the upper bound of the observed range in dust to gas ratio at fixed stellar mass, although they again do not reproduce the scatter in this relation. The discrepancy between the dwarf galaxy simulations and observations is therefore less pronounced at fixed stellar mass than at fixed metallicity. 

\section{The transition from atomic to molecular gas}\label{HI_H2_sect}

Molecular gas is a vital component of the ISM that is typically found to correlate with the star formation rate \citep[e.g.][]{bigiel11, leroy13, tacconi20}, although this correlation is not necessarily a causal relationship and may arise because molecules and star formation both require the gas to be shielded \citep{glover12}. Nevertheless, the transition from atomic to molecular gas is important for understanding the multiphase structure of the ISM and how the different ISM phases fuel star formation, and has been the topic of many studies both from an observational \citep[e.g.][]{gillmon06, shull21} and a theoretical \citep[e.g.][]{krumholz08, sternberg14} perspective.

The interstellar chemistry that drives this transition is sensitive to the local UV radiation field, which destroys molecules via photodissociation. Dust grains also play an important role as they can shield molecules from dissociating radiation, and can promote the formation of H$_{2}$ on grain surfaces. In this section, we explore how the variations in our models for local stellar radiation and dust depletion affect the properties of the atomic and molecular ISM phases, and compare our simulation predictions to observations.

\subsection{Thermodynamic properties of the interstellar gas}\label{Trho_sect}

\begin{figure}
\centering
\mbox{
	\includegraphics[width=84mm]{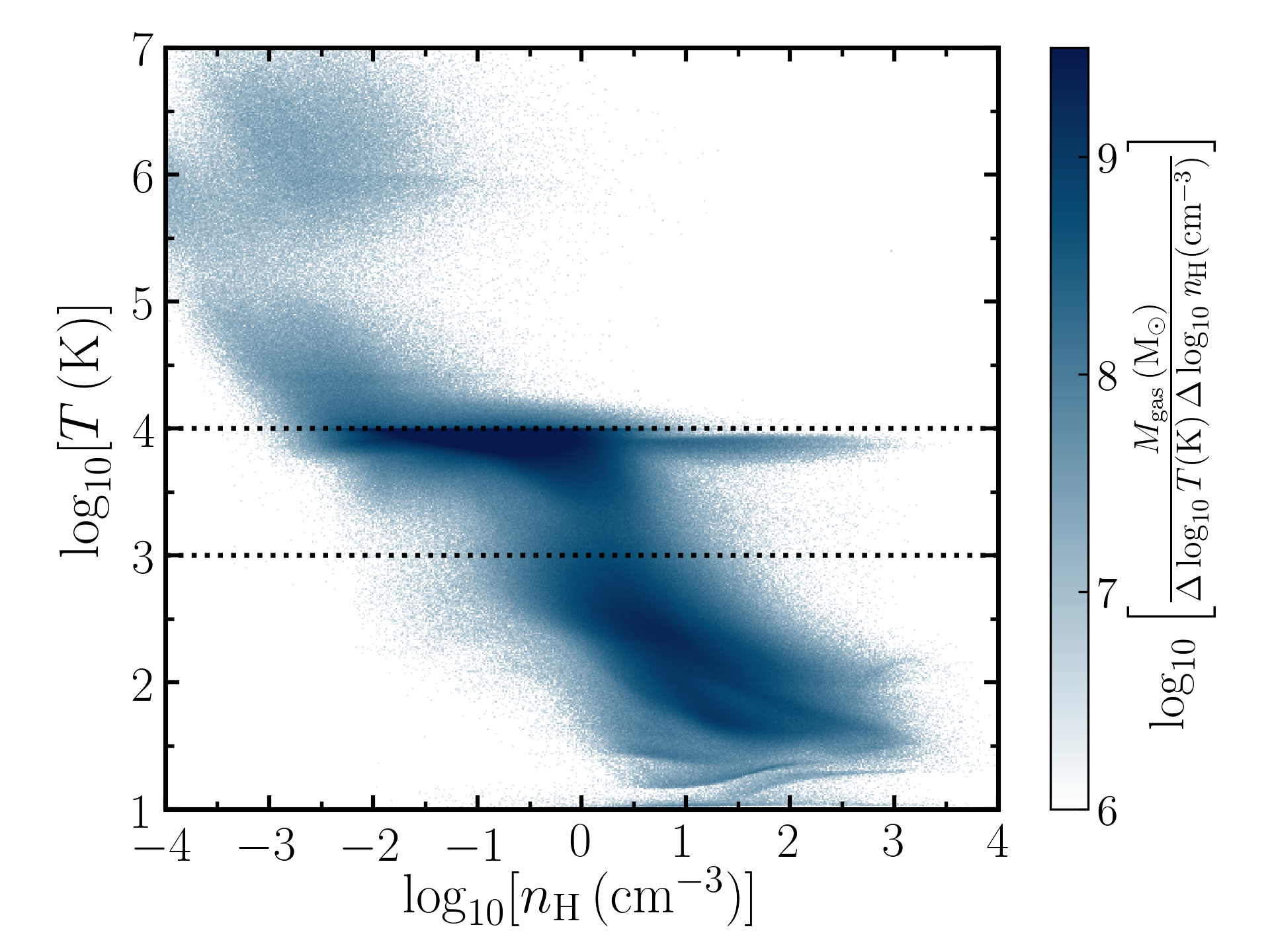}}
\vspace{-0.15 in}
\caption{Plot showing the distribution of gas mass as a function of temperature $T$ and density $n_{\rm{H}}$ in the simulation m1e12 using our fiducial model. We include all gas particles within the galaxy disc in the final snapshot of the simulation, after 500~Myr. To aid the discussion later in this section, we divide the gas into three temperature ranges: $T \! \leq \! 10^{3} \, \rm{K}$, $10^{3} \! < \! T \! \leq \! 10^{4} \, \rm{K}$ and $T \! > \! 10^{4} \, \rm{K}$, as delineated by the horizontal dotted lines.
\vspace{-0.15 in}} 
\label{Trho_fig}
\end{figure}

We start by looking at the temperature and density distribution of interstellar gas in our simulations. Fig.~\ref{Trho_fig} shows the temperature versus density for all gas in the disc of the galaxy in the simulation of m1e12 with the fiducial model. We see that the gas forms different phases in temperature-density space. At densities $n_{\rm{H}} \! \sim \! 0.01 \! - \! 1 \, \rm{cm}^{-3}$ most of the gas mass lies in a warm phase with temperatures close to $T \! \sim \! 10^{4} \, \rm{K}$. Thermal instabilities enable this warm phase to cool starting at $n_{\rm{H}} \! \sim \! 0.1 \, \rm{cm}^{-3}$, and by $n_{\rm{H}} \! \gtrsim \! 1 \, \rm{cm}^{-3}$ most of the gas mass is in the cold phase ($T \! < \! 10^{3} \, \rm{K}$). As we go to higher densities the ISM continues to cool, reaching temperatures $\lesssim \! 100 \, \rm{K}$ by $n_{\rm{H}} \! \gtrsim \! 10 \, \rm{cm}^{-3}$. Since lines of constant pressure follow a power law with a slope of $-1$ in this plot, we see that the different ISM phases are approximately in pressure equilibrium, albeit with a large scatter of more than 1 dex. This picture is qualitatively similar to other theoretical models for the thermodynamic structure of the multiphase ISM \citep[e.g.][]{wolfire03}.

The horizontal branch at temperatures just below $10^{4} \, \rm{K}$ and densities $\gtrsim \! 1 \, \rm{cm}^{-3}$ is due to H\textsc{ii} regions, where gas has been identified within the Str\"{o}mgren radius of a star particle and is there photoionised and photoheated by the stellar radiation (see Section~\ref{flux_sect}). The hot phase ($T \! > \! 10^{4} \, \rm{K}$) is created by stellar feedback in these simulations, as we do not include feedback from AGN, nor do we include a hot gaseous halo.

At low temperatures ($\lesssim$30~K) the gas forms distinct tracks in temperature-density space. We find that gas particles currently in this region have recently undergone a period of rapid cooling within the preceding few Myr, typically from temperatures of a few hundred Kelvin or more. In \citet{richings14a} we showed that non-equilibrium effects can enhance the cooling rate below $10^{4} \, \rm{K}$, which allows gas to initially cool below the thermal equilibrium temperature expected at the given density, before heating back up towards thermal (and chemical) equilibrium. The distinct tracks seen at low temperatures in Fig.~\ref{Trho_fig} are therefore likely due to particles piling up at the minimum temperature in this non-equilibrium evolution. Indeed, we will see below that the chemical abundances in this region are out of equilibrium. 

\begin{figure}
\centering
\mbox{
	\includegraphics[width=84mm]{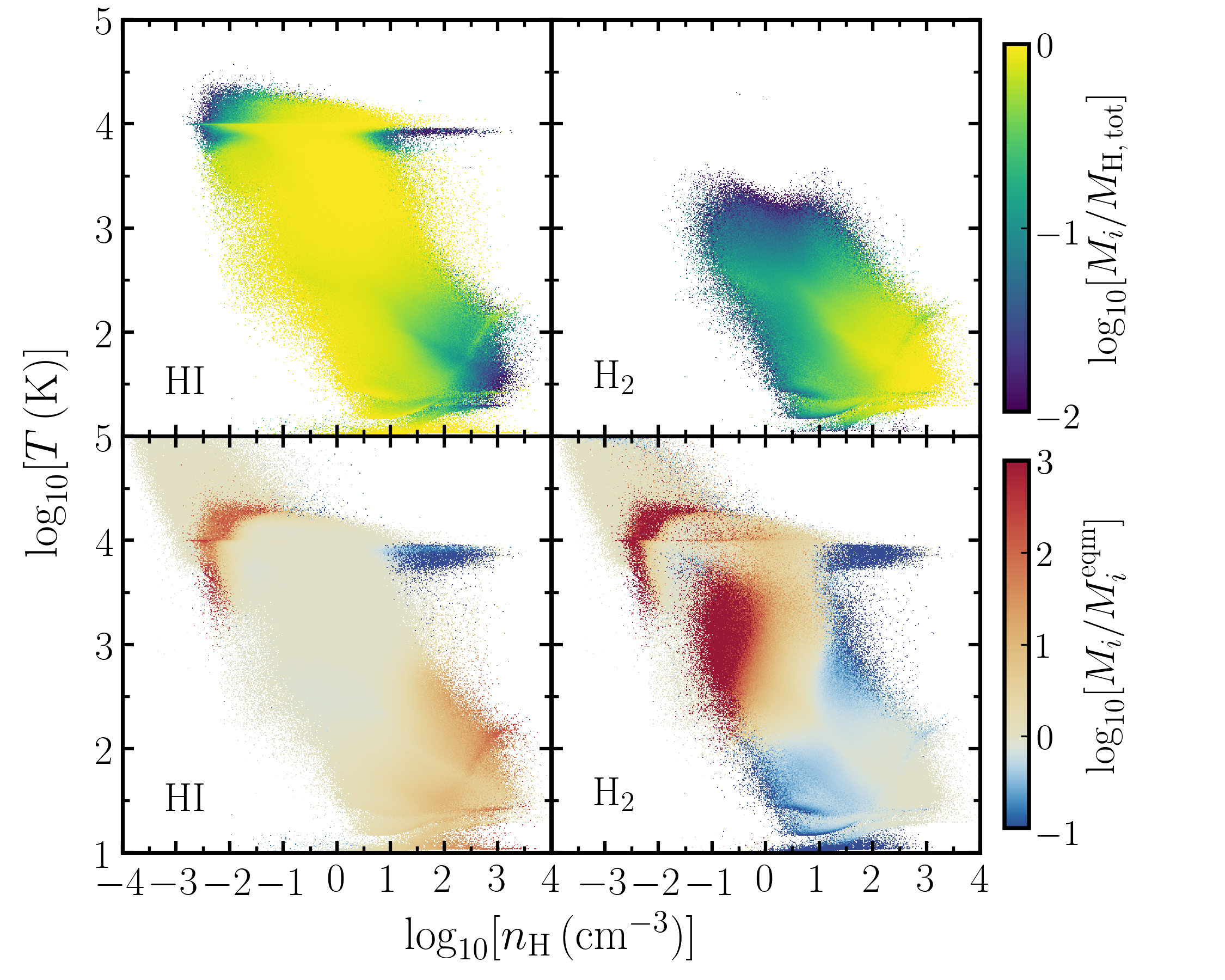}}
\vspace{-0.15 in}
\caption{\textit{Top row:} The fraction of total hydrogen mass in H\textsc{i} (left panel) and H$_{2}$ (right panel) as a function of temperature, $T$, and density, $n_{\rm{H}}$, in the disc of m1e12 using the fiducial model. \textit{Bottom row:} The ratio of non-equilibrium to equilibrium H\textsc{i} (left panel) and H$_{2}$ (right panel) as a function of $T$ and $n_{\rm{H}}$. Molecular hydrogen in particular shows strong non-equilibrium effects in certain regions of the $T \! - \! n_{\rm{H}}$ space, with an enhancement of more than 3 orders of magnitude in non-equilibrium at low densities ($n_{\rm{H}} \! \sim \! 0.1 \, \rm{cm}^{-3}$) and temperatures $T \! \sim \! 10^{3} \, \rm{K}$. 
\vspace{-0.15 in}} 
\label{Trho_HI_H2_fig}
\end{figure}

We can now look at the distribution of atomic and molecular hydrogen in the temperature-density phase space. The top row of Fig.~\ref{Trho_HI_H2_fig} shows the fraction of the total hydrogen mass in each species ($M_{i} / M_{\rm{H, \, tot}}$) as a function of temperature and density for m1e12 using the fiducial model, including all gas within the galaxy disc. We see that H\textsc{i} (left-hand panel) dominates at densities $n_{\rm{H}} \! \sim \! 0.01 \! - \! 10 \, \rm{cm}^{-3}$, covering a broad range of temperatures $T \! \sim 10 \! - \! 10^{4} \, \rm{K}$. Molecular hydrogen (right-hand panel) dominates at high densities, $n_{\rm{H}} \! \gtrsim \! 10 \, \rm{cm}^{-3}$, and is mostly found at temperatures $T \! \lesssim \! 300 \, \rm{K}$. However, non-negligible H$_{2}$ fractions can also be found at lower densities and higher temperatures than this. 

The species fractions in the top row of Fig.~\ref{Trho_HI_H2_fig} use the non-equilibrium chemical abundances from the simulations. However, as many simulations of galaxy formation assume that the species are in chemical equilibrium, it is interesting to explore the impact of non-equilibrium effects on our chemical predictions. We therefore ran the \textsc{chimes} chemistry solver on each gas particle from the simulation snapshots in post-processing to calculate the equilibrium chemical abundances.

The bottom row of Fig.~\ref{Trho_HI_H2_fig} shows the ratio of non-equilibrium to equilibrium species masses ($M_{i} / M_{i}^{\rm{eqm}}$) in m1e12 as a function of temperature and density. Molecular hydrogen (right-hand panel) shows particularly strong non-equilibrium effects. There is an enhancement of more than three orders of magnitude in the non-equilibrium H$_{2}$ fraction at $n_{\rm{H}} \! \sim \! 0.1 \, \rm{cm}^{-3}$ and $T \! \sim \! 10^{3} \, \rm{K}$, which is due to colder, denser molecular gas that was recently heated for example by stellar feedback but has not yet had sufficient time for the molecules to be fully destroyed. While H$_{2}$ does not dominate the total hydrogen budget in this region of the $T \! - \! n_{\rm{H}}$ space, the non-equilibrium H$_{2}$ fraction still reaches $\sim \! 1 \! - \! 10$ per cent here. At this temperature the ro-vibrational transitions of the H$_{2}$ molecule can be collisionally excited, so this non-equilibrium enhancement may have important consequences for observational predictions of the infrared H$_{2}$ emission lines. We will explore these predictions further in a future work. 

At higher densities, there are two distinct regions where the non-equilibrium H$_{2}$ fraction is suppressed by up to an order of magnitude, at $n_{\rm{H}} \! \sim \! 1 \! - \! 10 \, \rm{cm}^{-3}$, $T \! \lesssim \! 100 \, \rm{K}$ and $n_{\rm{H}} \! \sim 100 \, \rm{cm}^{-3}$, $T \! \sim \! 100 \! - \! 10^{3} \, \rm{K}$. These regions also coincide with enhancements in the non-equilibrium H\textsc{i} abundances (left-hand panel). These effects may be due to gas that was previously at higher temperatures and has recently cooled but has not yet had sufficient time for molecules to fully form. 

Gas in H\textsc{ii} regions, with $n_{\rm{H}} \! \gtrsim \! 1 \, \rm{cm}^{-3}$ and temperatures just below $10^{4} \, \rm{K}$, exhibits strongly suppressed H\textsc{i} and H$_{2}$ abundances in non-equilibrium. This is perhaps unsurprising given that H\textsc{ii} regions evolve on relatively short time-scales of a few~Myr \citep[e.g.][]{kim19}. However, even in equilibrium the mean H\textsc{i} and H$_{2}$ fractions in H\textsc{ii} regions are low ($\approx \! 0.05$ and $\approx \! 2 \! \times \! 10^{-8}$, respectively), since the hydrogen is predominantly ionised. 

Other studies using galaxy simulations with time-dependent models for the H$_{2}$ chemistry have also found strong non-equilibrium effects in the H$_{2}$ abundances \citep[e.g.][]{dobbs08, pelupessy09, richings16}, although \citet{gnedin11} find that an equilibrium treatment is sufficient to capture the atomic to molecular transition. 

\begin{figure}
\centering
\mbox{
	\includegraphics[width=84mm]{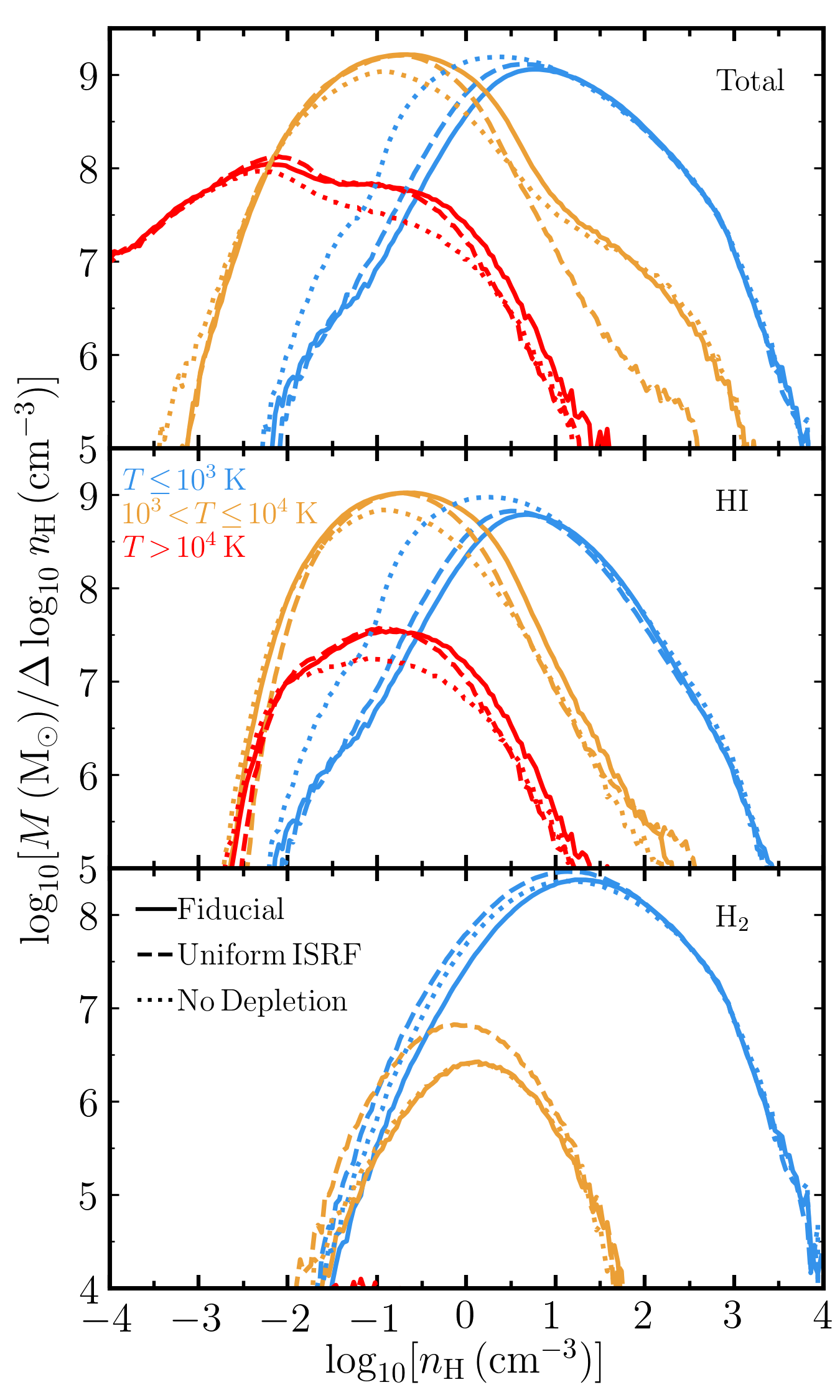}}
\vspace{-0.15 in}
\caption{Mass-weighted density distributions in low-temperature ($T \! \leq \! 10^{3} \, \rm{K}$; blue), intermediate-temperature ($10^{3} \! < \! T \! \leq \! 10^{4} \, \rm{K}$; orange) and high-temperature ($T \! > \! 10^{4} \, \rm{K}$; red) gas in m1e12 for the fiducial (solid), uniform ISRF (dashed) and no depletion (dotted) models. The top panel shows the total gas distribution, while the middle and bottom panels show the distributions of H\textsc{i} and H$_{2}$, respectively. Compared to the fiducial model, the uniform ISRF model contains less intermediate-temperature gas at $n_{\rm{H}} \! \gtrsim \! 10 \, \rm{cm}^{-3}$ and more intermediate-temperature H$_{2}$ at $n_{\rm{H}} \! \lesssim \! 10 \, \rm{cm}^{-3}$, while the no depletion model exhibits an enhancement in low-temperature H\textsc{i} gas at $n_{\rm{H}} \! \lesssim \! 10 \, \rm{cm}^{-3}$.   
\vspace{-0.15 in}} 
\label{nH_hist_fig}
\end{figure}

Figs.~\ref{Trho_fig} and \ref{Trho_HI_H2_fig} focussed on the m1e12 simulation using our fiducial model. To quantitatively compare the effects of local stellar radiation and dust depletion, we show in Fig.~\ref{nH_hist_fig} the one-dimensional density distributions in low-temperature ($T \! \leq \! 10^{3} \, \rm{K}$; blue), intermediate-temperature ($10^{3} \! < \! T \! \leq \! 10^{4} \, \rm{K}$; orange) and high-temperature ($T \! > \! 10^{4} \, \rm{K}$; red) gas. These temperature ranges are also illustrated in Fig.~\ref{Trho_fig}, and were chosen as they highlight the different phases and transitionary stages that are relevant to the ISM. We compare the fiducial (solid), uniform ISRF (dashed) and no depletion (dotted) models, using the final snapshot after 500~Myr from m1e12 in each case. The top, middle and bottom panels show the total gas, H\textsc{i} and H$_{2}$ distributions, respectively.

The overall gas distributions are fairly similar in the three models, with a few exceptions. In the uniform ISRF model, there is less intermediate-temperature gas at densities $n_{\rm{H}} \! \gtrsim \! 10 \, \rm{cm}^{-3}$ in the top panel of Fig.~\ref{nH_hist_fig}. This is due to H\textsc{ii} regions created by the photoionisation of gas within the Str\"{o}mgren radius around star particles, as we do not include the subgrid H\textsc{ii} region model in the uniform ISRF runs (see Section~\ref{flux_sect}). This difference is not seen in H\textsc{i} and H$_{2}$ (middle and bottom panels), as hydrogen is predominantly ionised in these regions.

In the bottom panel of Fig.~\ref{nH_hist_fig}, the distribution of intermediate-temperature H$_{2}$ is enhanced in the uniform ISRF model compared to the other two models. Given that in this temperature range the total available gas mass at these densities ($n_{\rm{H}} \! \lesssim \! 10 \, \rm{cm}^{-3}$) is similar between all three models, this suggests that the local treatment of stellar fluxes is more efficient at dissociating molecules in this regime. This has little impact on the total H$_{2}$ mass, which is dominated by low-temperature gas, but will be important for observational predictions of infrared H$_{2}$ emission, which arises from molecular gas at these intermediate temperatures.

Comparing the dotted and solid curves in the top and middle panels, we see that the no depletion model enhances the low-temperature H\textsc{i} component at $n_{\rm{H}} \! \lesssim \! 1 \, \rm{cm}^{-3}$. This is due to increased metal cooling when we do not reduce metals in the gas phase to account for depletion onto dust grains, which makes it easier for gas from the high- and intermediate-temperature phases to cool at these intermediate densities.

\subsection{Transition column density}\label{transition_sect}

\begin{figure}
\centering
\mbox{
	\includegraphics[width=84mm]{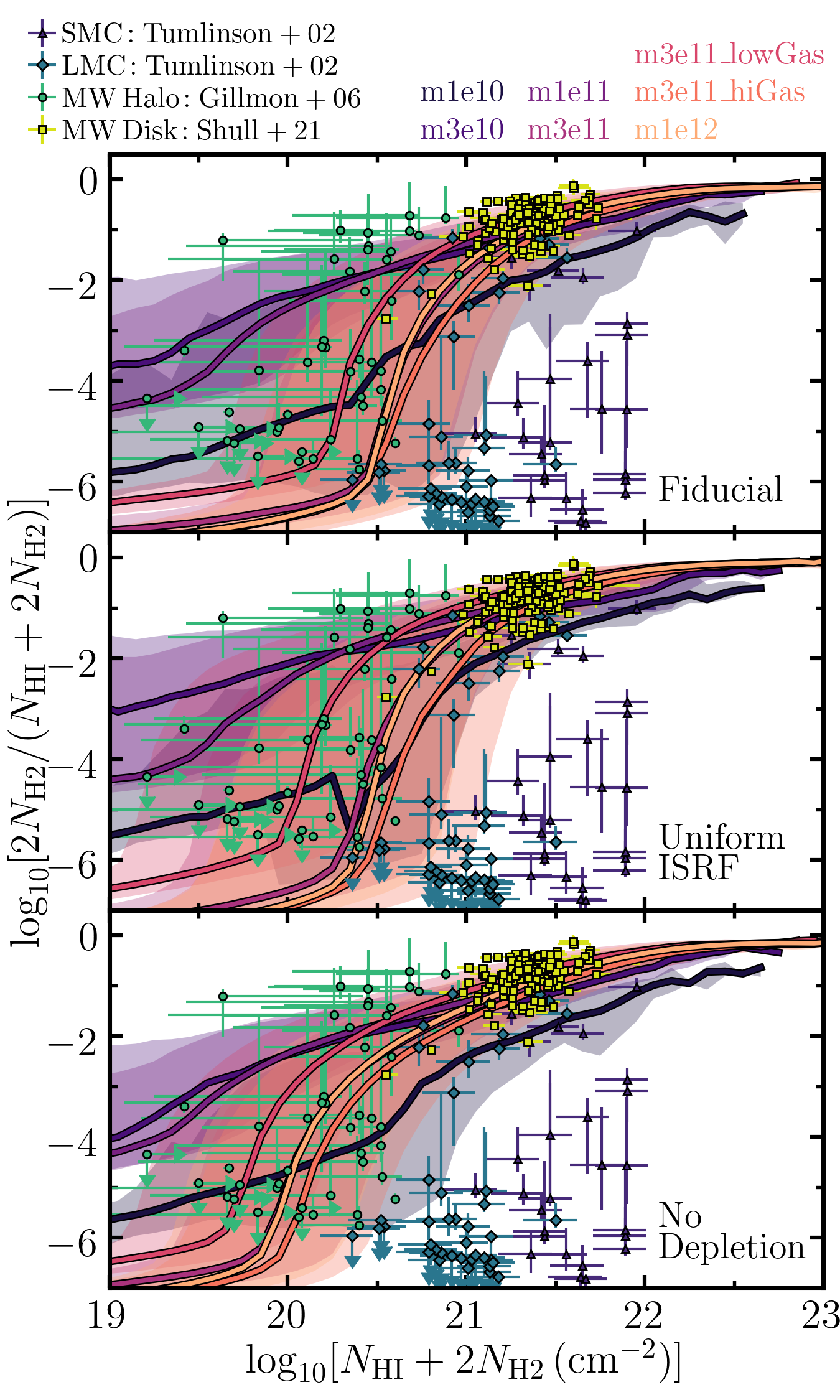}}
\vspace{-0.15 in}
\caption{The H$_{2}$ fraction plotted against total neutral hydrogen column density. The solid curves show the median H$_{2}$ fraction in bins of total column density from the simulations, measured in pixels 4~pc across viewing the galaxy disc face-on, while the shaded regions indicate the tenth to ninetieth percentile range in each bin. Different simulations are represented by different colours as shown in the legend. The top, middle and bottom panels show the fiducial, uniform ISRF, and no depletion models, respectively. The data points show observations from the Small Magellanic Cloud (SMC; \citealt{tumlinson02}; purple), Large Magellanic Cloud (LMC; \citealt{tumlinson02}; blue), the Milky Way at galactic latitudes $|b| \! > \! 20^{\circ}$ (MW Halo; \citealt{gillmon06}; green), and the Milky Way at $|b| \! < \! 10^{\circ}$ (MW Disk; \citealt{shull21}; yellow). While the high-mass simulations broadly overlap the Milky Way observations, the dwarfs only reproduce the highest H$_{2}$ fractions seen in the SMC or LMC.
\vspace{-0.15 in}} 
\label{H2_fraction_fig}
\end{figure}

The formation of the molecular phase requires that the gas becomes shielded from dissociating UV radiation. As the shielding is sensitive to the column density of the gas cloud, with higher column densities able to absorb a greater proportion of the incident radiation, it is therefore useful to look at the transition from atomic to molecular gas as a function of column density.

For each galaxy simulation, we create maps of the H\textsc{i} and H$_{2}$ column densities with pixels 4~pc across, viewing the galaxy disc face-on. We then bin the pixels according to the total neutral hydrogen column density, $N_{\rm{HI}} + 2 N_{\rm{H}2}$, combining five snapshots at 100~Myr intervals for each simulation, and calculate the median and tenth to ninetieth percentile H$_{2}$ fraction, $2 N_{\rm{H}2} / (N_{\rm{HI}} + 2 N_{\rm{H}2})$, in each bin. Fig.~\ref{H2_fraction_fig} shows the median (solid curves) and tenth to ninetieth percentile range (shaded regions) of the H$_{2}$ fraction versus neutral hydrogen column density in simulations using the fiducial (top panel), uniform ISRF (middle panel) and no depletion (bottom panel) models. Different galaxies are indicated by different colours, as shown in the legend.

The data points in Fig.~\ref{H2_fraction_fig} show observed H\textsc{i} and H$_{2}$ column densities along lines of sight in the Small and Large Magellanic Clouds (SMC and LMC; purple and blue respectively) from \citet{tumlinson02}, and in the Milky Way at galactic latitudes $|b| \! > \! 20^{\circ}$ (MW Halo; green) from \citet{gillmon06} and at $|b| \! < \! 10^{\circ}$ (MW Disk; yellow) from \citet{shull21}. These observational studies all measure H$_{2}$ column densities from far-UV absorption lines in the Lyman Werner bands using the Far Ultraviolet Spectrographic Explorer (FUSE) telescope. The H\textsc{i} column densities used in \citet{shull21} were obtained by fitting Ly$\alpha$ absorption, while \citet{tumlinson02} and \citet{gillmon06} derive H\textsc{i} column densities from 21~cm emission. 

In the galaxy simulations with the highest masses (m3e11 to m1e12), the molecular phase dominates at neutral column densities $\gtrsim \! 10^{22} \, \rm{cm}^{-2}$. The H$_{2}$ fraction declines steeply at lower column densities, reaching median fractions below $10^{-6}$ at column densities $\lesssim \! 10^{20.5} \, \rm{cm}^{-2}$. The median and tenth to ninetieth percentile range of H$_{2}$ fractions in these high-mass simulations broadly overlap the two observational samples in the Milky Way (green and yellow points), particularly at the higher column densities of the MW Disk sample. 

In the three simulations with a halo mass of $3 \! \times \! 10^{11} \, \rm{M}_{\odot}$, the transition from H\textsc{i} to H$_{2}$ moves towards higher column densities as the disc gas fraction increases. This may be due to the increasing star formation rate with increasing gas fraction, which increases the strength of the interstellar radiation field in the galaxy disc and hence tends to increase the column density of the atomic to molecular transition \citep[e.g.][]{schaye04, sternberg14}. 

Compared to the high-mass galaxies, the simulations m3e10 and m1e11 exhibit larger H$_{2}$ fractions at column densities $\lesssim \! 10^{20.5} \, \rm{cm}^{-2}$, resulting in a more gradual transition from atomic to molecular gas, whilst in the lowest mass galaxy in our sample, m1e10, the H$_{2}$ fractions are lower than the intermediate-mass galaxies at all column densities. We therefore do not see a monotonic trend in the H\textsc{i} to H$_{2}$ transition with halo mass. However, we caution that, while the H$_{2}$ fractions in m3e11 and m1e12 exhibit good numerical convergence, those in the lower-mass galaxies show significant differences at low column densities ($<$10$^{21} \, \rm{cm}^{-2}$) between runs with different resolutions (see Appendix~\ref{resolution_sect}). In m1e11 the H$_{2}$ fractions at low column densities increase from low to standard resolution, while in m3e10 they decrease from standard to high resolution. The low-column density trends that we see for the dwarf galaxies in Fig.~\ref{H2_fraction_fig} are therefore not robust. 

The structural properties of the LMC are closest to our m1e11 simulated galaxy, while the SMC is nearest to m1e10. If we compare these simulations to the observational data from \citet{tumlinson02} in Fig.~\ref{H2_fraction_fig}, the simulations lie close to the highest H$_{2}$ fractions measured in the LMC and SMC. However, there are many sight lines in these two observational samples with much lower H$_{2}$ fractions than are found in the simulations, by up to 4 orders of magnitude at the same column density.

This discrepancy suggests that our simulations do not correctly capture the atomic to molecular transition in the dwarf galaxy regime. However, in the simulations we calculate the column densities by projecting the gas onto a grid viewing the galaxy disc face on, rather than modelling mock observations of H2 absorption spectra along lines of sight. We therefore do not determine column densities in the same way as the observations, and we do not capture the same selection effects that may be present in the observational samples. The latter approach of using mock absorption spectra would allow for a more direct comparison between the simulations and observations, but such an analysis is beyond the scope of this work. 

As noted above, the low-column density H$_{2}$ fractions in dwarf galaxies are not well converged when we vary the numerical resolution. However, running m3e10 at 8 times higher mass resolution did not improve the agreement with observational data at high column densities ($\gtrsim$10$^{21} \, \rm{cm}^{-2}$; see Appendix~\ref{resolution_sect}). The discrepancies between our dwarf galaxies and observations of the LMC and SMC are therefore unlikely to be caused by limited numerical resolution alone. 

In Section~\ref{flux_sect}, we cautioned that our shielding model uses a single, average column density for the local gas cloud, which will tend to overestimate the strength of the shielding as the photochemical rates are typically dominated by the lines of sight at low column densities. This would lead to higher molecular abundances, which could contribute to the discrepancy in the atomic to molecular transitions that we see between our model predictions for dwarf galaxies and the observations of the LMC and SMC. 

In the top and middle panels of Fig.~\ref{H2_fraction_fig} we see that the trends of H$_{2}$ fraction with column density are similar in the fiducial and uniform ISRF models. This suggests that the large local variations in stellar flux seen in Fig.~\ref{m1e12_flux_fig} with the fiducial model do not have a strong impact on the atomic to molecular transition. This result is consistent with the theoretical studies of \citet{schaye04} and \citet{krumholz09}, who concluded that the transition from H\textsc{i} to H$_{2}$ is driven primarily by column density and secondarily by metallicity, with a weaker dependence on the incident radiation field.

Comparing the top and bottom panels of Fig.~\ref{H2_fraction_fig}, we find that the no depletion model exhibits higher H$_{2}$ fractions at low column densities ($\lesssim \! 10^{21} \, \rm{cm}^{-2}$) in high-mass galaxies compared to the fiducial model, resulting in a shallower transition between the atomic and molecular phases. This is due to the increased dust abundance at low densities in the no depletion model. However, at higher column densities the H$_{2}$ fraction is similar in the two models, as it is dominated by high-density gas for which the dust to metals ratio is identical in both cases. 

\subsection{Global H\textsc{i} and H$_{2}$ properties}\label{global_HI_H2_sect} 

In the previous section we studied the H$_{2}$ fraction along individual lines of sight through each galaxy. We now consider how the total mass of atomic and molecular hydrogen in each galaxy ($M_{\rm{HI}}$ and $M_{\rm{H2}}$, respectively) vary with stellar mass. While the total gas mass in the galaxy simulations was determined from the stellar mass in the initial conditions according to observed redshift zero scaling relations (see Section~\ref{IC_sect}), the partitioning of the gas between atomic and molecular phases remains a prediction of the chemical modelling. 

\begin{figure}
\centering
\mbox{
	\includegraphics[width=84mm]{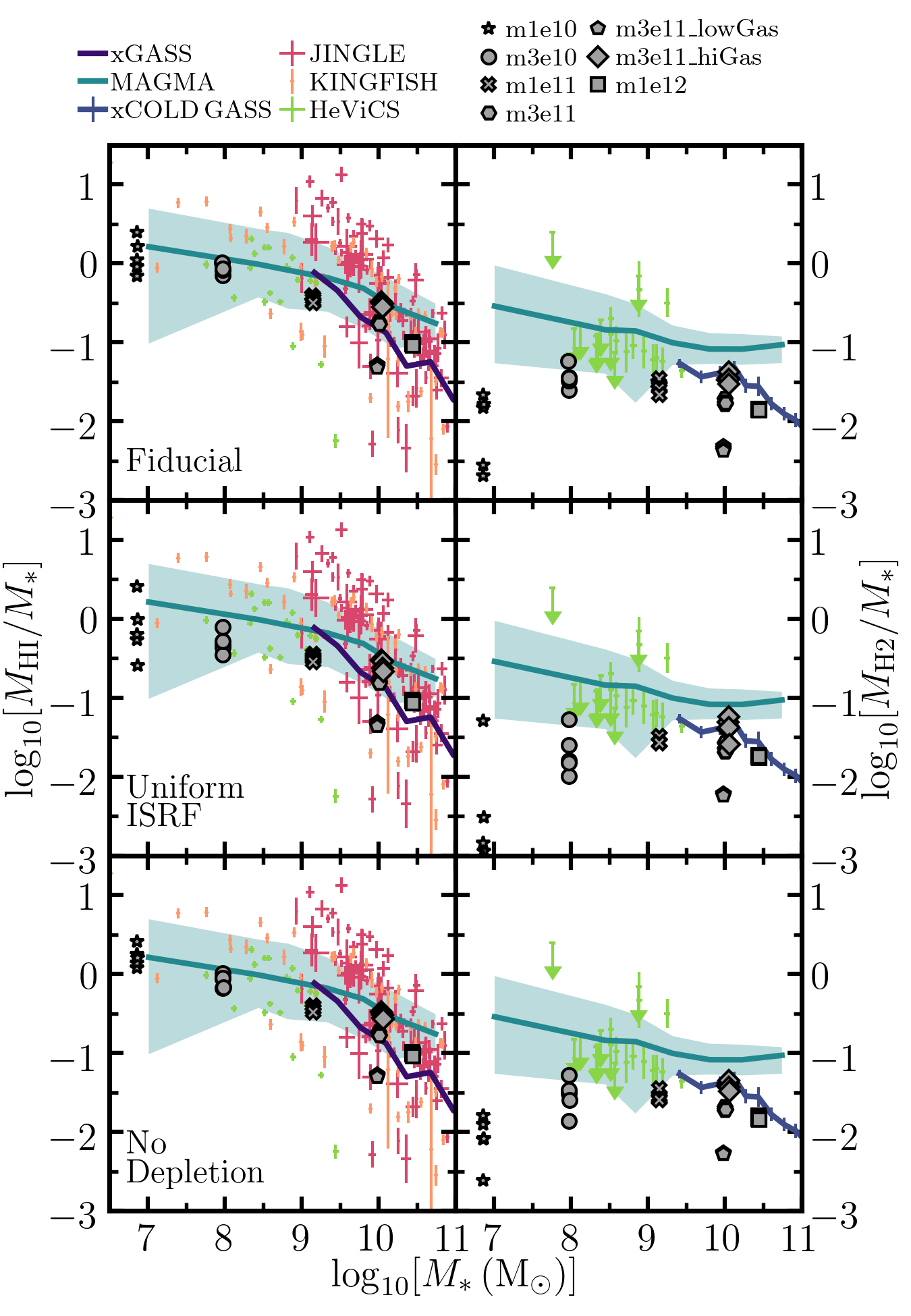}}
\vspace{-0.15 in}
\caption{The H\textsc{i} to stellar mass ratio ($M_{\rm{HI}} / M_{\ast}$; left-hand column) and H$_{2}$ to stellar mass ratio ($M_{\rm{H2}} / M_{\ast}$; right-hand column) plotted against stellar mass. The grey symbols show five snapshots from each simulated galaxy at 100~Myr intervals, using the fiducial (top row), uniform ISRF (middle row) and no depletion (bottom row) models. The solid curves show observed median relations from the xGASS \citep{catinella18}, MAGMA \citep{hunt20} and xCOLD GASS \citep{saintonge17} surveys. For the MAGMA sample we also show the $\pm 1 \sigma$ deviations in each stellar mass bin, indicated by the corresponding shaded regions. The coloured data points show individual galaxies from the JINGLE \citep{saintonge18}, KINGFISH \citep{kennicutt11} and HeViCS \citep{grossi16} surveys. The simulations are in good agreement with observed H\textsc{i} ratios. While the high-mass simulations also reproduce the median H$_{2}$ ratios, the simulated dwarf galaxies underpredict the molecular fraction.
\vspace{-0.15 in}} 
\label{global_HI_H2_fig}
\end{figure}

Fig.~\ref{global_HI_H2_fig} shows the ratios $M_{\rm{HI}} / M_{\ast}$ (left-hand column) and $M_{\rm{H2}} / M_{\ast}$ (right-hand column) plotted against $M_{\ast}$ in our simulated galaxies (grey symbols). We include five snapshots from each galaxy at intervals of 100~Myr. The top, middle and bottom rows show the fiducial, uniform ISRF and no depletion models, respectively. 

The scaling between H\textsc{i}, H$_{2}$ and stellar components of galaxies has also been the subject of many observational studies. The solid curves in Fig.~\ref{global_HI_H2_fig} show observed scaling relations from three low-redshift galaxy surveys. For the xGASS survey, we plot the median H\textsc{i} ratios in bins of stellar mass reported in Table~1 of \citet{catinella18}. For the MAGMA survey, we show median H\textsc{i} and H$_{2}$ ratios in stellar mass bins from figure~6 of \citet{hunt20}, together with the $\pm 1 \sigma$ deviations in each bin as indicated by the shaded region. For the xCOLD GASS survey, we show the median H$_{2}$ ratios for the whole sample given in Table~6 of \citet{saintonge17}, where the error bars denote the uncertainty in the median ratio for each bin. The H$_{2}$ masses reported in \citet{saintonge17} include the contribution from helium, however in our simulations we only show the H$_{2}$ mass. We have therefore divided the H$_{2}$ masses from \citet{saintonge17} by a factor of 1.36 to remove the helium correction. Finally, the coloured data points in Fig.~\ref{global_HI_H2_fig} show individual galaxies from the JINGLE \citep{saintonge18}, KINGFISH \citep{kennicutt11, hunt19}, and HeViCS \citep{grossi16} surveys. As noted in Section~\ref{dust_properties_sect}, we multiply the stellar masses from the JINGLE and KINGFISH surveys by 1.06 to convert from the \citet{chabrier03} IMF assumed in the observations to the \citet{kroupa01} IMF used in the simulations \citep{speagle14}. We also applied this conversion to the stellar masses in the xGASS, xCOLD GASS and MAGMA samples, which also assumed a \citet{chabrier03} IMF. 

The simulated H\textsc{i} fractions in the left-hand column of Fig.~\ref{global_HI_H2_fig} are in good agreement with the observational data, lying well within the scatter of the observed relations. The simulations reproduce the trend of decreasing $M_{\rm{HI}} / M_{\ast}$ with increasing stellar mass, with the H\textsc{i} fraction decreasing by approximately 1~dex from m1e10 to m1e12. 

In the right-hand column of Fig.~\ref{global_HI_H2_fig}, the H$_{2}$ fractions of the high-mass galaxy simulations ($M_{200, \, \rm{crit}} \! \geq \! 3 \times 10^{11} \, \rm{M}_{\odot}$, or $M_{\ast} \! \gtrsim \! 10^{10} \, \rm{M}_{\odot}$) are close to the observed median relation from \citet{saintonge17}, except for m3e11\_lowGas which exhibits lower H$_{2}$ masses due to the lower total gas fraction. 

At lower masses, the simulations increasingly appear to underpredict the H$_{2}$ mass expected for their stellar mass, with the dwarf galaxies generally lying below the $\pm 1 \sigma$ scatter of the MAGMA sample. However, as the observed molecular masses were derived from CO luminosities, the observational samples are more sensitive to high-H$_{2}$ fraction galaxies. For example, the MAGMA sample only includes galaxies that have been detected in CO, while 9 out of the 20 dwarfs in the HeViCS sample are upper limits. We do not account for these selection effects in our simulations, so it is unclear whether these apparent discrepancies are a true failing of the model or arise simply because the handful of dwarf galaxies in our simulated sample would not be included in the observed surveys. There are also uncertainties in converting CO luminosity to H$_{2}$ mass \citep[e.g.][]{bolatto13,chiang21}, particularly at low metallicities. As the \textsc{chimes} chemistry network includes CO, we will explore the relations between CO luminosity and H$_{2}$ mass further in a future work. 

The apparent discrepancy between simulations and observations in the dwarf galaxies in the right-hand column of Fig.~\ref{global_HI_H2_fig} may also seem to contradict the results of Fig.~\ref{H2_fraction_fig}, in which we saw that the dwarf galaxy simulations cannot reproduce the very low H$_{2}$ fractions seen in many sight lines through the LMC and SMC. However, the LMC and SMC observations from \citet{tumlinson02} measure H$_{2}$ in absorption rather than from CO emission, and so the observational data sets in Figs.~\ref{H2_fraction_fig} and \ref{global_HI_H2_fig} are subject to different selection effects and may be probing different regimes. 

Comparing the three rows in Fig.~\ref{global_HI_H2_fig}, we find that the total atomic and molecular components are similar in the fiducial, uniform ISRF and no depletion models. This agrees with our results from Section~\ref{transition_sect}, in which we saw that the transition from atomic to molecular hydrogen is not strongly affected by the treatment of local versus uniform stellar fluxes, while the inclusion of metal depletion from the gas phase onto dust grains only affects H$_{2}$ fractions at low column densities ($\lesssim$$10^{21} \, \rm{cm}^{-2}$) which do not dominate the total molecular mass. 

\section{Emission line tracers of the star formation rate}\label{emission_line_sect} 

There are a wide range of observational diagnostics that are commonly used to determine the star formation rate \citep{kennicutt98, kennicutt12, davies16}. As young, massive stars emit predominantly at UV wavelengths, observations of continuum UV emission can measure recent star formation activity, although such observations are sensitive to dust. These can be supplemented with observations of the infrared continuum to account for UV radiation that has been absorbed by dust grains and re-emitted at longer wavelengths. The star formation rate can also be inferred from emission lines produced by species that are photoionised by massive stars. The H$\alpha$ line is perhaps the most famous example \citep[e.g.][]{kennicutt94}, although as it lies at optical wavelengths it is also affected by dust attenuation. Far infrared (FIR) lines from metal ions have also been shown to correlate with the star formation rate \citep{delooze14}, and are less sensitive to dust effects. As these observational tracers probe star formation on different time-scales, they can also depend on the recent star formation history in the galaxy \citep[e.g.][]{sparre17, floresvelazquez21}. 

In this section we study emission line predictions from our simulations and how they correlate with the total star formation rate, which we compare to observed galaxy surveys. The detailed interstellar chemistry modelled in these simulations will be important for calculating these emission lines, as it determines the relative abundances of the ions involved. As the star formation rate is inferred from the emission lines based on how the young stars photoionise the surrounding gas, we would expect that the treatment of the stellar radiation will also play a vital role. 

Other studies have explored predictions for line emission from cosmological simulations based on subgrid models \citep[e.g.][]{hirschmann17, olsen21}. These approaches have the advantage that they do not need to explicitly resolve the regions that produce the emission, as they are treated in a subgrid fashion, which is particularly important for large-scale simulations of the Universe. However, they rely on assumptions for the structure of the unresolved components, and they do not capture effects of non-equilibrium chemistry. Our simulation predictions in this work do account for the non-equilibrium chemistry, but rely on explicitly resolving the emitting regions, and so they offer a complementary approach to these subgrid models. 

\subsection{Modelling line emission in post-processing}\label{line_emission_method_sect} 

We calculate the emission lines from our simulations by post-processing the simulation outputs with the publicly available radiative transfer code \textsc{radmc-3d} \citep{dullemond12}, which follows the emission, propagation and absorption of spectral lines together with stellar emission and the absorption, scattering and thermal emission from dust grains. 

As \textsc{radmc-3d} operates on a grid, we first construct an Adaptive Mesh Refinement (AMR) grid from the particle distribution in the simulation. Each cell is refined until it contains no more than 8 gas and/or star particles. The non-equilibrium ion and molecule abundances of each gas particle, together with the ion-weighted temperatures and velocities, are then projected onto the AMR grid, using the same smoothing kernel as the MFM hydro solver. The star particles are also smoothed and projected onto the grid, split between the eight stellar age bins with spectra shown in Fig.~\ref{SB99_fig}. 

We include graphite and silicate grains in our \textsc{radmc-3d} calculations. We take the abundance of graphite and silicate grains at solar metallicity from the `ISM' grain abundances in v13.01 of the \textsc{cloudy} photoionisation \citep{mathis77, ferland13}, which are typical of the ISM in the Milky Way. For each gas particle we scale these grain abundances by the total metallicity relative to solar $(Z / \rm{Z}_{\odot})$. We then scale these by the density-dependent dust to metals ratio predicted by our empirical dust depletion model ($DTM / DTM_{\rm{MW}}$, i.e. the bottom panel of Fig.~\ref{depl_fig}), except for simulations run with the no depletion model for which we assume $DTM / DTM_{\rm{MW}} = 1$. We thus obtain a dust to gas ratio of $2.4 \times 10^{-3} (DTM / DTM_{\rm{MW}}) (Z / \rm{Z}_{\odot})$ and $4.0 \times 10^{-3} (DTM / DTM_{\rm{MW}}) (Z / \rm{Z}_{\odot})$ for graphite and silicate grains, respectively. The dust temperature in each cell of the AMR grid is calculated by \textsc{radmc-3d} using the stellar radiation. 

The level populations of ions and molecules in each cell are calculated in \textsc{radmc-3d} from the gas properties and non-equilibrium species abundances. We use the Local Velocity Gradient (LVG) method to calculate the level populations, as this approximates the effects of non-Local Thermodynamic Equilibrium (non-LTE). We use atomic data and collisional excitation rates from the \textsc{lamda}\footnote{\url{https://home.strw.leidenuniv.nl/~moldata/}} \citep{schoier05} and \textsc{chianti}\footnote{\url{https://www.chiantidatabase.org}} \citep{dere97, landi13} databases. The line emissivity in each cell can then be calculated from the level populations. As H$\alpha$ emission is typically dominated by recombination (which is not accounted for in the calculation of level populations in \textsc{radmc-3d}), we instead calculate the H$\alpha$ emissivities due to recombination of H\textsc{ii} and collisional excitation of H\textsc{i} for each cell in the \textsc{radmc-3d} AMR grid using rates from \citet{raga15}.

For each emission line we produce a 3D data cube in position-position-velocity space, with velocities spanning $\pm 200 \, \rm{km} \, \rm{s}^{-1}$ about the line centre at a spectral resolution of $2 \, \rm{km} \, \rm{s}^{-1}$, and a spatial resolution of $20 \, \rm{pc}$. We repeat the \textsc{radmc-3d} calculation with emission lines disabled to determine the continuum spectrum from thermal dust emission and starlight, which we subtract from the total emission to obtain the line emission. 

\subsection{Synthetic emission line predictions}\label{line_prediction_sect} 

\begin{figure}
\centering
\mbox{
	\includegraphics[width=84mm]{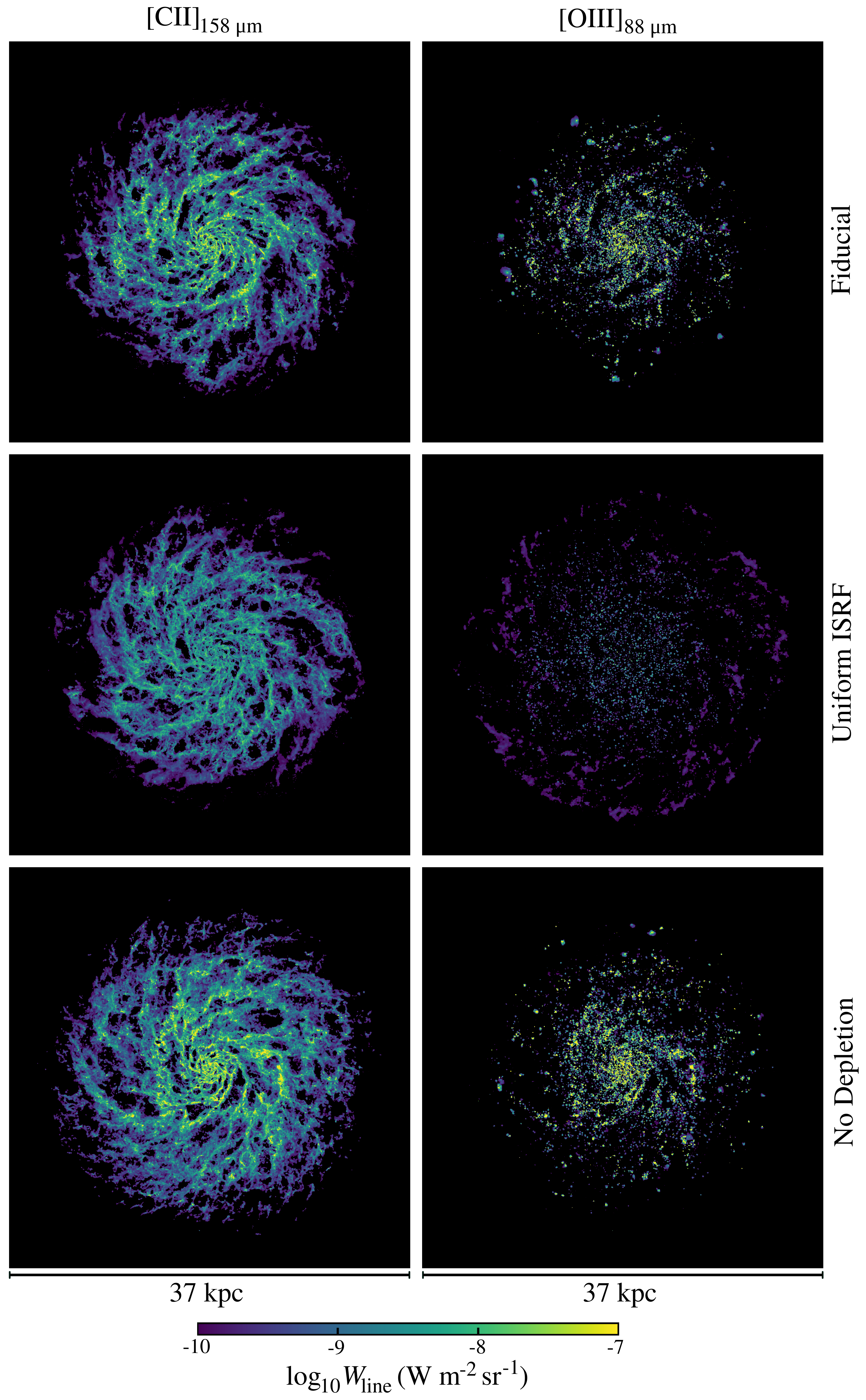}}
\vspace{-0.15 in}
\caption{Velocity-integrated maps of the continuum-subtracted FIR line emission from [C\textsc{ii}]$_{158 \rm{\mu m}}$ (left-hand column) and [O\textsc{iii}]$_{88 \rm{\mu m}}$ (right-hand column) in the m1e12 simulation run with the fiducial (top row), uniform ISRF (middle row) and no depletion (bottom row) models. These emission lines are noticeably weaker in the uniform ISRF model compared to the fiducial model, particularly for [O\textsc{iii}]$_{88 \rm{\mu m}}$, as the uniform ISRF model misses the contribution from H\textsc{ii} regions. 
\vspace{-0.15 in}} 
\label{emission_map_fig}
\end{figure}

Fig.~\ref{emission_map_fig} shows velocity-integrated maps of the continuum-subtracted line emission from the FIR lines [C\textsc{ii}]$_{158 \rm{\mu m}}$ and [O\textsc{iii}]$_{88 \rm{\mu m}}$ in the left- and right-hand columns, respectively. The three rows from top to bottom show the m1e12 simulation using the fiducial, uniform ISRF and no depletion models. 

In the fiducial model, [C\textsc{ii}]$_{158 \rm{\mu m}}$ emission is strongest along the spiral arms, but there remains a significant diffuse component in between. In contrast, [O\textsc{iii}]$_{88 \rm{\mu m}}$ is more strongly concentrated in small, bright regions along the arms, with very little diffuse emission. Comparing these to the image of stellar light and H$\alpha$ emission in Fig.~\ref{m1e12_morph_fig}, we find that [O\textsc{iii}]$_{88 \rm{\mu m}}$ predominantly arises from H\textsc{ii} regions around young stars in our simulations. This is unsurprising, as high-energy photons produced by young stars are required to photoionise oxygen to O\textsc{iii}. 

The [C\textsc{ii}]$_{158 \rm{\mu m}}$ emission in the uniform ISRF model is somewhat weaker and misses the brightest intensities seen along the spiral arms in the fiducial model, although there is still a significant diffuse component. The difference is more dramatic in [O\textsc{iii}]$_{88 \rm{\mu m}}$, which is much weaker in the uniform ISRF model, due to the lack of H\textsc{ii} regions in this case. 

The [C\textsc{ii}]$_{158 \rm{\mu m}}$ and [O\textsc{iii}]$_{88 \rm{\mu m}}$ morphology in the no depletion model, which does include H\textsc{ii} regions, is very similar to the fiducial model. As we will see below, the total luminosity of these lines is stronger with the no depletion model, as the gas-phase abundances of carbon and oxygen are higher when we do not account for the depletion onto dust grains. 

\begin{figure*}
\centering
\mbox{
	\includegraphics[width=168mm]{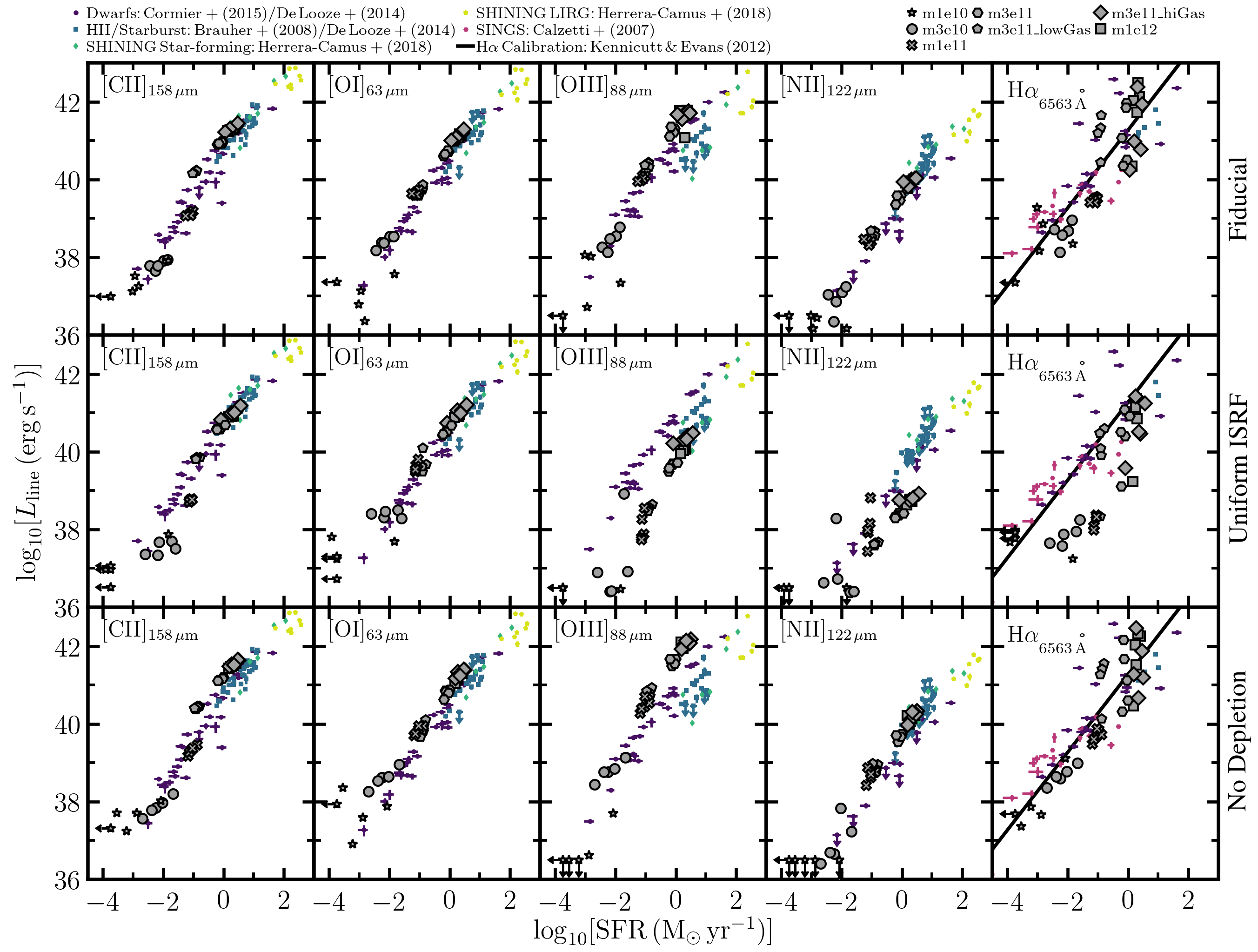}}
\vspace{-0.15 in}
\caption{Total line luminosity ($L_{\rm{line}}$) versus total star formation rate (SFR) over the whole galaxy for emission lines used as observational tracers of the SFR. The columns from left to right show [C\textsc{ii}]$_{158 \rm{\mu m}}$, [O\textsc{i}]$_{63 \rm{\mu m}}$, [O\textsc{iii}]$_{88 \rm{\mu m}}$, [N\textsc{ii}]$_{122 \rm{\mu m}}$ and H$\alpha_{6563 \text{\AA}}$. The simulations are shown by grey symbols, with five snapshots at intervals of 100~Myr for each galaxy. The star formation rates in the simulations are averaged over the preceding 10~Myr. The top, middle and bottom rows show simulations with the fiducial, uniform ISRF and no depletion models, respectively. The coloured symbols show observational data from the \textit{Herschel} Dwarf Galaxy Survey \citep{cormier15} and the starburst galaxies from \citet{brauher08}, with star formation rates from \citet{delooze14} in both cases, together with star-forming galaxies and Luminous Infrared Galaxies (LIRGs) from the SHINING survey \citep{herreracamus18} and galaxies from the SINGS survey \citep{calzetti07}. We also show the calibration of H$\alpha$ luminosity versus SFR from \citet{kennicutt12} in the right-hand panels (solid lines). The fiducial model broadly reproduces the observed correlations between emission line luminosity and star formation rate. 
\vspace{-0.15 in}} 
\label{SFR_tracers_fig}
\end{figure*}

To compare our simulation predictions to observations, we calculate the total line luminosity, $L_{\rm{line}}$, integrated over the disc of the galaxy. We consider four fine-structure FIR metal lines that are important for metal cooling and have been found observationally to correlate with star formation rate: [C\textsc{ii}]$_{158 \rm{\mu m}}$, [O\textsc{i}]$_{63 \rm{\mu m}}$, [O\textsc{iii}]$_{88 \rm{\mu m}}$ and [N\textsc{ii}]$_{122 \rm{\mu m}}$, together with the optical line H$\alpha_{6563 \text{\AA}}$. These are plotted in Fig.~\ref{SFR_tracers_fig} versus the total star formation rate. The grey symbols show the simulation predictions for the fiducial (top row), uniform ISRF (middle row), and no depletion (bottom row) models. We include five snapshots from each simulation, at intervals of 100~Myr. In the simulations, the star formation rate is averaged over the preceding 10~Myr.

The coloured symbols in Fig.~\ref{SFR_tracers_fig} show observed line luminosities, with star formation rates derived from continuum measurements, as detailed below. We include FIR emission line measurements from the \textit{Herschel} Dwarf Galaxy Survey \citep{cormier15} and the galaxies in the \citet{brauher08} sample of ISO observations that are identified as starbursts. We take the star formation rates for these samples from \citet{delooze14}, which were derived from FUV and 24~$\rm{\mu m}$ emission using the calibrations of \citet{kennicutt09} and \citet{hao11}. Measurements of H$\alpha_{6563 \text{\AA}}$ emission in these galaxies are taken from \citet{gildepaz03}, \citet{moustakas06}, \citet{kennicutt08} and \citet{ostlin09}, where available. 

We also show FIR observations of star-forming galaxies and Luminous Infrared Galaxies (LIRGs) from the SHINING survey \citep{herreracamus18}. We calculate the star formation rates in this sample from the $63 \, \rm{\mu m}$ continuum flux densities, based on the calibration between the $70 \, \rm{\mu m}$ luminosity and star formation rate from \citet{calzetti10}. At optical wavelengths, we also include H$\alpha_{6563 \text{\AA}}$ observations from \citet{calzetti07} for galaxies selected from the SINGS survey, with star formation rates derived from $24 \, \rm{\mu m}$ continuum emission using the calibration from \citet{rieke09}. Finally, we show the calibration between H$\alpha_{6563 \text{\AA}}$ luminosity and star formation rate from \citet{kennicutt12} in the right-hand column (solid lines).

For the observational data we use continuum-derived star formation rates as they are independent from the emission line measurements. However, simple estimators such as these might be subject to uncertainties, for example \citet{utomo14} find systematic differences between the star formation rates from UV plus IR estimators compared to those derived from modelling the composite spectral energy distributions with stellar population synthesis models. Any such uncertainties will affect our comparisons to the simulation data, for which we use the true star formation rate. 

In the top row of Fig.~\ref{SFR_tracers_fig} we see that the fiducial model broadly reproduces the observed correlations between luminosity and star formation rate for these emission lines. There are some deviations between the simulation predictions and the observations though. Most notably, the simulations of the most massive galaxies overpredict the [O\textsc{iii}]$_{88 \rm{\mu m}}$ luminosity by up to a factor $\approx$2 compared to observations at the same star formation rate. Furthermore, the dwarf galaxy m1e10 exhibits greater scatter in [O\textsc{iii}]$_{88 \rm{\mu m}}$ and [O\textsc{i}]$_{63 \rm{\mu m}}$, with some snapshots lying up to an order of magnitude below the observations, while other snapshots from m1e10 are close to the observed correlation. Nevertheless, the simulation predictions overall are in reasonably good agreement with the observational data. 

In the uniform ISRF model, the [O\textsc{iii}]$_{88 \rm{\mu m}}$ and H$\alpha_{6563 \text{\AA}}$ luminosities are up to an order of magnitude lower than in the fiducial model. As noted above, this model does not include the subgrid prescription for H\textsc{ii} regions, which dominate the total emission of these lines in the fiducial model. The decrease in [C\textsc{ii}]$_{158 \rm{\mu m}}$ and [O\textsc{i}]$_{63 \rm{\mu m}}$ luminosities in the uniform ISRF model is more modest. As we saw in Fig.~\ref{emission_map_fig}, the [C\textsc{ii}]$_{158 \rm{\mu m}}$ emission includes a significant diffuse component outside H\textsc{ii} regions, which is still present in the uniform ISRF model. 

The no depletion model exhibits stronger FIR metal line luminosities, by up to a factor $\approx$2, compared to the fiducial model. This is due to the increased elemental abundances of carbon, oxygen and nitrogen in the gas phase at fixed total metallicity when we do not account for the depletion of these metals onto dust grains. This leads to increased tension between the simulation predictions and observational data, particularly for the [O\textsc{iii}]$_{88 \rm{\mu m}}$ luminosity in the most massive galaxies. However, the H$\alpha_{6563 \text{\AA}}$ luminosity remains unaffected by the model for metal depletion.

Other elements such as iron are depleted more strongly than carbon or oxygen, with the gas phase abundance of iron reduced by two orders of magnitude at high densities (see Fig.~\ref{depl_fig}). We therefore expect dust depletion could have an even greater effect on emission lines from species such as Fe\textsc{ii} and Fe\textsc{iii} \citep[e.g.][]{osterbrock92, rodriguez02, delgadoinglada09}. We will explore emission from heavily depleted species such as this in a future work.

In Appendix~\ref{resolution_sect} we compare the emission line luminosities in the fiducial model from simulations run at different resolutions. While many of these luminosity predictions show good numerical convergence, there are some galaxies for which particular emission lines differ significantly between resolution levels. For example, in m3e10 the luminosities of [C\textsc{ii}]$_{158 \rm{\mu m}}$, [O\textsc{i}]$_{63 \rm{\mu m}}$ and [N\textsc{ii}]$_{122 \rm{\mu m}}$ increase by up to an order magnitude from standard to high resolution, while the [C\textsc{ii}]$_{158 \rm{\mu m}}$ luminosity of m1e11 decreases by an order of magnitude from low to standard resolution. However, we do not see any trends of particular emission lines always increasing or decreasing systematically with resolution. It is unclear to what extent these differences may be due to stochastic variations between runs. 

% Define colours to highlight individual
% cells in the following table.
\definecolor{red1_1}{rgb}{0.962, 0.933, 0.933}
\definecolor{red1_2}{rgb}{0.979, 0.962, 0.962} 
\definecolor{red1_3}{rgb}{0.669, 0.408, 0.408}
\definecolor{red1_5}{rgb}{0.944, 0.901, 0.901}
\definecolor{red2_2}{rgb}{0.897, 0.816, 0.816}
\definecolor{red2_3}{rgb}{0.934, 0.882, 0.882}
\definecolor{red2_5}{rgb}{0.941, 0.894, 0.894} 
\definecolor{red3_1}{rgb}{0.986, 0.976, 0.976} 
\definecolor{red3_2}{rgb}{0.902, 0.825, 0.825} 
\definecolor{red3_3}{rgb}{0.904, 0.828, 0.828}
\definecolor{red3_4}{rgb}{0.818, 0.674, 0.674}
\definecolor{red3_5}{rgb}{0.857, 0.744, 0.744}
\definecolor{red4_1}{rgb}{0.994, 0.990, 0.990}
\definecolor{red4_2}{rgb}{0.875, 0.778, 0.778}
\definecolor{red4_3}{rgb}{0.589, 0.265, 0.265}
\definecolor{red4_5}{rgb}{0.598, 0.283, 0.283}
\definecolor{red5_1}{rgb}{0.997, 0.994, 0.994}
\definecolor{red5_3}{rgb}{0.312, 0.030, 0.030}
\definecolor{red5_5}{rgb}{0.878, 0.782, 0.782}
\definecolor{red6_1}{rgb}{0.993, 0.988, 0.988}
\definecolor{red6_2}{rgb}{0.903, 0.826, 0.826}
\definecolor{red6_3}{rgb}{0.640, 0.358, 0.358}
\definecolor{red6_5}{rgb}{0.917, 0.852, 0.852}
\definecolor{red7_2}{rgb}{0.921, 0.859, 0.859}
\definecolor{red7_3}{rgb}{0.422, 0.030, 0.030}
\definecolor{red7_5}{rgb}{0.901, 0.823, 0.823}

\definecolor{blue2_1}{rgb}{0.936, 0.954, 0.964}
\definecolor{blue2_4}{rgb}{0.700, 0.784, 0.832}
\definecolor{blue4_4}{rgb}{0.180, 0.409, 0.541}
\definecolor{blue5_2}{rgb}{0.747, 0.818, 0.859}
\definecolor{blue5_4}{rgb}{0.061, 0.324, 0.474}
\definecolor{blue6_4}{rgb}{0.428, 0.588, 0.679} 
\definecolor{blue7_1}{rgb}{0.977, 0.984, 0.987}
\definecolor{blue7_4}{rgb}{0.030, 0.201, 0.379}

\begin{table}
\begin{minipage}{84mm}
\centering
\caption{Ratios of emission line luminosities calculated with non-equilibrium and equilibrium abundances, $L_{\rm{noneq}} / L_{\rm{eqm}}$, at 500~Myr using the fiducial model. Values highlighted in red and blue correspond to an enhancement and reduction, respectively, of the luminosity when non-equilibrium abundances are used.}
\label{noneq_vs_eqm_L}
\begin{tabular}{cccccc}
  \hline
   & \multicolumn{5}{c}{$L_{\rm{noneq}} / L_{\rm{eqm}}$} \\ 
  \cline{2-6} 
  Galaxy & [C\textsc{ii}] & [O\textsc{i}] & [O\textsc{iii}] & [N\textsc{ii}] & H$\alpha$ \\ 
   & $158 \rm{\mu m}$ & $63 \rm{\mu m}$ & $88 \rm{\mu m}$ & $122 \rm{\mu m}$ & $6563 \rm{\AA}$ \\ 
\hline
m1e10 & \cellcolor{red1_1} 1.08 & \cellcolor{red1_2} 1.04 & \cellcolor{red1_3} \textcolor{white}{1.66} & 1.00 & \cellcolor{red1_5} 1.11 \\
m3e10 & \cellcolor{blue2_1} 0.93 & \cellcolor{red2_2} 1.21 & \cellcolor{red2_3} 1.13 & \cellcolor{blue2_4} 0.75 & \cellcolor{red2_5} 1.12 \\
m1e11 & \cellcolor{red3_1} 1.03 & \cellcolor{red3_2} 1.20 & \cellcolor{red3_3} 1.19 & \cellcolor{red3_4} 1.36 & \cellcolor{red3_5} 1.29 \\
m3e11 & \cellcolor{red4_1} 1.01 & \cellcolor{red4_2} 1.25 & \cellcolor{red4_3} \textcolor{white}{1.82} & \cellcolor{blue4_4} \textcolor{white}{0.52} & \cellcolor{red4_5} \textcolor{white}{1.80} \\
m3e11\_lowGas & \cellcolor{red5_1} 1.01 & \cellcolor{blue5_2} 0.78 & \cellcolor{red5_3} \textcolor{white}{2.38} & \cellcolor{blue5_4} \textcolor{white}{0.49} & \cellcolor{red5_5} 1.24 \\
m3e11\_hiGas & \cellcolor{red6_1} 1.01 & \cellcolor{red6_2} 1.19 & \cellcolor{red6_3} \textcolor{white}{1.72} & \cellcolor{blue6_4} \textcolor{white}{0.61} & \cellcolor{red6_5} 1.17 \\
m1e12 & \cellcolor{blue7_1} 0.97 & \cellcolor{red7_2} 1.16 & \cellcolor{red7_3} \textcolor{white}{2.16} & \cellcolor{blue7_4} \textcolor{white}{0.45} & \cellcolor{red7_5} 1.20 \\ 
\hline
\end{tabular}
\end{minipage}
\end{table}

The emission line luminosities shown in Fig.~\ref{SFR_tracers_fig} were computed using the non-equilibrium ion abundances from the simulations. To study the impact of non-equilibrium chemistry on the predicted luminosities in the fiducial model, we also repeated the \textsc{radmc-3d} calculations from the final snapshot at 500~Myr using ion abundances in chemical equilibrium. The equilibrium abundances were determined by integrating the \textsc{chimes} reaction network to equilibrium at constant density and temperature for each gas particle in the snapshot. 

Table~\ref{noneq_vs_eqm_L} summarises the ratios of luminosities calculated with non-equilibrium and equilibrium abundances, $L_{\rm{noneq}} / L_{\rm{eqm}}$, for each of the five emission lines that we considered in Fig.~\ref{SFR_tracers_fig}. Values highlighted in red indicate where non-equilibrium effects enhance the luminosity, while blue values show where the luminosity is suppressed in non-equilibrium. The ratios in Table~\ref{noneq_vs_eqm_L} were calculated from the final snapshot of each simulation, however they may also vary in time, particularly in the dwarf galaxies which exhibit strong variations in the star formation rate (see Fig.~\ref{SFH_fig}). 

The luminosity of [O\textsc{iii}]$_{88 \rm{\mu m}}$ shows the greatest enhancement in non-equilibrium, by up to a factor of 2.38 in m3e11\_lowGas. The photoionisation of O\textsc{ii} to O\textsc{iii} requires the presence of high energy photons ($>$35~eV) produced by young, massive stars. This emission line is therefore particularly sensitive to short time-scale variations in the local star formation rate, which could drive these non-equilibrium effects. In contrast, [N\textsc{ii}]$_{122 \rm{\mu m}}$ typically exhibits a lower luminosity in most of our simulations when we use non-equilibrium abundances, by up to a factor of 0.45 in m1e12. The [C\textsc{ii}]$_{158 \rm{\mu m}}$ luminosity is least affected by the non-equilibrium chemistry, differing by less than 10 per cent compared to equilibrium in all of our simulations with the fiducial model. 

\section{Conclusions}\label{conclusions_sect} 

We have presented a suite of simulations of isolated disc galaxies ranging from dwarfs to Milky Way-mass, with a mass resolution of $400 \, \rm{M}_{\odot}$ per particle, and structural properties initially set according to observed scaling relations at redshift zero. These simulations combine the \textsc{fire-2} subgrid galaxy formation models with the \textsc{chimes} non-equilibrium chemistry and cooling module. 

In our fiducial model, we coupled the chemical reaction network to the local stellar fluxes computed from star particles using the approximate \textsc{lebron} radiative transfer method. This method assumes that the absorption of stellar radiation occurs locally around the star particle producing the radiation and around the receiving gas particle. We also implemented an empirical density-dependent model for the depletion of metals from the gas phase onto dust grains, based on observed depletion factors. We then repeated each simulation with two model variations. First, we replaced the local stellar fluxes with a spatially uniform interstellar radiation field normalised according to the star formation rate surface density of the galaxy disc, which we averaged over the preceding 10~Myr (the uniform ISRF model). Second, we applied a constant dust to metals ratio and disabled the depletion of metals from the gas phase due to dust grains (the no depletion model). 

By comparing these model variations to the fiducial runs, we explored the impact of local stellar fluxes and metal depletion on the non-equilibrium chemistry, and resulting observational diagnostics, of the ISM. We particularly focus on observations of the H\textsc{i} to H$_{2}$ transition and emission line tracers of the star formation rate. Our main results are as follows: 

\begin{enumerate}[leftmargin=\parindent]
  \item Dwarf galaxies run with the uniform ISRF model exhibit stronger outflows, resulting in disc gas fractions up to 30 per cent lower than the fiducial model, which may be due to the lack of H\textsc{ii} region feedback in the uniform ISRF runs, while the total mass of stars formed over 500~Myr is up to 27 per cent higher with the uniform ISRF model. In contrast, the model variations have little effect on the total star formation rates and disc gas fractions in galaxies with halo masses $M_{200, \, \rm{crit}} \! \geq \! 10^{11} \, \rm{M}_{\odot}$ (see Fig.~\ref{SFH_fig}). 
  \item Non-equilibrium effects can lead to strong enhancement and suppression of H\textsc{i} and H$_{2}$ abundances in certain regions of density-temperature space. At densities $n_{\rm{H}} \! \sim \! 0.1 \, \rm{cm}^{-3}$ and temperatures $T \! \sim \! 10^{3} \, \rm{K}$ the H$_{2}$ abundance is enhanced by more than 3 orders of magnitude compared to chemical equilibrium, due to recent heating of cold, dense gas that has had insufficient time to fully destroy the molecules. This may have important consequences for predictions of infrared emission lines produced by rovibrational transitions of H$_{2}$ at these temperatures. In contrast, the non-equilibrium H$_{2}$ fraction is suppressed by up to an order of magnitude at $n_{\rm{H}} \! \sim \! 1 \! - \! 10 \, \rm{cm}^{-3}$, $T \! \lesssim \! 100 \, \rm{K}$ and $n_{\rm{H}} \! \sim 100 \, \rm{cm}^{-3}$, $T \! \sim \! 100 \! - \! 10^{3} \, \rm{K}$, with a corresponding enhancement in H\textsc{i}. This may be due to recently cooling gas that has had insufficient time to fully form molecules (Fig.~\ref{Trho_HI_H2_fig}).
  \item Compared to the fiducial model, the m1e12 simulation run with a uniform ISRF produces less intermediate-temperature ($10^{3} \! < \! T \! \leq \! 10^{4} \, \rm{K}$) gas at $n_{\rm{H}} \! \gtrsim \! 10 \, \rm{cm}^{-3}$, due to the lack of H\textsc{ii} regions, and more intermediate-temperature H$_{2}$ gas at $n_{\rm{H}} \! \lesssim \! 10 \, \rm{cm}^{-3}$. Meanwhile, the no depletion model enhances the low-temperature ($T \! \leq \! 10^{3} \, \rm{K}$) H\textsc{i} component at $n_{\rm{H}} \! \lesssim \! 10 \, \rm{cm}^{-3}$, due to an increase in metal cooling (Fig.~\ref{nH_hist_fig}).
  \item Our simulation predictions for the H$_{2}$ fraction versus total neutral hydrogen column density in high-mass galaxies ($M_{200, \, \rm{crit}} \! \gtrsim \! 3 \times 10^{11} \, \rm{M}_{\odot}$) broadly overlap with absorption line observations in the Milky Way \citep{gillmon06, shull21}. However, our dwarf galaxy simulations can only reproduce the highest H$_{2}$ fractions observed by \citet{tumlinson02} in the LMC and SMC (Fig.~\ref{H2_fraction_fig}).
  \item The ratio of total H\textsc{i} to total stellar mass as a function of stellar mass in our simulations is in good agreement with observations. However, while the simulated H$_{2}$ to stellar mass fractions agree with observations at high stellar masses ($M_{\ast} \! \gtrsim \! 10^{10} \, \rm{M}_{\odot}$), our dwarf galaxy simulations underpredict the observations by $\sim \! 1 - 2$ orders of magnitude (Fig.~\ref{global_HI_H2_fig}). This may be due to selection effects in the observational samples. 
  \item In our fiducial model, [C\textsc{ii}]$_{158 \rm{\mu m}}$ emission in m1e12 is brightest along the spiral arms, but with a significant diffuse component from inter-arm regions. The [O\textsc{iii}]$_{88 \rm{\mu m}}$ line is more strongly concentrated in compact regions along the spiral arms, with very little diffuse emission, as it is produced mostly within the Str\"{o}mgren radii of young stars. The morphology of the [O\textsc{iii}]$_{88 \rm{\mu m}}$ emission differs dramatically in the uniform ISRF model, which lacks these H\textsc{ii} regions, but the C\textsc{ii} emission still retains the diffuse component in this case (Fig.~\ref{emission_map_fig}).
  \item The fiducial model broadly reproduces observed correlations between line luminosity and star formation rate for the emission lines [C\textsc{ii}]$_{158 \rm{\mu m}}$, [O\textsc{i}]$_{63 \rm{\mu m}}$, [O\textsc{iii}]$_{88 \rm{\mu m}}$, [N\textsc{ii}]$_{122 \rm{\mu m}}$ and H$\alpha_{6563 \text{\AA}}$ (Fig.~\ref{SFR_tracers_fig}). The most significant deviation between our simulation predictions and observations is for the [O\textsc{iii}]$_{88 \rm{\mu m}}$ line, which is overpredicted by up to a factor $\approx$2 in the simulations of the most massive galaxies in our sample. The line luminosities are lower in the uniform ISRF model, by up to an order of magnitude for [O\textsc{iii}]$_{88 \rm{\mu m}}$ and H$\alpha_{6563 \text{\AA}}$, due to the lack of H\textsc{ii} regions. The no depletion model predicts up to a factor $\approx$2 higher luminosities for the FIR metal lines due to the increase in gas-phase metal abundances, but H$\alpha_{6563 \text{\AA}}$ is unaffected. Non-equilibrium effects enhance the luminosity of [O\textsc{iii}]$_{88 \rm{\mu m}}$ in our fiducial model by up to a factor of 2.38, while the [N\textsc{ii}]$_{122 \rm{\mu m}}$ luminosity is typically suppressed by up to a factor of 0.45. In contrast, [C\textsc{ii}]$_{158 \rm{\mu m}}$ differs by less than 10 per cent when comparing luminosities calculated with non-equilibrium and equilibrium abundances. 
\end{enumerate} 

We have thus shown that the treatment of local stellar fluxes and depletion of metals onto dust grains affects the synthetic emission line predictions from our simulations, particularly for lines commonly used as star formation rate tracers. In the case of stellar radiation, this is primarily because we need to capture the irradiation of H\textsc{ii} regions within the Str\"{o}mgren radius surrounding young stars, as these regions contribute to, and in many cases dominate, the emission from these lines. Correctly accounting for metal depletion is important as it reduces the gas-phase abundance of metal species available to produce line emission. 

However, the local stellar fluxes and metal depletion generally have little impact on the overall galaxy evolution, for example in terms of total star formation rate, except in the case of dwarf galaxies. In \citet{richings16} we compared simulations of isolated galaxies run using spatially uniform radiation fields with constant normalisations of different strengths. We found that weaker radiation fields led to higher star formation rates and stronger galactic outflows, as they enabled more gas to cool to the cold, star-forming ISM phase. We therefore conclude that the evolution of the galaxy depends on the average strength of the interstellar radiation field throughout the galactic disk, but is not sensitive to local variations of stellar fluxes within the disk, although this conclusion may depend on the subgrid treatment of star formation and stellar feedback employed in the simulation. It may also depend on the treatment of radiative transfer, as the assumption of local extinction in the \textsc{lebron} method may lead to an overestimate of the importance of distant sources as the intervening extinction is not fully accounted for. 

While we have only focussed on a handful of comparisons between the simulations and observations in this paper, the detailed chemical modelling in this simulation suite will enable us to make predictions for many additional observational diagnostics, such as FIR line deficits, nebular emission line ratios from individual star-forming regions, and emission line tracers of molecular gas. We will explore these aspects further in future works. 

\section*{Acknowledgements}

We thank Caleb Choban and Jonathan Stern for useful comments and suggestions. AJR was supported by a COFUND/Durham Junior Research Fellowship under EU grant 609412; and by the Science and Technology Facilities Council [ST/T000244/1]. CAFG was supported by NSF through grants AST-1715216, AST-2108230, and CAREER award AST-1652522; by NASA through grants 17-ATP17-0067 and 21-ATP21-0036; by STScI through grants HST-AR-16124.001-A and HST-GO-16730.016-A; by CXO through grant TM2-23005X; and by the Research Corporation for Science Advancement through a Cottrell Scholar Award. ABG was supported by an NSF-GRFP under grant DGE-1842165. This work used the DiRAC@Durham facility managed by the Institute for Computational Cosmology on behalf of the STFC DiRAC HPC Facility (www.dirac.ac.uk). The equipment was funded by BEIS capital funding via STFC capital grants ST/K00042X/1, ST/P002293/1, ST/R002371/1 and ST/S002502/1, Durham University and STFC operations grant ST/R000832/1. DiRAC is part of the National e-Infrastructure. 

\section*{Data availability}

The data underlying this article will be shared on reasonable request to the corresponding author. A public version of the \textsc{gizmo} code can be found at \url{http://www.tapir.caltech.edu/~phopkins/Site/GIZMO.html}, and a public version of the \textsc{chimes} code can be found at \url{https://richings.bitbucket.io/chimes/home.html}.

{}

\appendix

\section{Calibration of escape fraction parameters}\label{esc_fraction_sect}

As discussed in Section~\ref{flux_sect}, our model for local stellar fluxes contains two free parameters, the escape fractions of FUV and EUV radiation from H\textsc{ii} regions. These parameters determine how much radiation from star particles can propogate beyond the surrounding Str\"{o}mgren radius around the star and contribute to the diffuse interstellar radiation field. To understand how the resulting stellar fluxes depend on these parameters, we repeated the m3e11\_lowRes08 galaxy simulations with FUV and EUV escape fractions varied independently between 0.005 and 0.5. 

\begin{figure}
\centering
\mbox{
	\includegraphics[width=84mm]{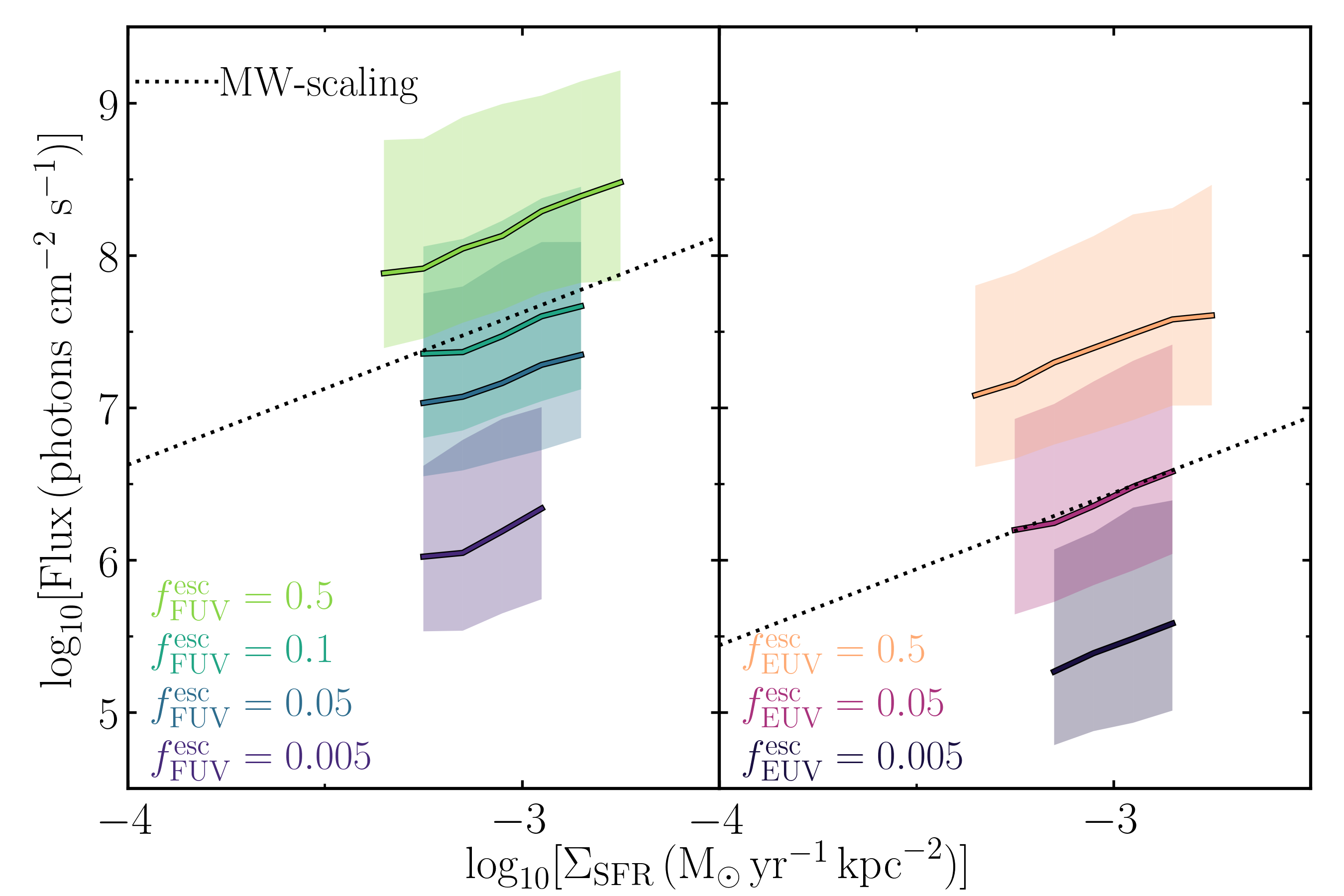}}
\vspace{-0.15 in}
\caption{Flux of diffuse stellar radiation in the FUV (left-hand panel) and EUV (right-hand panel) bands versus the star formation rate surface density averaged over the disc of the galaxy ($\Sigma_{\rm{SFR}, \, \rm{disc}}$) in m3e11\_lowRes08. The solid curves show the median flux of all gas particles in the galaxy disc that lie outside H\textsc{ii} regions, in bins of $\Sigma_{\rm{SFR}, \, \rm{disc}}$, and the shaded regions indicate the 10$^{\rm{th}}$ to 90$^{\rm{th}}$ percentiles. The colours denote different escape fractions from H\textsc{ii} regions in the FUV and EUV bands. The black dotted lines show a linear scaling between the flux and $\Sigma_{\rm{SFR}, \, \rm{disc}}$ normalised to Milky Way values. In both panels, we see that the strength of the stellar radiation increases with increasing escape fraction. We therefore choose escape fractions of 0.1 and 0.05 in the FUV and EUV bands, respectively, to match the Milky Way scaling. 
\vspace{-0.15 in}} 
\label{fesc_cal_fig}
\end{figure}

Fig.~\ref{fesc_cal_fig} shows the median (solid curves) and tenth to ninetieth percentile (shaded regions) fluxes in the FUV (left-hand panel) and EUV (right-hand panel) bands, plotted against the star formation rate surface density averaged over the whole disc in the preceding 10~Myr ($\Sigma_{\rm{SFR, \, disc}}$). As we saw in Fig.~\ref{flux_fullDisc_fig}, the fluxes scale linearly with $\Sigma_{\rm{SFR, \, disc}}$. However, the normalisation increases with increasing escape fraction, as more radiation propagates into the diffuse component. 

The dotted lines in Fig.~\ref{fesc_cal_fig} indicate a linear scaling normalised to the flux of the interstellar radiation field of the Milky Way in the local solar neighbourhood \citep{black87} at $\Sigma_{\rm{SFR,} \, \rm{MW}} = 4 \times 10^{-3} \, \rm{M}_{\odot} \, \rm{kpc}^{-2}$ \citep{robertson08}. For our fiducial model, we therefore chose escape fractions of 0.1 and 0.05 in the FUV and EUV bands, respectively, as these best reproduce the normalisation expected from observations of the radiation field in the Milky Way. 

\section{Variations in numerical resolution}\label{resolution_sect}

To test the numerical convergence of our results we repeated some of our simulations with different resolutions. Our main runs use baryonic and dark matter particle masses of $m_{\rm{b}} = 400 \, \rm{M}_{\odot}$ and $m_{\rm{DM}} = 1910 \, \rm{M}_{\odot}$, respectively. The gravitational softening of gas particles is adaptive and set to the mean inter-particle spacing at the particle's density, with a minimum of 0.08~pc. This results in a softening length of 2.2~pc at the star formation density threshold $n_{\rm{H}} = 10^{3} \, \rm{cm}^{-3}$. The star and dark matter particles use constant gravitational softenings of 1.6~pc and 2.8~pc, respectively. We then repeated the five galaxy models with varying halo mass using 8 times lower mass resolution and gravitational softenings increased by a factor of 2. We also repeated the m3e10 dwarf galaxy with 8 times higher mass resolution and gravitational softenings reduced by a factor of 2. See Table~\ref{galaxy_pars} for a summary of the simulation parameters. The simulations at lower and higher resolution were only run using the fiducial model. 

\begin{figure}
\centering
\mbox{
	\includegraphics[width=84mm]{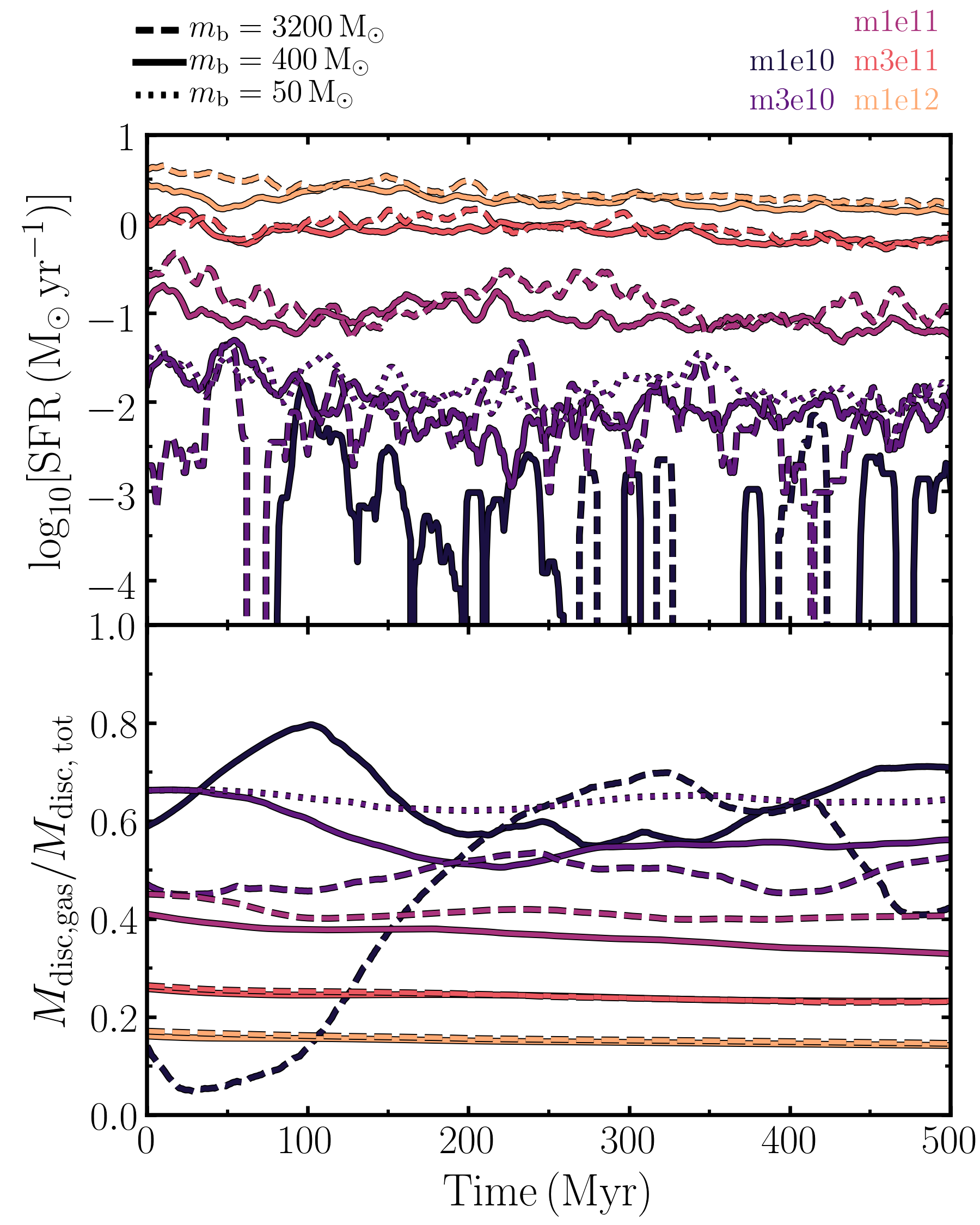}}
\vspace{-0.15 in}
\caption{Star formation rate in the galaxy disc averaged over the preceding 10~Myr (top panel) and gas mass fraction in the galaxy disc (bottom panel), plotted versus time. The dashed, solid and dotted line styles show simulations run at low, standard and high resolution, respectively, while the line colours indicate the different galaxies. The dwarf galaxies exhibit different evolutions of star formation rate and gas fraction particularly between low and standard resolution, although this may simply reflect stochastic variations between runs due to the bursty nature of star formation in this regime. 
\vspace{-0.15 in}} 
\label{SFH_resTest_fig}
\end{figure}

\begin{figure}
\centering
\mbox{
	\includegraphics[width=84mm]{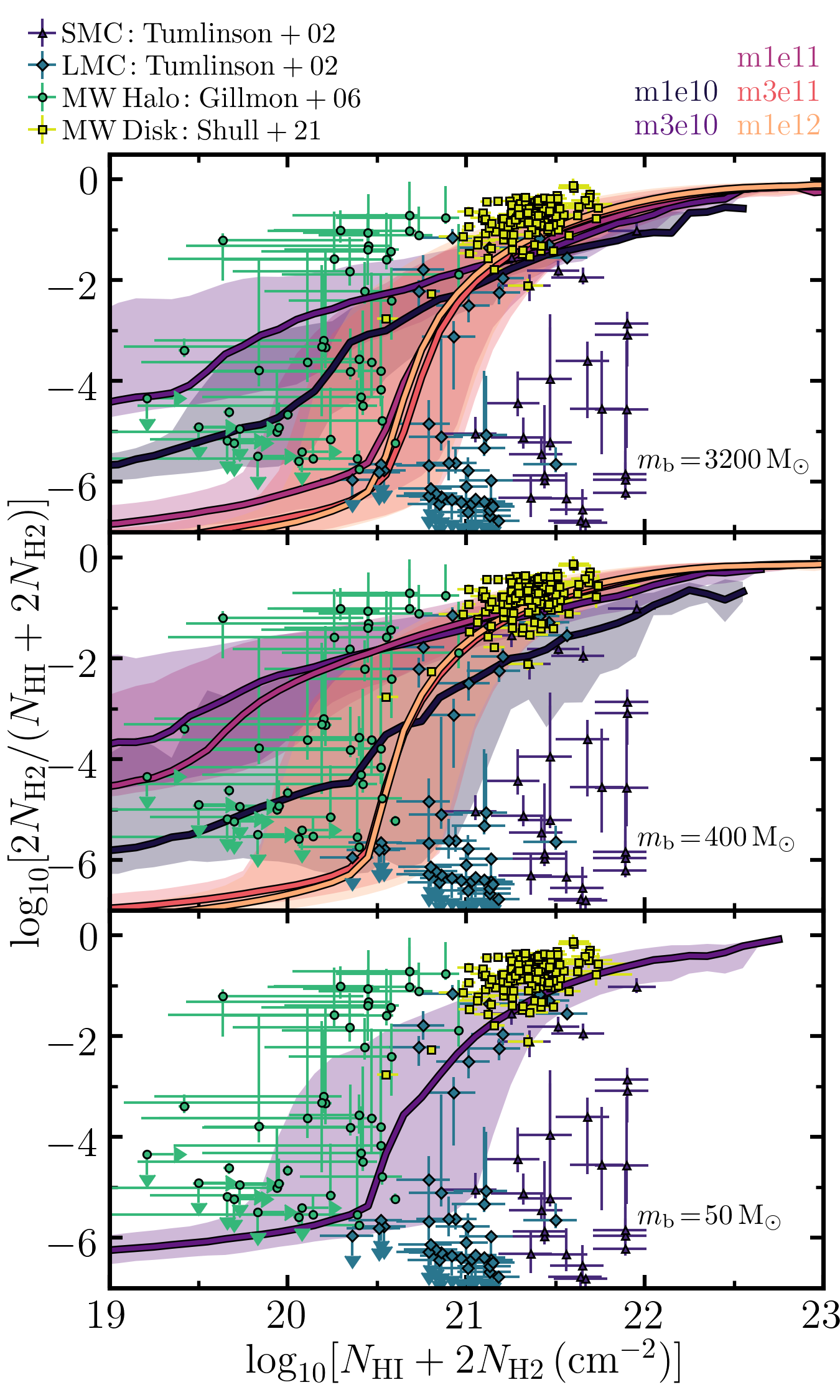}}
\vspace{-0.15 in}
\caption{H$_{2}$ fraction versus total neutral hydrogen column density. The median H$_{2}$ fractions in bins of total column density from the simulations are shown by the solid lines, and the tenth to ninetieth percentile ranges in each bin are indicated by the shaded regions. Different colours represent different galaxy models as shown in the legend. The top, middle and bottom panels show simulations run at low, standard and high resolution, respectively. The data points show observations from the Small Magellanic Cloud (SMC; \citealt{tumlinson02}; purple), Large Magellanic Cloud (LMC; \citealt{tumlinson02}; blue), the Milky Way at galactic latitudes $|b| \! > \! 20^{\circ}$ (MW Halo; \citealt{gillmon06}; green), and the Milky Way at $|b| \! < \! 10^{\circ}$ (MW Disk; \citealt{shull21}; yellow). While the high-mass galaxies show good agreement between low and standard resolution, the H$_{2}$ fractions at low column densities ($<$10$^{21} \, \rm{cm}^{-2}$) in dwarf galaxies are not well converged. 
\vspace{-0.15 in}} 
\label{H2_fraction_resTest_fig}
\end{figure}

\begin{figure}
\centering
\mbox{
	\includegraphics[width=84mm]{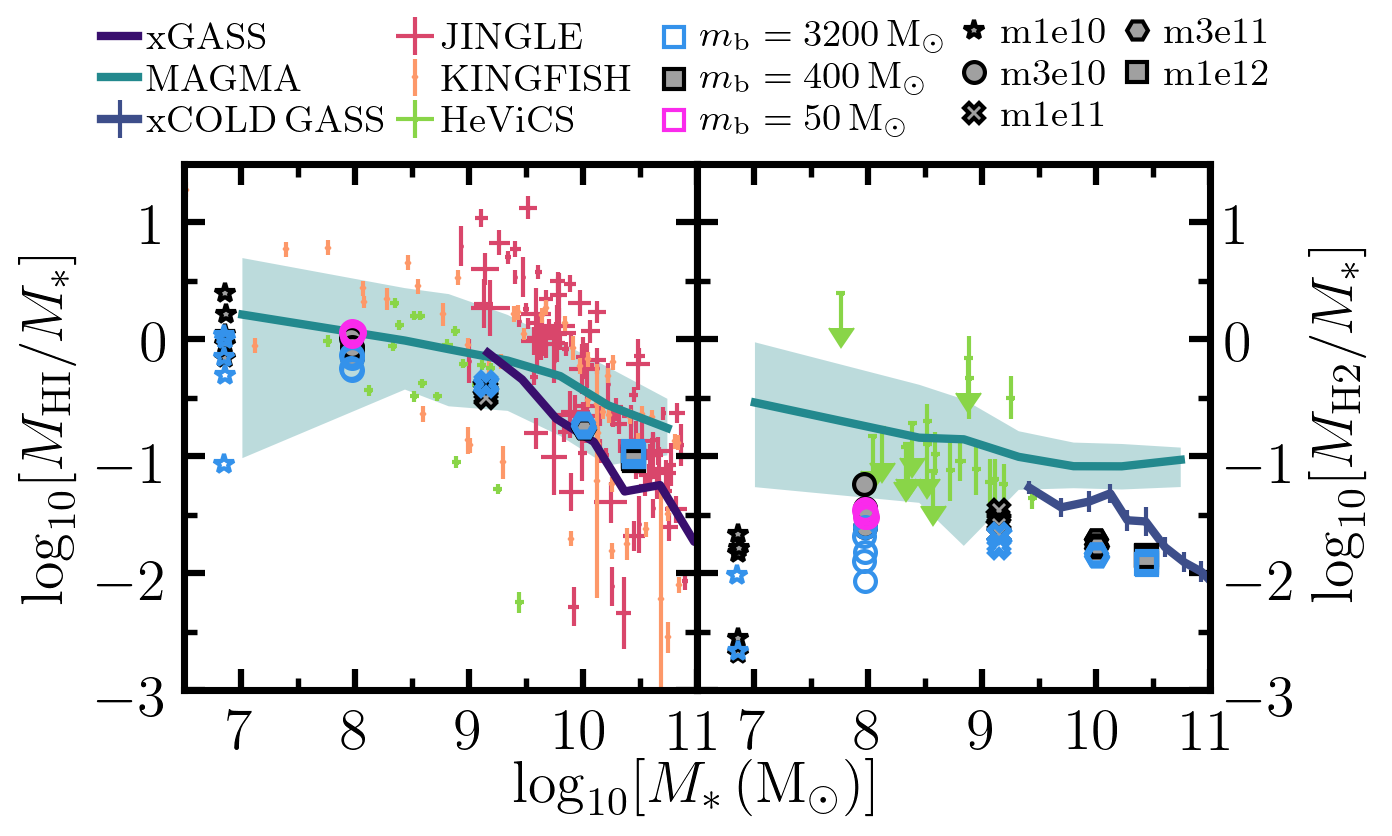}}
\vspace{-0.15 in}
\caption{The ratios of H\textsc{i} to stellar mass ($M_{\rm{HI}} / M_{\ast}$; left-hand column) and H$_{2}$ to stellar mass ($M_{\rm{H2}} / M_{\ast}$; right-hand column) versus stellar mass. The simulations at low, standard and high resolution are shown by the blue, grey and pink symbols, respectively. Observed median relations from the xGASS \citep{catinella18}, MAGMA \citep{hunt20} and xCOLD GASS \citep{saintonge17} surveys are shown by the solid curves. We also show the $\pm 1 \sigma$ deviations in each stellar mass bin for the MAGMA sample, indicated by the corresponding shaded regions. Individual galaxies from the JINGLE \citep{saintonge18}, KINGFISH \citep{kennicutt11} and HeViCS \citep{grossi16} surveys are indicated by the remaining data points. The dwarf galaxies show a large spread of up 1~dex in H\textsc{i} and H$_{2}$ masses between resolution levels, broadly reflecting the differences in total disc gas fraction. Convergence between low and standard resolution is better in the high-mass galaxies, although the H$_{2}$ masses are still slightly higher at standard resolution. 
\vspace{-0.15 in}} 
\label{global_HI_H2_resTest_fig}
\end{figure}

\begin{figure*}
\centering
\mbox{
	\includegraphics[width=168mm]{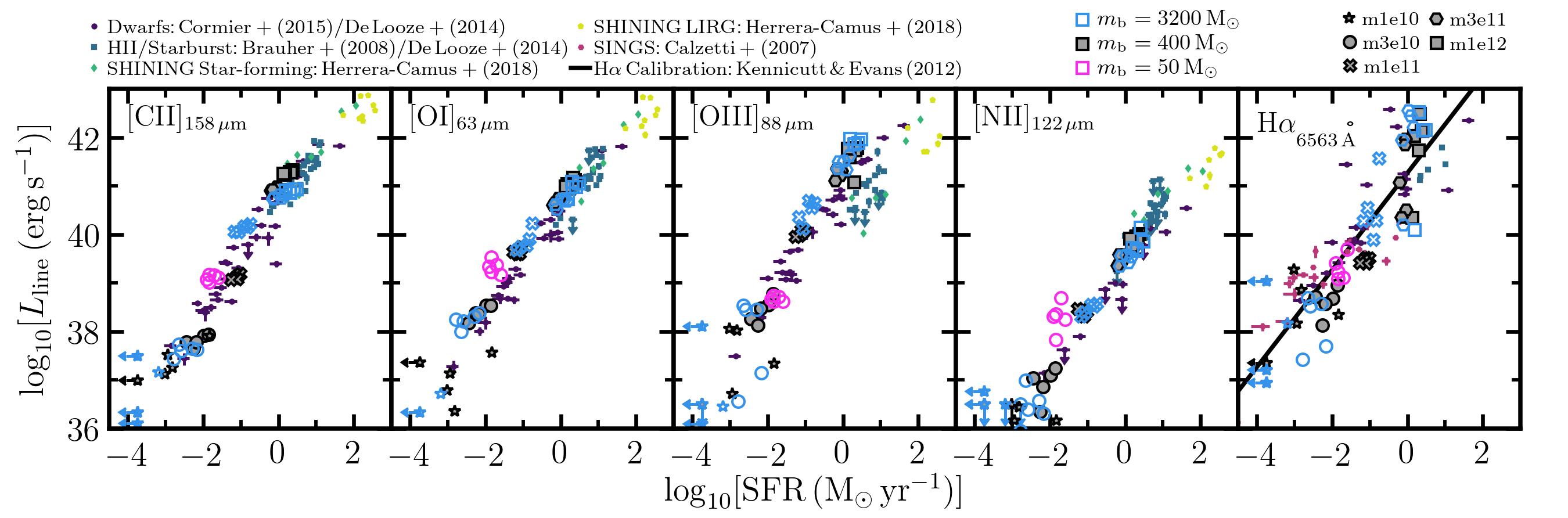}}
\vspace{-0.15 in}
\caption{Emission line luminosity ($L_{\rm{line}}$) plotted against star formation rate (SFR). From left to right the panels show [C\textsc{ii}]$_{158 \rm{\mu m}}$, [O\textsc{i}]$_{63 \rm{\mu m}}$, [O\textsc{iii}]$_{88 \rm{\mu m}}$, [N\textsc{ii}]$_{122 \rm{\mu m}}$ and H$\alpha_{6563 \text{\AA}}$. The blue, grey and pink symbols show the simulations at low, standard and high resolution, respectively. The remaining symbols show observational data from the \textit{Herschel} Dwarf Galaxy Survey \citep{cormier15} and the starburst galaxies from \citet{brauher08}, with star formation rates from \citet{delooze14} in both cases, together with star-forming galaxies and Luminous Infrared Galaxies (LIRGs) from the SHINING survey \citep{herreracamus18} and galaxies from the SINGS survey \citep{calzetti07}. The right-hand panel also shows the calibration of H$\alpha$ luminosity versus SFR from \citet{kennicutt12} as a solid line. 
\vspace{-0.15 in}} 
\label{SFR_tracers_resTest_fig}
\end{figure*}

Fig.~\ref{SFH_resTest_fig} shows the evolution of star formation rate (top panel) and disc gas mass fraction (bottom panel), calculated as in Fig.~\ref{SFH_fig}, comparing simulations run at low (dashed lines), standard (solid lines), and high (dotted lines) resolution. In the dwarf galaxies (m1e10 and m3e10) we see large differences in the gas fraction between the low and standard resolution runs. This might be due to the bursty nature of star formation in this regime \citep[e.g.][]{fauchergiguere18}, as dwarf galaxies can easily lose a significant proportion of their gas content via outflows driven by stellar feedback after periods of intense star formation, which must then be re-accreted before further star formation can continue. The differences we see here may therefore simply reflect stochastic variations between runs. The m1e10 dwarf galaxy in particular has relatively few baryonic particles at low resolution (initially 1.9$\times$10$^{4}$ gas and 2.1$\times$10$^{3}$ star particles), and so the timing of individual feedback events can be somewhat random due to poor sampling. 

These differences in the gas evolution of dwarf galaxies between low and standard resolution may also be caused by the difficulty in setting up the initial conditions in a stable disc configuration. As described in Section~\ref{IC_sect}, we first run each galaxy model for an initial 300~Myr settling in period, during which the supernova feedback time-scales have been reduced. This enables the gas to settle into a stable disc, without an initial burst of star formation destroying the gas disc altogether. The simulation snapshot after 300~Myr is then used as the initial conditions for the main run with the full \textsc{chimes} chemistry model, and so the evolution shown in Fig.~\ref{SFH_resTest_fig} starts from this point and does not include the initial settling in period. However, we see that the gas fraction at time $t = 0 \, \rm{Myr}$ in m1e10 at low resolution (10 per cent) is much less than at standard resolution (60 per cent). This indicates that the initial gas disc in m1e10 at low resolution was disrupted during the initial settling in phase, and then re-accretes onto the galaxy over the following 200~Myr. 

At high resolution, the evolution of star formation rate and disc gas fraction in m3e10 is much closer to those at standard resolution, although the high resolution run retains higher gas fractions by up to $\approx$10 per cent. However, for the high resolution run we used the snapshot from the standard resolution simulation after the 300~Myr settling in period and increased the resolution by splitting each particle into eight, to avoid the computational expense of re-running the initial 300~Myr period again at higher resolution. This is why the high resolution run starts from the same gas fraction at $t = 0 \, \rm{Myr}$, which also reduces uncertainties in how we set up the initial conditions. 

The galaxy models at higher masses ($M_{200}$$\geq$$10^{11} \, \rm{M}_{\odot}$) show much closer agreement in the star formation rates and disc gas fraction between low and standard resolution. 

Fig.~\ref{H2_fraction_resTest_fig} compares the H$_{2}$ fraction versus neutral hydrogen column density, calculated as in Fig.~\ref{H2_fraction_fig}, from simulations at low (top panel), standard (middle panel), and high (bottom panel) resolution. The solid lines and shaded regions show the median and tenth to ninetieth percentile range of the H$_{2}$ fraction, respectively, from the simulations, while the data points show the observational data as described in Section~\ref{transition_sect}. 

The H$_{2}$ fractions in the high-mass galaxies (m3e11 and m1e12) are in good agreement between the low and standard resolution runs. However, in the m1e11 galaxy the H$_{2}$ fraction decreases more strongly towards low column densities ($<$10$^{21} \, \rm{cm}^{-2}$) at low resolution than at standard resolution. The dwarf galaxies m1e10 and m3e10 exhibit similar behaviour between the low and standard resolution runs. However, at high resolution the H$_{2}$ fraction in m3e10 is lower at low column densities ($<$10$^{21} \, \rm{cm}^{-2}$) than at standard resolution. 

These trends suggest that at lower galaxy masses (and with lower metallicities) we require higher resolution to correctly capture the H$_{2}$ fractions at low column densities. 

Fig.~\ref{global_HI_H2_resTest_fig} shows the H\textsc{i} to stellar mass and H$_{2}$ to stellar mass ratios in the left- and right-hand columns, respectively, as calculated in Fig.~\ref{global_HI_H2_fig}. The blue, grey and pink symbols show simulations run at low, standard and high resolution, respectively, while the solid curves, shaded regions and remaining data points indicate the observational data as described in Section~\ref{global_HI_H2_sect}. 

The H\textsc{i} and H$_{2}$ masses in the dwarf galaxies, m1e10 and m3e10, show large spreads of up to an order of magnitude between different resolutions, and also between different snapshots of the same resolution run. This broadly reflects the differences in the total disc gas fractions of the dwarf galaxies, as discussed above. In the high-mass galaxies m3e11 and m1e12, the H\textsc{i} masses are similar between the low and standard resolution. The H$_{2}$ masses of these galaxies are slightly higher at standard resolution than at low resolution, suggesting that the total molecular content may not be fully converged, however these differences are smaller than those seen in the dwarf galaxies. 

In Fig.~\ref{SFR_tracers_resTest_fig} we compare the emission line luminosity versus star formation rate relation, as in Fig.~\ref{SFR_tracers_fig}, for simulations run at low (blue symbols), standard (grey symbols) and high (pink symbols) resolution. The observational data, shown by the remaining symbols, are described in Section~\ref{line_prediction_sect}. 

In many cases the emission line predictions show good agreement between the different resolutions. However, there are several examples where this is not the case. In m3e10, the high resolution run exhibits luminosities of [C\textsc{ii}]$_{158 \rm{\mu m}}$, [O\textsc{i}]$_{63 \rm{\mu m}}$ and [N\textsc{ii}]$_{122 \rm{\mu m}}$ that are up to an order of magnitude higher than those at low and standard resolution. While the star formation rate in the high resolution run is also somewhat higher, this does not fully account for the increased luminosities of these three lines as the high resolution predictions lie above the observational data, in contrast to the low and standard resolution runs which are broadly in agreement with observations. The H$\alpha_{6563 \text{\AA}}$ luminosity is also higher in the high resolution run of m3e10, however this appears to be primarily due to the increased star formation run, as the high resolution predictions still follow the relation expected from observations in this case. 

In the high-mass galaxies, m3e11 and m1e12, the largest discrepancies can be seen in the [C\textsc{ii}]$_{158 \rm{\mu m}}$ luminosity, which is up to a factor $\approx$4 times higher at standard resolution than at low resolution, despite the star formation rate being unchanged. In contrast, the [C\textsc{ii}]$_{158 \rm{\mu m}}$ luminosity in m1e11 decreases by an order of magnitude from low to standard resolution. The luminosities of the other emission lines show better agreement between the low and standard resolution runs in the high-mass galaxies. 

Thus while some galaxies do exhibit significant differences in some of the emission lines between resolution levels, there is no systematic trend of particular emission lines increasing or decreasing in luminosity with numerical resolution for all galaxies. It is therefore unclear to what extent these differences might be driven, at least partially, by stochastic variations between runs. Extending our sample of galaxy models would help reduce these uncertainties. 

\label{lastpage}

\end{document}